# Exploiting Cross-Lingual Knowledge in Unsupervised Acoustic Modeling for Low-Resource Languages

FENG, Siyuan

A Thesis Submitted in Partial Fulfilment

of the Requirements for the Degree of

Doctor of Philosophy

in

Electronic Engineering

The Chinese University of Hong Kong

May 2020

*To my family.*

# Abstract

This thesis describes an investigation on unsupervised acoustic modeling (UAM) for automatic speech recognition (ASR) in the zero-resource scenario, where only untranscribed speech data is assumed to be available. UAM is not only important in addressing the general problem of data scarcity in ASR technology development but also essential to many non-mainstream applications, for examples, language protection, language acquisition and pathological speech assessment. The present study is focused on two research problems. The first problem concerns unsupervised discovery of basic (subword level) speech units in a given language. Under the zero-resource condition, the speech units could be inferred only from the acoustic signals, without requiring or involving any linguistic direction and/or constraints. The second problem is referred to as unsupervised subword modeling. In its essence a frame-level feature representation needs to be learned from untranscribed speech. The learned feature representation is the basis of subword unit discovery. It is desired to be linguistically discriminative and robust to non-linguistic factors. Particularly extensive use of cross-lingual knowledge in subword unit discovery and modeling is a focus of this research.

For unsupervised subword modeling, a two-stage system framework is adopted. The two stages are known as frame labeling and deep neural network bottleneck feature (DNN-BNF) modeling. Various approaches to speaker adaptation are applied to produce robust input features for frame labeling and DNN training. These approaches include feature-space maximum likelihood linear regression (fMLLR) assisted by out-of-domain (OOD) ASR, disentangled speech representation learning and speaker adversarial training. Experimental results on the Zero Resource Speech Challenge (ZeroSpeech) 2017 show that the fMLLR approach achieves the most significant performance improvement against the baseline system and further improvement could be attained by combining multiple adaptation approaches.

It is noted that the quality of frame labels has a significant impact on the DNN-BNF framework. The frame labeling approaches proposed in the thesis include Dirichlet process Gaussian mixture model-hidden Markov model (DPGMM-HMM) and OOD ASR decoding. A label filtering algorithm is developed to improve the quality of DPGMM-HMM frame labels. Multi-task learning with the DPGMM-HMM labels and OOD ASR labels is investigated. Experimental evaluation on the ZeroSpeech 2017 tasks demonstrates the advantage of DPGMM-HMM labels over DPGMM clustering labels. The best performance, which is achieved by combining multiple types of labels and BNFs, is comparable to the best submitted system to ZeroSpeech 2017.

Unsupervised unit discovery is tackled with the acoustic segment modeling (ASM) approach. Multiple language-mismatched phone recognizers are used to generate the initial segmentation of speech and the phone posteriorgram features of speech segments. A symmetric Kullback–Leibler (KL) divergence based distance metric is employed to analyze the linguistic relevance of discovered subword units. Experiments on a multilingual speech database show that the proposed methods achieve comparable performance to a previous study, with a simpler implementation. The KL divergence metric is consistent with the conventional measure of purity. The discovered subword units are found to provide a good coverage of the linguistically-defined phones. A few exceptions, e.g., /er/ in Mandarin, could be explained by the limited modeling capability of the adopted system. The confusion of a discovered unit between ground-truth phones can be alleviated by increasing the number of clusters.

# 摘要

本論文描述了針對零資源條件的非監督性聲學建模的研究。零資源條件假定語音數據不帶有任何標註。非監督性聲學建模不僅有助於解決語音識別技術發展中的數據量不足的問題，對於語音技術的很多非主流的應用也至為關鍵，例如語言保護，語言習得，病理語音診斷等。本項研究主要針對兩個問題。第一個問題是在制定目標語言中的基本語音單元的自動發現。在零資源條件下，由於缺乏最基本的語言學信息，基本語音單元的推斷只能依賴聲學信號本身。第二個問題是如何從不帶標註的語音數據學習一種幀層次的語音特征。該種特征會作為語音單元發現的基礎。理想的語音特征應當具有語言學區分度和對於非語言學因素的穩定性。本研究的重點之一為如何充分利用跨語言學知識幫助語音單元的發現和建模。

對於非監督性語音單元建模，本文建議採取的結構包含兩個步驟，即幀標註和深度神經網絡瓶頸特征（神經特征）建模。為了給幀標註和神經特征建模提供穩定的輸入特征，本文探討了多種說話人適應的方法，包括：利用非目標語言的自動語音識別系統獲得特征空間最大似然線性回歸特征，語音分離表達學習，和說話人對抗訓練。本文在零資源條件語音競賽 2017 的數據集上進行了實驗驗證。實驗結果表明，線性回歸特征獲得了最顯著的性能提升，而透過融合多種說話人適應的方法，系統性能可得到進一步提升。

幀標註的質量對於神經特征學習的成效有顯著的影響。本文提出了兩種幀標註的方法，包括：狄利克雷過程高斯混合模型-隱馬爾科夫模型（狄馬模型），和非目標語言的自動語音識別系統解碼。本文建議了一個過濾算法以提升狄馬標註的質量，並將上述兩種標註融合在多任務學習結構中。實驗結果證實狄馬標註相比於狄利克雷過程高斯混合模型標註具有優勢。本文中最優系統的性能與參加零資源條件語音競賽 2017 的最佳系統相當。

對於非監督性語音單元發現，本文採用聲學片段建模的方法，利用多個非目標語言的音素識別系統對目標語言進行片段分割和後驗概率特征提取。本文提出並利

摘要

用一種基於 KL 散度的距離度量，對自動發現的語音單元進行語言學關聯性分析。通過在多語言語音數據集上的實驗驗證，我們發現所提出的片段分割方法和過去的研究方法性能相近，而方法的實現更簡單。基於 KL 散度的距離標準和傳統的純度標準具有一致性。自動發現的語音單元相對語言學定義的語音單元整體有很好的涵蓋。對於一些未能發現的語音單元，例如普通話的元音/er/，我們認為這反映了目前系統建模的能力還有所不足。部分自動發現的語音單元在語言學角度造成混淆。而在非監督聚類中增加類別數可以有效降低混淆性。

# Acknowledgement

I would like to express my sincere gratitude to my supervisor, Prof. Tan Lee, for his patient guidance, thoughtful ideas and continued support and trust during my Ph.D. research. I would thank Prof. P. C. Ching and Prof. Wing-Kin Ma for their advice on my thesis research during group meetings and progress presentations.

I wish to thank Haipeng for introducing me into the area of speech processing and keeping giving me advice over the past four years, even though we literally had no overlap in time at the DSP Lab. I also would like to thank David for giving an excellent tutorial to the HTK toolkit and on the fundamentals of automatic speech recognition during my undergraduate internship. Thanks to Raymond for sharing experience in conducting scientific research in the academia and industry.

I would like to thank numerous DSP Lab colleagues. It was a wonderful time working with them. Special thanks to Michael for sharing his expertise in Linux operation and server machine maintanence. I am proud to be the one in charge of the server machine administration after Michael graduated. Thanks to Yuanyuan and Carol for their research advice as senior colleagues when I joined as a fresh student. Thanks to Ying and Xurong for keeping sharing thoughts and giving encouragement during the study period. I also express my thanks to Sammi, Herman, Shuiyang, Matthew, Zhiyuan, Yuzhong, Jiarui, Ryan, Mingjie, Dehua, Guangyan, Yatao, Gary, Wilson, etc. I sincerely appreciate Arthur, our technician, for taking care of equipment, software, schedule.

I am very grateful to be a member of DSP coffee and basketball interest groups. It was a pleasant time enjoying coffee with Herman, Sammi, Michael, Gary, etc. They taught me a lot in how to make excellent coffee. I will not forget the basketball games with Prof. Lee, Herman, Shuiyang, Dehua, Mingjie, Yatao etc.

Acknowledgement

I express my sincerest gratitude to my family, for their love and support.

# Contents

Contents

Contents

# List of Figures

# List of Tables

List of Tables

# Chapter 1

# Introduction

## 1.1  Low-resource speech modeling

Speech is an important and the most natural means of human communication. Speech technology has been one of the core areas in artificial intelligence (AI), paralleling computer vision and natural language processing. The major components of speech technology are automatic speech recognition (ASR), text-to-speech (TTS), speaker identification (SID) and speaker verification (SV), language identification (LID), voice conversion (VC), etc.

ASR refers to a computational process of identifying the linguistic content of human speech from an acoustic signal. It could be described as a non-linear mapping process from continuous speech to discrete linguistic items such as words. ASR is closely related to other topics of speech technology research, e.g. VC, LID and spoken term detection. A typical ASR system is made up of an acoustic model (AM), a language model (LM), a pronunciation lexicon and a decoder. The AM calculates the probability of an observed a speech utterance conditioned on a sequence of predefined speech units (phonemes). The LM and lexicon jointly produce the probability of a sequence of words, each comprising a string of phonemes, for the given language. The decoder is implemented of a search algorithm to determine the best-matching word sequence for the input utterance based on the constraints and probabilities given jointly by the AM, the LM and the lexicon.

AM is the core component in an ASR system. It connects speech and phonemes.

## 1. Introduction

Traditionally, hidden Markov models with state observation probability being modeled by mixture of Gaussian are adopted to represent individual phonemes or other subword units in speech [2]. This type of model is commonly referred to as GMM-HMM. In recent years, the deep neural network (DNN) is shown to be more powerful than the GMM in modeling acoustic variation in speech, especially when abundant training are data available [3]. More recently, there is an increasing trend in modeling speech directly at word level [4]. This approach does not require the use of pronunciation lexicon and explicit phoneme-level modeling. The AM and LM probabilities[1] are estimated jointly with a single network structure known as the connectionist temporal classification (CTC) [5]. It is also referred to as one type of end-to-end approaches [6].

AM training is usually considered to be a process of supervised learning. The training involves a large number of speech utterances and their transcriptions. The transcription tells what is being spoken in the respective utterance, and hence facilitates an ordered alignment between speech and the model. Training a high-performance DNN-HMM AM for a large-vocabulary application requires hundreds to thousands of hours of transcribed speech. In the Big Data era, collecting a good amount of speech data for a popular language like English and Mandarin is not difficult. However, acquiring or preparing transcriptions for training is far less straightforward and may involve unaffordable human effort. There exists an excessive amount of untranscribed speech for popular languages, and they could be exploited to further improve the AM.

There are about $7,000$ spoken languages in the world [7]. For most of them, the amount of transcribed speech data is very limited, or even non-existent. Many of these languages, e.g. ethnic minority languages in China and languages in Africa, may have never been formally studied. Many languages have very limited amount of transcribed speech data, or have no transcribed data. Knowledge about their linguistic properties is usually incomplete, or unavailable. Transcribing speech of these languages is difficult and impractical. Without transcription, conventional supervised acoustic modeling could not be applied straightforwardly. Therefore the investigation of unsupervised acoustic modeling methods has become a focal point

---

[1]Here we borrow terminologies from GMM-HMM/DNN-HMM to denote the probabilities. This framework does not have individual acoustic or language model anymore.

of research with significant application impact.

Low-resource speech modeling has gained increasing research interest in recent years. The term *low-resource* is used to refer to a range of scenarios when limited linguistic knowledge and limited amount of transcribed speech are available. In the related research literature, the low-resource assumption generally refers to the following two cases,

1. Limited transcribed data and large amount of untranscribed data;

2. Untranscribed data only.

In the first scenario, a straightforward idea is to train an initial AM with the transcribed data, followed by refining inferred transcriptions and re-training AM and/or LM in an iterative manner [8]. The inferred transcriptions are obtained by decoding with the updated models, while model retraining is carried out with the new transcriptions. A key research problem is how to effectively utilize the untranscribed data. Previous studies investigated semi-supervised learning methods to improve the seed AM with untranscribed data [9, 10].

The second scenario, which presents a highly restrictive data condition, is the major focus of this thesis. It is commonly known as the *zero-resource* condition, as virtually no knowledge about the language concerned is assumed to be available [11]. Languages to be modeled satisfying the zero-resource assumption are named as *zero-resource* languages.

Building AMs the zero-resource language is much more difficult than in the first scenario. Since there is no prior knowledge on the phoneme inventory, i.e. its size and constituents[2], the first step is to construct a hypothesized set of fundamental speech units in an unsupervised manner. It is expected that these hypothesized units are closely related to those linguistically-defined phonemes of the target language. Once the speech units are defined and modeled, they can be used to tokenize input speech for the target language, and generate phoneme-like pseudo transcriptions for downstream tasks e.g. zero-resource ASR, spoken term discovery.

Speech modeling in the zero-resource scenario has significant impact in the broad area of speech and language research. It is widely acknowledged that infants start to learn the first language primarily from speech interaction with their parents.

---

[2]Some closely related studies [12, 13] assume phoneme inventory size of a zero-resource language is known. However, this thesis assumes both size and components are unknown.

This learning process is considered to be minimally supervised or unsupervised in nature [14]. Research in zero-resource speech modeling could potentially help understand infant language acquisition mechanism. Zero-resource speech modeling could potentially be meaningful to the protection of endangered languages. With automatically discovered speech units, it would become possible to systematically document raw speech data of an endangered language. With properly organized speech data, linguists could analyze and derive the linguistic structure of the language in an objective and evidence-based way.

The application of low-resource speech modeling is not limited to those unpopular languages that genuinely lack general linguistic knowledge. One possible extension is toward the modeling of pathological speech, e.g. caused by voice, articulation and cognitive disorders. Pathological speech could be regarded as being low-resource for several reasons. First, collection of pathological speech data is more difficult than that of normal speech. As a matter of fact, in the research area of automatic assessment of pathological speech, shortage of training data has long been a major concern. Second, people with pathological symptoms tend to produce speech sounds that deviate greatly from the "norm". Some of the sounds may not even carry linguistic meanings. Third, there exist numerous types of diseases that may cause speech and language impairments. These diseases may also be co-existing. It is difficult, if not impossible, to develop feasible annotation or transcription schemes to properly deal with the specificities of impairments, and perform standardized transcriptions for acoustic modeling training. From this perspective, low-resource speech modeling approaches can be applied to tackle problems in pathological speech processing.

## 1.2   Focus of this research

The research presented in this thesis is focused on acoustic modeling for ASR in the zero-resource scenario, i.e. only untranscribed speech data are available, while linguistic knowledge about the target language is completely absent.

Unsupervised discovery of basic speech units is one of the key problems in this investigation. It is the first step in acoustic modeling when the modeling units are unknown or uncertain. We aim at phone-level acoustic modeling similar to conven-tional GMM-HMM and DNN-HMM models in ASR. Since the granularity and the size of phoneme inventory for a zero-resource language are unknown, supervised acoustic modeling algorithms cannot be directly applied to zero-resource languages.

Separating linguistic information from linguistically-irrelevant information solely based on raw untranscribed speech is the major difficulty in robust unsupervised unit discovery. In the ideal case, automatically discovered speech units could constitute the ground-truth phonemes of the target language, and are in good consistency with these phonemes. In practice, a wide variety of linguistically-irrelevant variations co-exist in the speech signal, e.g. speaker, emotion, channel. They are encoded with linguistic content in the acoustic signal, and are not easily separable. In supervised acoustic modeling, manually annotated transcription can be relied on to support robust discovery of acoustic units towards these irrelevant variations. In the concerned zero-resource scenario, speech units can only be inferred from acoustic signals. This may significantly affect the accuracy of unit discovery. For instance, speech sounds carrying the same phoneme but produced by different speakers might be mistakenly modeled as different speech units, due to the effect of speaker difference [15]. To alleviate this problem, a possible research direction would be to learn a feature representation that could support subword or word identification both across- and within-speakers.

Another focus of the thesis is on unsupervised learning of frame-level speech features that could discriminate fundamental speech units and are robust to linguistically-irrelevant variations. A good feature representation that differentiates linguistic information from non-linguistic information is essential and crucial in unit discovery. It is expected that the learned feature representation is superior to conventional spectral feature representations such as MFCCs or PLPs for unsupervised discovery of fundamental speech units.

Increasing the amount of data is known of being beneficial to acoustic modeling. For a resource-rich language, training data is by nature from the target language. For a zero-resource language, while in-domain data is scarce, there exist abundant resources for out-of-domain languages. This thesis investigates a variety of approaches to exploiting out-of-domain mismatched language resources in improving unsupervised unit discovery and feature representation learning. While each language has its distinctive linguistic properties, the speech sounds that can be produced in different languages may have significant overlap, because the basic mechanism of speech production is largely language-independent. Open-source speech corpora containing hundreds of speakers and hundreds to thousands of hours of transcribed speech are available for English and Mandarin, etc. The use of cross-lingual speaker and language resources to facilitate unsupervised acoustic modeling for zero-resource languages is extensively studied.

In a strict sense, out-of-domain resources are assumed unavailable in zero-resource scenarios. One may argue that unsupervised acoustic modeling for ASR should assume no access to any out-of-domain speech and language resources. From this perspective, 'unsupervised learning' is different from what is being considered in this thesis. Nevertheless, in practice, certain types of resources from major languages are often available. The *Unsupervised acoustic modeling* problem tackled in this thesis can be considered as an extension to the strict-sense unsupervised task.

## 1.3 Problem statement

### 1.3.1 Unsupervised subword modeling

Unsupervised subword modeling refers to the problem of learning frame-level feature representation that is discriminative to fundamental speech units, and robust to linguistically-irrelevant variations, under the assumption that only untranscribed speech data is available for training.

The problem was formally defined in the Zero Resource Speech Challenge (ZeroSpeech) 2015, a world-wide challenge encouraging low-resource speech modeling. The follow-up challenge, ZeroSpeech 2017 [14], continued to focus on this problem. This problem encourages direct performance comparison of different subword models and makes the performance unaffected by the quality of back-end decoders. It is noted that previous studies on unsupervised subword modeling were evaluated in different means, such as transcriptions, lattices, posteriorgrams, thus direct comparison was unavailable.

It was shown that speaker variation is the major difficulty in robust unsupervised subword modeling. This thesis considers speaker change as the main component of linguistic-irrelevant variation.

## 1.3.2 Unsupervised unit discovery

Unsupervised unit discovery is the problem of automatically discovering basic speech units of a language, assuming only untranscribed speech are available. The goal is to build AMs to cover the entire phoneme inventory of the target language. The outcome of a unit discovery system is the tokenization of untranscribed speech data with time alignments, using the discovered units as tokens. This problem was extensively studied. It was known in various terms, such as unsupervised acoustic modeling [16], self-organized unit (SOU) modeling [17, 18], acoustic unit discovery (AUD) [19, 20], acoustic model discovery [21] or unsupervised lexicon discovery [22, 23].

The results of unsupervised speech unit discovery are symbolic indices that do not convey explicit linguistic functions or meanings, e.g. vowels, fricatives etc. These units may represent either speech or non-speech patterns such as phonemes, noises and pauses. Performance measurements used in supervised acoustic modeling, such as word error rates (WERs) and phoneme error rates (PERs), are seldom used in unsupervised unit discovery, since there is no knowledge on the definition of word or phoneme. In practice, unsupervised unit discovery is measured in terms of the relevance of tokenization to golden time-aligned phoneme transcription [24]. Evaluation metrics such purity, F-score and normalized mutual information (NMI) are commonly used [25].

## 1.3.3 Relation between the two problems

The two problems mentioned above are closely related. Unsupervised subword modeling can be considered as a front-end optimization process for robust unsupervised unit discovery. Unsupervised unit discovery is the goal of performing unsupervised subword modeling. Besides, Unsupervised unit discovery could also be relied on to improve unsupervised subword modeling.

Specifically, a good feature representation that captures subword discriminative information and is robust to speaker variation has been shown beneficial to unit discovery [26]. On the other hand, a set of discovered units having good consistency with true phonemes of a language could provide phoneme-like speech transcriptions to assist subword discriminative feature learning [27].

### 1.3.4 Relation to other problems

The two problems tackled in this thesis are closely related to other speech tasks. For instance, unsupervised subword modeling can be applied to query-by-example spoken term detection (QbE-STD) for low-resource languages. QbE-STD is aimed at detecting audio documents from an archive which contain a specified spoken query [28]. Typically a QbE-STD system involves two steps, constructing acoustic feature representation for both query and audio documents, followed by computing the likelihood of the query occurring somewhere in the audio archive [29]. Feature representation plays an important role in QbyE-STD performance. Under the low-resource scenario, linguistic knowledge about the audio archive is assumed unknown. Phonetically-discriminative features learned by unsupervised subword modeling are expected to provide a desired representation for QbyE-STD.

Unsupervised unit discovery is considered as a building block for developing a complete unsupervised ASR system. The basic acoustic units learned by unit discovery serve as the basis for lexical modeling, where repetitive sequences of acoustic units are modeled by a hypothesized word-like unit. After generating work-like patterns and lexicon, an ASR system for the target language can be built.

## 1.4 Thesis outline

Chapter 2 provides a review on the problems of unsupervised subword modeling and unsupervised unit discovery. For unsupervised subword modeling, considerations on model structures, input feature types and frame labeling approaches are discussed. Three modeling frameworks for unsupervised unit discovery are also reviewed.

Chapter 3 describes the DNN-bottleneck feature (BNF) framework for unsupervised subword modeling. It defines the baseline system in this study. The objectives, database and evaluation metric of ZeroSpeech 2017 Challenge are introduced. The database and the evaluation metric are used throughout the thesis.

Chapter 4 is focused on applying speaker adaption to unsupervised subword modeling. Experiments on ZeroSpeech 2017 are described, with comparison to the baseline system in Chapter 3. Combination of these approaches is also studied.

Chapter 5 describes the proposed approaches to improving clustering-based frame labeling in unsupervised subword modeling. Two different types of frame labels are generated. Experimental results on ZeroSpeech 2017 are given.

Chapter 6 deals with the problem of unsupervised unit discovery with the acoustic segment modeling (ASM) framework, and presents proposed evaluation metric to analyze linguistic relevance of discovered subword units. Experiments on OGI multilingual telephone speech corpus are presented.

Chapter 7 concludes this thesis, summarizes the main contributions and suggests directions for future work.

# Chapter 2

# Related works

This chapter provides a detailed literature review on the two research problems concerned, namely, unsupervised subword modeling and unsupervised unit discovery.

## 2.1   Unsupervised subword modeling

Unsupervised subword modeling is formulated as a problem of feature representation learning. The key issue is how to retain linguistic-relevant information and suppress non-linguistic variation of speech signals by learning from a large amount of data. This problem has gained increasing research interest in recent years. Relevant works were mostly conducted on the ZeroSpeech 2015 and 2017 challenge datasets and evaluation metrics, which facilitate informative performance comparison.

Deep neural networks (DNNs) are widely investigated in unsupervised subword modeling. Typically a DNN model is trained using given speech data. The learned features are obtained either from a designated low-dimension hidden layer of the DNN, known as the bottleneck features (BNFs) [30], or from the softmax output layer, known as the posterior features or posteriorgram [31]. Various DNN structures have been investigated, as discussed in Section 2.1.1.

Feature learning can be realized with other machine learning techniques. This approach demonstrated competitive performance to DNN-based methods [27, 32]. Clustering of frame-level features is the key step, aiming at generating a number of frame clusters. Each cluster desirably corresponds to a discovered subword unit. By representing each cluster with a probability distribution, such as Gaussian distribution, a cluster posteriorgram can be constructed to be the learned representation [27].

Previous studies suggested that there are two key issues contributing to the performance of unsupervised subword modeling. They are the discriminability of input features, and the fitness of labels. The issues are crucial to DNN- and non-DNN-based models. Investigations on these two issues in previous studies are reviewed in Sections 2.1.2 and 2.1.3, respectively.

### 2.1.1 DNN structures

The DNN model structures proposed for feature learning are divided into three categories based on the training strategy, namely, supervised, unsupervised and weakly/pair-wise supervised.

Supervised models, e.g. multi-layer perceptron (MLP), were widely used in unsupervised subword modeling [1,30,31,33]. Training of these models requires labels of speech frames or segments, which are typically phone identities. In the resource-rich scenario, transcriptions can be used to generate frame labels. Whilst the acquisition of frame labels becomes a challenging problem in the zero-resource scenario. It is highly desired to derive some kinds of initial frame labels that have good correspondence to ground-truth phone alignments. With these initial labels, supervised model training can be applied and the trained model can be used to generate updated labels of input speech. BNFs or posteriorgram extracted from the trained models could be used as subword-discriminative representation. There were also attempts to eliminating the need for frame label acquisition [1, 34]. Typically, a DNN AM was trained with transcribed speech from an out-of-domain language, and used to generate BNFs or posteriorgrams.

Unsupervised neural network models, e.g. auto-encoder (AE) [31,35], denoising AE (dAE) [36] and variational AE (VAE) [37], do not require any kind of target labels for training. These models are trained to learn a compact intermediate-layer representation, which is capable of reconstructing the input representation (or denoised input representation as in dAE). The encoded representation is known as *embedding* and is able to retain linguistic information that are needed to reconstruct input speech. The embedding is also expected to be free of linguistically-irrelevant variation, and hence intrinsically suitable for subword modeling. In [36], AE and dAE were compared on the unsupervised subword modeling task. In [31], the AE was compared with supervised models. In [37], the vector-quantized VAE (VQ-VAE) was used to disentangle linguistic content and speaker characteristics and achieved better performance than AE. Although training of unsupervised models relies on less stringent data requirement, as it does not need transcription, the performance of unsupervised models is in general not as good as supervised ones. This is probably due to the fact that in the absence of supervision, it is relatively harder to separate subword-discriminative information from non-linguistic information.

Pair-wise supervised models, such as correspondence AE (cAE) [36] and siamese network [38], are useful because, by leveraging the knowledge about speech segment pairs which correspond to the same linguistic unit (word or subword) [39–41], such pair-wise relational information provides top-down constraints to guide linguistic unit discrimination [40]. Compared with self-supervision in AEs, pair-wise supervision provides additional information on linguistically-irrelevant variations . This knowledge is beneficial to robust subword modeling. In [36], the cAE was trained with pairs of segments representing the same linguistic units. In [38], the siamese network was trained with pairs of input features. The objective was to determine whether the two segments correspond to the same linguistic unit or not. In practice, pair-wise information may not be directly available for low-resource languages. Unsupervised term discovery (UTD) [42] was suggested as a feasible approach to obtaining such information.

## 2.1.2 Input features

Input feature representation plays a critical role in unsupervised subword modeling. Early studies used conventional spectral features like MFCCs [27], filter banks (FBanks) [35] and PLPs [43]. These features have been widely used in ASR acoustic modeling. Nevertheless, it is widely understood that spectral features are not optimized for unsupervised subword modeling [14], as they contain abundant linguistically-irrelevant variations caused by, e.g. speaker and noise, which may negatively impact the modeling [30, 44].

It was found that linear transforms estimated from spectral features are useful to improve frame clustering for subword modeling to a great extent. These transforms reduce linguistically-irrelevant variations encoded in speech, meanwhile retaining linguistic information. In [44, 45], linear discriminant analysis (LDA) and maximum likelihood linear transform (MLLT) demonstrated noticeable improvement. LDA and MLLT are widely used in supervised acoustic modeling. LDA minimizes intra-class discriminability and maximize inter-class discriminability of the speech features, while MLLT de-correlates feature components [44]. In [44, 46], feature-space maximum likelihood linear regression (fMLLR) based speaker adaptive training (SAT) were found to achieve further improvement over LDA+MLLT transforms. The estimation of LDA, MLLT and fMLLR require transcribed speech. In [44–46], supervised data were obtained by a two-pass frame clustering. In [30], speaker-normalized MFCC with vocal tract length normalization (VTLN) [47] was found better than raw MFCCs.

Input feature representation also plays a critical role in DNN-based unsupervised subword modeling. In [48], deep scattering features were shown to outperform FBanks as input features to siamese network training. Deep scattering features are stable and can be efficiently exploited by classifiers, while retaining richer information than FBanks [48]. In [1, 34], fMLLRs estimated by an out-of-domain language-mismatched ASR system were shown to be better than MFCCs as input features. The observations in [1, 34, 48] demonstrated the importance of input feature selection in DNN acoustic modeling. DNN acoustic models are preferred for its high-level representation learning capability, i.e., retaining subword-discriminative information in input features and suppressing non-linguistic information to a great extent. As a result, little attention was paid to the selection of DNN input features for unsupervised acoustic modeling, except for the aforementioned studies. In this thesis, approaches to learning input features to improve DNN acoustic modeling as well as frame clustering are extensively studied.

### 2.1.3 Frame labeling

Frame labeling is an important step in unsupervised subword modeling. It aims to create subword-like tokenization of untranscribed speech, with which feature learning models can be trained. Frame labels are mainly used for supervised DNN acoustic model training [30], as well as for posteriorgram generation [32, 49]. Frame labeling approaches are divided into two categories, namely, frame clustering and out-of-domain ASR decoding.

Frame clustering is a process of grouping together speech frames that have similar feature representations. Clustering-based frame labeling approaches assume that speech frames belonging to the same ground-truth subword unit are closer in the frame-level feature space than those belonging to different units. Under this assumption, cluster indices assigned to speech frames are expected to constitute subword-level time alignment of untranscribed speech. Various clustering algorithms were investigated in the literature. In [27], Dirichlet process Gaussian mixture model (DPGMM) [50] was applied to frame clustering, and demonstrated superior performance in unsupervised subword modeling. The system described in [27] achieved the best performance in ZeroSpeech 2015 [11]. DPGMM does not require a predefined number of clusters, which makes it suitable for frame clustering, as the number of subword units in a low-resource language is usually unknown. The success of DPGMM motivated subsequent studies on unsupervised subword modeling [30, 33, 44, 45]. Other clustering algorithms were also studied, e.g. GMM-universal background model (GMM-UBM) [49] and its variant, hidden Markov model-UBM (HMM-UBM) [49], where HMM training was applied after obtaining GMM clustering results. Note that the aforementioned DPGMM algorithm differs from GMM-UBM only in the prior distribution. Besides, $k$-means was also studied in the concerned task [51]. However, their results show that DPGMM is by far the most effective algorithm for speech frame clustering.

Exploiting out-of-domain resources for frame labeling is considered as a *transfer learning* approach. It can be realized in different manners. In [1], the ASR system for an out-of-domain resource-rich language was utilized to decode target speech in the language-mismatched manner, and generate frame labels from decoding lattices. In this way, target untranscribed speech are tokenized by the phoneme inventory of the out-of-domain language. While each language has its distinctive linguistic properties, the speech sounds that can be produced in different languages may have significant overlap, because the basic mechanism of speech production is largely language-independent [52].

## 2.2 Unsupervised unit discovery

Unsupervised unit discovery aims at finding acoustically homogeneous basic speech units, which are desirably equivalent to subword units or phonemes, from untranscribed speech data. An unsupervised unit discovery system typically involves three sub-problems, namely, speech segmentation, segment labeling, and subword modeling. Despite the variety of system frameworks reported in the literature, the formulation of these sub-problems is widely shared in common [21, 53, 54].

There are three types of models applied to unsupervised unit discovery, i.e., acoustic segment models (ASMs), nonparametric inference models and top-down constraint models. These models have all been investigated extensively.

### 2.2.1 Acoustic segment model (ASM)

The acoustic segment model (ASM) was first proposed by Lee et al. [53]. This work has shown far-reaching impact on subsequent studies in unsupervised unit discovery. In the ASM approach, initial segmentation, segment labeling and iterative acoustic modeling are performed in a sequential manner. During initial segmentation, each speech utterance is divided into a sequence of variable-length segments. The estimated segment boundaries are desirably in synchrony with ground-truth phoneme boundaries. Subsequently, speech segments with similar acoustic characteristics are grouped and labeled by the same index or symbol. In this way, the segment labels constitute a form of tokenization of input speech. The tokenization for a speech utterance is comprised of a sequence of symbols with time boundaries. It could be regarded as a phone-level time-aligned transcription of input speech. With this phone-level transcription, conventional techniques of acoustic model training can be applied to obtain hypothesized subword models. These subword models can in turn be used to decode the training speech into updated transcriptions. In short, the subword model training and the generation of transcription are carried out iteratively. In [18], a similar approach to ASM was proposed. The discovered subword units were referred to as self-organized units (SOUs).

There have been follow-up studies aiming to improve individual stages of the ASM framework. Most of them were focuses on better initial segmentation and seg-ment labeling. The task of initial segmentation is to obtain subword boundaries without using any prior knowledge about speech content of the input utterance. Lee et al. [53] applied a dynamic programming-based maximum-likelihood estimation approach to speech segmentation which was first proposed in [55]. Qiao et al. [56] developed a bottom-up hierarchical algorithm, which was later applied in other studies [25]. Pereiro Estevan et al. [57] proposed a maximum-margin clustering algorithm. Scharenborg et al. [58] extended [57] by proposing a two-step method that is able to use a mix of bottom-up information from speech signals and top-down information. Torbati et al. [59] suggested to use the Bayesian HMM with DP prior. In [60], segment boundaries were detected by locating the peaks on a spectral transition measure. Vetter et al. [61] and our previous study [52] exploited out-of-domain cross-lingual ASR systems to obtain hypothesized phoneme boundaries. These two studies were presented almost at the same time. Michel et al. [62] presented a blind phoneme segmentation method that designates peaks in the error curve of a model trained to predict speech frame by frame as potential boundaries.

Segment labeling is a clustering problem by nature. Speech segments derived from initial segmentation are first represented by fix-dimension feature vectors. These vectors are used as inputs to perform clustering. The resulting cluster indices are regarded as segment labels. In [53], the fixed-dimension representation of variable-length speech segments was obtained by performing vector quantization (VQ). Siu et al. [18] investigated a segmental GMM approach to speech segment clustering. Segmental GMMs use a polynomial function to approximate the trajectory of speech features within a segment. Wang et al. [63] applied GMM to cluster and label speech segments. In this approach, each segment is labeled with the index of the Gaussian component which provides the highest likelihood of the speech segment. This approach was further extended to the use of more sophisticated clustering algorithms, namely and direct segment clustering [64]. In [16], variants of GCC were investigated and discussed. Different objective functions, constraint formulations and similarity measures were compared. Wang et al. [25] investigated the use of spectral clustering and its combination with GCC in segment labeling, and reported improved performance compared to using GCC only.

At the stage of iterative acoustic modeling, GMM-HMM [18, 53] and DNN-

HMM [65] were adopted. With these models, Viterbi decoding can be performed to update the segment labels and time alignment. The iterative process generally results in better tokenization and subword models [25].

### 2.2.2 Nonparametric inference model

In addition to the ASM based approaches, nonparametric inference has been investigated extensively for unsupervised unit discovery [19–21, 66–68]. In [21], a Bayesian model was applied to perform speech segmentation, segment clustering and subword modeling, as a joint optimization process. These three tasks are arranged in a sequence in ASM. The nonparametric Bayesian model infers subword units, segment frames and cluster segments in an integrated manner. The Gibbs sampling (GS) method was used for parameter inference.

The Bayesian model [21] was refined in a broad range of investigations. For instance, in the work by Ondel et al. [20], variational Bayes (VB) was used to replace GS, so as to enable inference on a large-scale speech dataset. It was shown that VB is better than GS in terms of both speed and accuracy. A follow-up work [66] done by Ondel et al. extended [20] by applying a nonparametric Bayesian language model to better utilize contextual information in speech. Another follow-up work [67] investigated the use of *informative prior* in the VB model [20]. The informative prior is obtained from out-of-domain resource-rich language resource. Ebbers et al. [68] extended the works in [18, 20, 21] by replacing GMM with VAE in order to leverage the capability of VAE in modeling sophisticated emission distribution. The proposed model was named HMM-VAE, which was applied in a series of subsequent works [19, 67]. In [19], the HMM-VAE model is embedded in a Bayesian framework with a DP prior for the distribution of the acoustic units. In this way, the number of subword units to be discovered does not need to be pre-defined, as DP automatically determines the optimal number. In a most recent work by Ondel et al. [54], Bayesian Subspace GMM (SGMM) model was proposed to restrict the unit discovery system in modeling phonetic content in speech, and ignore linguistic-irrelevant variations such as speaker and noise. In order to estimate phonetic space, the models were trained with transcribed speech of resource rich languages. The underlying assumption is made that the phonetic subspace of the target low-resource language and a resource-rich language could be close, given that these two languages have phones in common.

### 2.2.3 Top-down constraint model

The ASM and the nonparametric inference models described above are related in that they both aim at discovering subword units, and start modeling at frame level. They could be regarded as *bottom-up* approaches [15], which clearly have drawbacks. In the review article [15], it was concluded that bottom-up methods tend to over-cluster the realization variants of the same phonetic identity. Such fine clusters may reflect the variation caused by speaker difference, change of speaking style and environment, etc. As a result, it is hard for bottom-up modeling approaches to discriminate whether two speech realizations belong to the same phone or not. On the other hand, unsupervised discovery of word-size units is considered less ambiguous than subword-like ones [69]. The similarity between realizations of a word spoken by different speakers is much more prominent than that of a phoneme [15]. It is expected that word-level information, if available, could be exploited as top-down supervision or constraints to subword unit discovery [40,69]. This reflects the intrinsic structure of speech, which embraces and encodes multi-scale information, ranging from low-level subwords (phonemes) to higher-level syllables, words etc.

There were a few attempts to incorporating word-level information as top-down constraints to assist unsupervised unit discovery [69–72]. Under the zero-resource assumption, word entity information is unavailable. One of the possible approaches is to assume that same-different word pair information is available or at least obtainable [69, 73]. Jansen et al. [69] tackled subword unit discovery by first generating same-different word pair information via a spoken term discovery system [42], followed by applying the word-pair supervision to guide the partition of a pre-estimated UBM into different GMMs. Each of the GMMs corresponded to a learned subword unit. In the UBM partitioning process, word-pair information was used to construct a similarity matrix of UBM components, with which spectral clustering was adopted to cluster UBM components. The results in [69] demonstrated the usefulness of word-pair information as top-down constraints in subword unit discovery.

Another approach to exploiting top-down constraints in unsupervised unit discovery is to simultaneously model subword- and word-like units. Chung et al. [71]

presented a two-level acoustic pattern discovery model which jointly discover sub-word and word units. The system comprises three stages that correspond to the optimization of acoustic model, language model and pronunciation lexicon respectively. Word-like acoustic patterns were discovered by identifying subword-like patterns appearing frequently together. Top-down constraints were constructed by learning a lexicon of word-like acoustic patterns so that each word unit consisted of a sequence of subword units. In this model, top-down constraints and subword unit discovery were iteratively updated. Their experimental results revealed that word-like unit constraints were useful in subword unit discovery. The approaches in [71] was later applied in Chung et al. [72] as speech tokenizers to tackle both unsupervised unit discovery and feature representation learning.

# Chapter 3

# Unsupervised subword modeling: a DNN-BNF framework

This chapter introduces a DNN-BNF modeling framework to tackle unsupervised subword modeling, and defines the task of frame-level feature representation learning. The DNN-BNF framework serves as the baseline in this thesis. Our proposed approaches to robust unsupervised subword modeling as discussed in Chapters 4 and 5 are all based on this framework.

This chapter also describes the database and evaluation metric for ZeroSpeech 2017 Track 1: unsupervised subword modeling [14], which are used in system performance evaluation throughout this thesis.

This Chapter is organized as follows. Section 3.1 provides an overview of the DNN-BNF modeling framework. Frame labeling and supervised DNN-BNF modeling are presented in Sections 3.2 and 3.3. In Section 3.4, an overview of the ZeroSpeech 2017 database is given. Section 3.5 introduces the evaluation metric of ZeroSpeech 2017. Experiments and discussion on the baseline system is shown in Section 3.6.

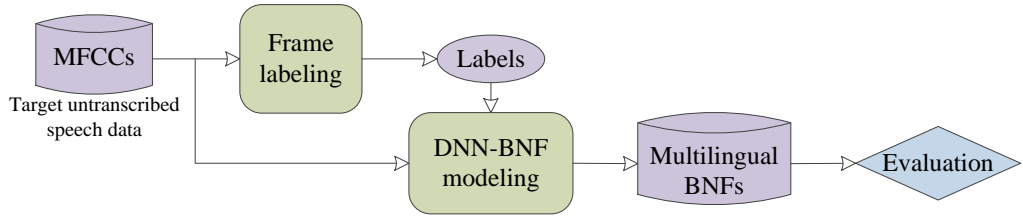

**Figure 3.1:** DNN-BNF framework for unsupervised subword modeling.

# 3.1 Overview of DNN-BNF framework

The DNN-BNF modeling framework consists of two stages, namely, frame labeling and supervised DNN-BNF modeling. Frame labels obtained in the first stage provide supervision for DNN training in the second stage. The trained DNN is used to extract BNFs as the subword-discriminative representation of target zero-resource speech. The general framework is illustrated in Figure 3.1. The DNN-BNF modeling framework was investigated in many previous studies [27, 30], and achieved very good performance in ZeroSpeech 2015 [11] and 2017 [14].

# 3.2 Frame labeling

Frame labeling is an essential step to prepare the target untranscribed speech for supervised DNN-based subword modeling. The generated frame labels that have good correspondence to ground-truth phoneme time-alignments are highly desired. A possible way to perform frame labeling is by clustering. DPGMM clustering [50] has been widely applied for frame labeling [27, 30, 45]. It is adopted in our system.

## 3.2.1 Definition of DPGMM

DPGMM is an an non-parametric Bayesian extension to GMM, where a Dirichlet process prior replaces the vanilla GMM. DPGMM does not require a pre-defined cluster number. This makes DPGMM intrinsically suitable for unsupervised subword modeling, as the number of subword units contained in a zero-resource language is usually unknown.

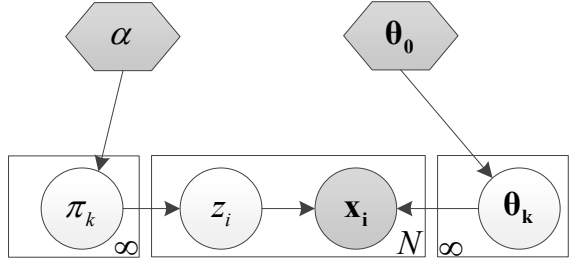

**Figure 3.2:** Graphical illustration of DPGMM.

The graphical illustration of DPGMM is shown in Figure 3.2. Given a set of observations $\{\boldsymbol{x_1}, \boldsymbol{x_2}, \ldots, \boldsymbol{x_N}\}$, DPGMM assumes that they are generated by a random process,

1. Mixture weights $\boldsymbol{\pi} = \{\pi_k\}_{k=1}^{\infty}$ are generated according to a stick-breaking process with concentration parameter $\alpha$ [74].

2. GMM parameters $\boldsymbol{\theta} = \{\boldsymbol{\theta_k}\}_{k=1}^{\infty}$ are generated according to Normal-inverse-Wishart (NIW) distribution with parameters $\boldsymbol{\theta_0}$ [75].

3. A discrete label $z_i$ is sampled from all the Gaussian components $\{k\}_{k=1}^{\infty}$ according to the distribution of $\boldsymbol{\pi}$.

4. $\boldsymbol{x_i}$ is drawn from the $z_i$-th Gaussian component $\boldsymbol{\theta_{z_i}}$.

Here, $\boldsymbol{\theta_k} = \{\boldsymbol{\mu_k}, \boldsymbol{\Sigma_k}\}$ consists of mean and covariance parameters of the $k$-th Gaussian component. $\boldsymbol{\theta_0} = \{\boldsymbol{m_0}, \boldsymbol{S_0}, \kappa_0, \nu_0\}$ consists of parameters of NIW where $\boldsymbol{m_0}$ and $\boldsymbol{S_0}$ are prior means for $\boldsymbol{\mu_k}$ and $\boldsymbol{\Sigma_k}$ respectively, $\kappa_0$ and $\nu_0$ are belief-strenths in $\boldsymbol{m_0}$ and $\boldsymbol{S_0}$ respectively.

## 3.2.2   Inference of DPGMM

Inference of DPGMM has been investigated by a number of studies [76–80]. The methods include Markov chain Monte Carlo (MCMC) based sampling [76, 77] and variational inference [78–80], etc. In this thesis, a parallelizable split and merge sampler [50] is adopted for two main reasons. First, this sampler explicitly represents the mixture weights $\boldsymbol{\pi}$ for GMM posteriorgram computation [27], which is needed in our frame labeling process. Second, this sampler is parallelizable, making it scalable to a large amount of speech frames (e.g. a 20-hour speech database typically comprises

$7.2 \times 10^6$ frames) [27, 32].

The parallelizable split and merge sampler conducts inference by alternating between **restricted DPGMM Gibbs sampling** and **split/merge sampling**.

## Restricted DPGMM Gibbs sampling

Restricted DPGMM Gibbs sampling assumes a discrete label $z$ is sampled from the finite number $(K)$ of Gaussian components. Let us denote the set of discrete labels as $\mathcal{Z} = \{z_1, z_2, \ldots, z_N\}$. The posterior sampling of $\boldsymbol{\pi}$ is denoted as,

$$\boldsymbol{\pi} = (\pi_1, \pi_2, \ldots, \pi_K, \pi'_{K+1}) \sim \text{Dir}(N_1, N_2, \ldots, N_K, \alpha), \tag{3.1}$$

$$\pi'_{K+1} = 1 - \sum_{k=1}^{K} \pi_k, \tag{3.2}$$

$$N_k = \sum_{i=1}^{N} \delta(z_i = k), \tag{3.3}$$

where $\alpha$ could be understood as the relative probability of assigning an observation with a new Gaussian component label, as opposed to an existing label. The sampling of $\{\boldsymbol{\theta_1}, \boldsymbol{\theta_2}, \ldots, \boldsymbol{\theta_K}\}$ is denoted as,

$$\boldsymbol{\theta_k} = \{\boldsymbol{\mu_k}, \boldsymbol{\Sigma_k}\} \stackrel{\propto}{\sim} \text{NIW}(\boldsymbol{m_k}, \boldsymbol{S_k}, \kappa_k, \nu_k), k \in \{1, 2, \ldots, K\}, \tag{3.4}$$

where $x \stackrel{\propto}{\sim} y$ represents sampling $x$ from distribution proportional to $y$. The parameters of NIW are computed as,

$$\kappa_k = \kappa_0 + N_k, \tag{3.5}$$

$$\nu_k = \nu_0 + N_k, \tag{3.6}$$

$$\boldsymbol{m_k} = \frac{\kappa_0 \boldsymbol{m_0} + N_k \bar{\boldsymbol{x}}_{\boldsymbol{k}}}{\kappa_k}, \tag{3.7}$$

$$\boldsymbol{S_k} = \boldsymbol{S_0} + \sum_{i=1}^{N} \delta(z_i = k)(\boldsymbol{x_i}\boldsymbol{x_i}^{\intercal} + \kappa_0 \boldsymbol{m_0}\boldsymbol{m_0}^{\intercal} - \kappa_k \boldsymbol{m_k}\boldsymbol{m_k}^{\intercal}), \tag{3.8}$$

where $\bar{\boldsymbol{x}}_{\boldsymbol{k}}$ is computed as,

$$\bar{\boldsymbol{x}}_{\boldsymbol{k}} = \frac{\sum_{i=1}^{N} \delta(z_i = k)\boldsymbol{x_i}}{\sum_{i=1}^{N} \delta(z_i = k)}, \tag{3.9}$$

and could be interpreted as the mean of observation data that are assigned by label $k$. Note that Equations (3.5) to (3.8) were applied in [81] to perform maximum a posteiori (MAP) estimation towards GMM parameters, with the application of model adaptation in GMM-HMM based ASR systems. The sampling of $z_i$ is denoted as,

$$z_i \overset{\propto}{\sim} \sum_{k=1}^{K} \pi_k \mathcal{N}(\boldsymbol{x_i}|\boldsymbol{\mu_k}, \boldsymbol{\Sigma_k}) \mathbb{1}(z_i = k), \tag{3.10}$$

where $\mathbb{1}(z_i = k)$ is a $K$-dimension one-hot vector whose $z_i$-th dimension is 1.

## Split/merge sampling

Split/merge sampling consists of two steps, namely, splitting Gaussian component into 2 sub-components, and Metropolis-Hastings split/merge.

In the first step, each Gaussian component is split into sub-components with mixture weights $\boldsymbol{\tilde{\pi}_k} = \{\tilde{\pi}_{kl}, \tilde{\pi}_{kr}\}$ and parameters $\boldsymbol{\tilde{\theta}_k} = \{\boldsymbol{\tilde{\theta}_{kl}}, \boldsymbol{\tilde{\theta}_{kr}}\}$. The observation data $\boldsymbol{x_i}$ is assigned with a sub-component label $\tilde{z}_i \in \{l, r\}$. The following steps are used to sample sub-components:

$$\{\tilde{\pi}_{kl}, \tilde{\pi}_{kr}\} \sim \text{Dir}(N_{kl} + \frac{\alpha}{2}, N_{kr} + \frac{\alpha}{2}), \tag{3.11}$$

$$\boldsymbol{\tilde{\theta}_{kl}} \overset{\propto}{\sim} \mathcal{N}(\boldsymbol{x_{kl}}|\boldsymbol{\tilde{\theta}_{kl}}) \text{NIW}(\boldsymbol{\tilde{\theta}_{kl}}|\boldsymbol{\theta_0}), \tag{3.12}$$

$$\boldsymbol{\tilde{\theta}_{kr}} \overset{\propto}{\sim} \mathcal{N}(\boldsymbol{x_{kr}}|\boldsymbol{\tilde{\theta}_{kr}}) \text{NIW}(\boldsymbol{\tilde{\theta}_{kr}}|\boldsymbol{\theta_0}), \tag{3.13}$$

$$\tilde{z}_i \overset{\propto}{\sim} \sum_{i=1}^{N} \delta(\tilde{z}_i = s) \tilde{\pi}_{z_i, s} \mathcal{N}(\boldsymbol{x_i}|\boldsymbol{\tilde{\theta}_{z_i, s}}), s \in \{l, r\}, \tag{3.14}$$

where $N_{kl}$ and $N_{kr}$ are defined as,

$$N_{kl} = \sum_{i=1}^{N} \delta(z_i = k)\delta(\hat{z}_i = l), \tag{3.15}$$

$$N_{kr} = \sum_{i=1}^{N} \delta(z_i = k)\delta(\hat{z}_i = r), \tag{3.16}$$

and could be interpreted as the number of observation data belonging to $z_i$ and assigned with sub-component label $l$ and $r$ respectively.

In the second step, split or merge moves in a Metropolis-Hastings (MH) fashion is proposed [27]. In the following description, the notation $\hat{A}$ denotes the proposal for the variable $A$. Let $Q \in \{Q_{split_c}, Q_{merge_{m,n}}\}$ denote proposal move selected randomly from *split* or *merge*. $Q_{split_c}$ denotes splitting Gaussian component $c$ into $m$ and $n$. $Q_{merge_{m,n}}$ denotes merging components $m$ and $n$ into $c$. Conditioned on $Q_{split_c}$, variables are sampled as,

$$(\hat{\mathcal{Z}}_m, \hat{\mathcal{Z}}_n) = \text{split}_c(\mathcal{Z}, \tilde{\mathcal{Z}}), \tag{3.17}$$

$$(\hat{\pi}_m, \hat{\pi}_n) = \pi_c \boldsymbol{\pi_{sub}}, \boldsymbol{\pi_{sub}} = (\pi_m, \pi_n) \sim \text{Dir}(\hat{N}_m, \hat{N}_n), \tag{3.18}$$

$$(\hat{\boldsymbol{\theta}_m}, \hat{\boldsymbol{\theta}_n}) \sim q(\hat{\boldsymbol{\theta}_m}, \hat{\boldsymbol{\theta}_n} | \mathcal{X}, \hat{\mathcal{Z}}, \tilde{\hat{\mathcal{Z}}}), \tag{3.19}$$

$$(\hat{\tilde{\boldsymbol{v}}_m}, \hat{\tilde{\boldsymbol{v}}_n}) \sim p(\hat{\tilde{\boldsymbol{v}}_m}, \hat{\tilde{\boldsymbol{v}}_n} | \mathcal{X}, \hat{\mathcal{Z}}). \tag{3.20}$$

Conditioned on $Q_{merge_{m,n}}$, variables are sampled as,

$$\hat{\mathcal{Z}}_c = \text{merge}_{m,n}(\mathcal{Z}), \tag{3.21}$$

$$\hat{\pi}_c = \hat{\pi}_m + \hat{\pi}_n, \tag{3.22}$$

$$\hat{\boldsymbol{\theta}_c} \sim q(\hat{\boldsymbol{\theta}_c} | \mathcal{X}, \hat{\mathcal{Z}}, \tilde{\hat{\mathcal{Z}}}), \tag{3.23}$$

$$\hat{\tilde{\boldsymbol{v}}_c} \sim p(\hat{\tilde{\boldsymbol{v}}_c} | \mathcal{X}, \hat{\mathcal{Z}}). \tag{3.24}$$

The function $\text{split}_c(\cdot)$ splits the labels of Gaussian component $c$ according to the assignment of sub-components, $\text{merge}_{m,n}(\cdot)$ merges labels of components $m$ and $n$. With the Hasting ratio $H$ computed as suggested in [50], the proposed split/merge moves are accepted with probability $\min\{1, H\}$ in a Metropolis-Hastings MCMC framework.

### 3.2.3 Frame labeling based on DPGMM clustering

DPGMM is applied to perform frame labeling towards untranscribed speech data. Let us consider $M$ zero-resource languages. For the $i$-th language, frame-level MFCC features are denoted as $\{\boldsymbol{o_1^i}, \boldsymbol{o_2^i}, \ldots, \boldsymbol{o_T^i}\}$, where $T$ is the total number of frames. By applying DPGMM clustering towards $T$ frames, $K$ Gaussian components $\boldsymbol{\theta}$ together with their mixture weights $\boldsymbol{\pi}$ are obtained to represent $K$ clusters of frame-level features. The frame-level labels $\{l_1^i, l_2^i, \ldots, l_T^i\}$ are obtained by

$$l_t^i = \underset{1 \leq k \leq K}{\arg\max} \, P(k | \boldsymbol{o_t^i}), \tag{3.25}$$

where $P(k|\boldsymbol{o_t^i})$ denotes the posterior probability of $\boldsymbol{o_t^i}$ with respect to the $k$-th Gaussian component, which is computed as,

$$P(k|\boldsymbol{o_t^i}) = \frac{\pi_k \mathcal{N}(\boldsymbol{o_t^i}|\boldsymbol{\mu_k}, \boldsymbol{\Sigma_k})}{\sum_{j=1}^{K} \pi_j \mathcal{N}(\boldsymbol{o_t^i}|\boldsymbol{\mu_j}, \boldsymbol{\Sigma_j})}. \tag{3.26}$$

These frame labels are referred to as *DPGMM labels* in this thesis.

## 3.3 Supervised DNN-BNF modeling

### 3.3.1 BNF representation by DNN

BNF representation refers to output representation of a designated low-dimension DNN hidden layer. This low-dimension layer is usually named as the BN layer [82]. BNF has been shown able to provide a compact and phonetically-discriminative representation of speech, and suppress linguistically-irrelevant variation e.g. speaker change [82]. It has been widely applied in conventional acoustic modeling tasks [83–85]. Recently, BNFs were also studied in the zero-resource scenario [30, 31, 33]. In our study, BNF representation is adopted as the frame-level feature representation for unsupervised subword modeling.

BNF representation is learned in the second stage of our DNN-BNF modeling framework. In this stage, a supervised DNN model is trained with speech features to predict their corresponding DPGMM labels. The training procedure is similar to that of a conventional DNN acoustic model [3]. The only difference is that DPGMM labels, instead of HMM forced-alignments, are used as the supervision for DNN training. During the extraction of BNFs, speech features are fed into the DNN input till the BN layer. The BNF representation is generated as subword-discriminative feature representation for target zero-resource languages.

In principle, the choice of hidden layer structures is flexible, such as feed forward, time-delay, convolutional, LSTM, etc. In practice, the feed forward structure is most commonly used in the concerned task [30, 31], probably because MLP models are easier to be trained than CNN and LSTM with limited amount of data. In this thesis, MLP is adopted as the DNN architecture in the baseline system.

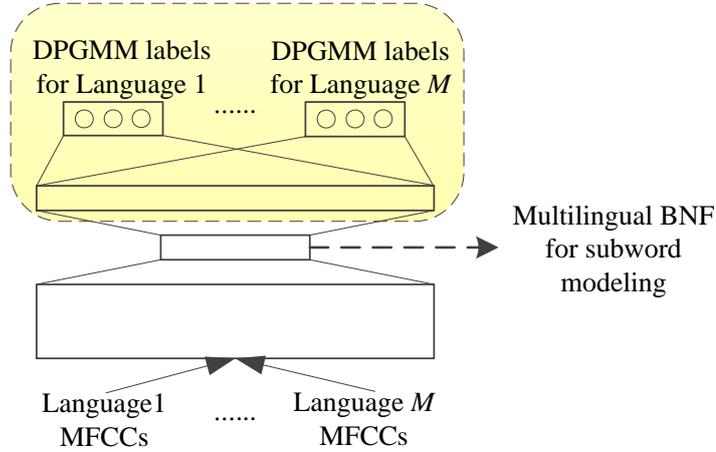

**Figure 3.3:** MTL-DNN model used to extract multilingual BNF.

## 3.3.2 Multilingual BNF representation

If there are multiple zero-resource languages' untranscribed speech to be modeled, the DNN model can be trained in a multilingual manner [86]. Past works have shown that multilingual BNFs outperform monolingual ones in representing speech [85]. This can be partially explained that multilingual DNN-BNF modeling leverages a wider range of phonetic diversity, and that the amount of data for DNN training is increased.

Multilingual BNFs can be generated through multi-task learning DNN (MTL-DNN) modeling. MTL [87] is a commonly adopted strategy to learn multilingual acoustic models [86] and BNFs [30]. The structure of MTL-DNN adopted in the baseline system is illustrated in Figure 3.3. There are in total $M$ tasks in the MTL-DNN, each corresponding to a target zero-resource language. The hidden layers, including the BN layer, are shared across all tasks, while the output layers are task (language)-specific. The input features for the $M$ languages are merged before training the MTL-DNN. The loss function of MTL-DNN is weighted cross-entropy, which is defined as,

$$\mathcal{L} = \sum_{i=1}^{M} \omega^i \mathcal{L}^i(\boldsymbol{o_{1 \to T}^i}, l_{1 \to T}^i, \theta), \tag{3.27}$$

where $\omega^i$ and $\mathcal{L}^i$ are the weight and the cross-entropy loss of the $i$-th task, $\theta$ denotes parameters of the MTL-DNN model, $\boldsymbol{o_{1 \to T}^i}$ and $l_{1 \to T}^i$ are input speech features and corresponding labels related to the $i$-th task. $\mathcal{L}^i$ of an arbitrary $T$-frame speech segment $\{\boldsymbol{o_1^i}, \boldsymbol{o_2^i}, \ldots, \boldsymbol{o_T^i}\}$ is computed as,

$$\mathcal{L}^i = -\sum_{t=1}^{T}\sum_{k=1}^{K} \delta(l_t^i = k) \log P_{\text{DNN}}(k|\boldsymbol{o_t^i}), \tag{3.28}$$

where $\delta(\cdot)$ is the indicator function, $P_{\text{DNN}}(k|\boldsymbol{o_t^i})$ is the $k$-th element of the MTL-DNN softmax output corresponding to the $i$-th task. The probability distribution $P_{\text{DNN}}(\cdot|\cdot)$ is parameterized by the MTL-DNN.

During training, given a pair of input feature $\boldsymbol{o_t^i}$ and its DPGMM label $l_t^i$, shared hidden layers and the output layer corresponding to the $i$-th language are updated, while the other output layers are unchanged. After training, layers of the MTL-DNN inside the dashed box in Figure 3.3 are discarded, and the remaining structure of the MTL-DNN is used as the feature extractor to generate multilingual BNFs.

## 3.4 ZeroSpeech 2017 database

The database in ZeroSpeech 2017 Track 1 [14] consists of *development data* and *surprise data*. The development data are provided with open-source evaluation software and ground-truth phoneme alignment information, which enables immediate performance evaluation. The surprise data are not provided with such software and information publicly. In this thesis, experiments are conducted on development data.

The development data of ZeroSpeech 2017 covers three target languages, namely English, French and Mandarin. Although the ultimate goal of our research is to develop systems that perform well on real-world zero-resource languages, it is nevertheless reasonable to adopt resource-rich languages for experimental purposes, provided that during training phase transcriptions and linguistic knowledge for these languages are assumed unavailable.

For each of the three languages in ZeroSpeech 2017 development data, there are separate training set and test set of untranscribed speech. Speaker identity information is provided for the train sets but not available for the test sets. The test data are organized into subsets of different utterance lengths: 1 second, 10 second and 120 second. Detailed information about the dataset are given as in Table 3.1.

**Table 3.1:** Development data of ZeroSpeech 2017 Track 1.

| | Training | | | Test |
|---|---|---|---|---|
| | Duration | No. speakers-L[1] | No. speakers-R[1] | Duration |
| English | 45 hrs | 60 | 9 | 27 hrs |
| French | 24 hrs | 18 | 10 | 18 hrs |
| Mandarin | 2.5 hrs | 8 | 4 | 25 hrs |

Note that the amounts of training data for the three target languages cover a large diversity. This is aimed to measure the degree to which the proposed systems are sensitive to the amount of training data.

## 3.5  ZeroSpeech 2017 evaluation metric

The evaluation metric adopted for ZeroSpeech 2017 Track 1 task is the ABX subword discriminability. Inspired by the match-to-sample task in human psychophysics, it is a simple method to measure the discriminability between two categories of speech units [11]. The basic ABX task is to decide whether $X$ belongs to $x$ or $y$, if $A$ belongs to $x$ and $B$ belongs to $y$, where $A$, $B$ and $X$ are three data samples, $x$ and $y$ are the two pattern categories concerned. The performance evaluation in ZeroSpeech 2017 is carried out on the triphone minimal-pair task. A triphone minimal pair comprises two triphone sequences, which have different center phones and identical context phones, for examples, "beg"-"bag", "api"-"ati". Discriminating triphone minimal pairs is a non-trivial task. The performance of a feature representation on the triphone minimal-pair ABX task is considered a good indicator of its efficacy in speech modeling [48].

Let $x$ and $y$ denote a pair of triphone categories. Consider three speech segments $A$, $B$ and $X$, where $A$ and $X$ belong to category $x$ and $Y$ belongs to $y$. The ABX discriminability of $x$ from $y$ is measured in terms of the ABX error rate $\epsilon(x, y)$, which is defined as the probability that the distance of $A$ from $X$ is greater than that of

---

[1]speakers-L/-R denotes speakers with rich/limited amount of speech data

$B$ from $X$, i.e.,

$$\epsilon(x,y) = \frac{1}{|S(x)|(|S(x)|-1)|S(y)|} \sum_{A \in S(x)} \sum_{B \in S(y)} \sum_{X \in S(x) \setminus \{A\}}$$

$$(\mathbb{1}_{d(A,X)>d(B,X)} + \frac{1}{2}\mathbb{1}_{d(A,X)=d(B,X)}), \tag{3.29}$$

where $S(x)$ and $S(y)$ denote the sets of features that represent triphone categories $x$ and $y$, respectively. $d(\cdot, \cdot)$ denotes the dissimilarity between two speech segments, which is computed by dynamic time warping (DTW) in our study. The frame-level dissimilarity measure used for DTW scoring is the cosine distance. The function $\mathbb{1}_{d(A,X)>d(B,X)}$ has the value 1 if $d(A,X) > d(B,X)$ is satisfied, otherwise its value is 0. Note that $\epsilon(x,y)$ is asymmetric to $x$ and $y$. A symmetric form can be defined by taking average of $\epsilon(x,y)$ and $\epsilon(y,x)$. The overall ABX error rate is obtained by averaging over all triphone categories and speakers in the test set. A high ABX error rate means that the feature representation is not discriminative, and vice versa. Intuitively, the error rate should be no larger than 50%, as by random decision, the expectation of ABX error rate is 50%.

There are two evaluation conditions defined in ZeroSpeech 2017, namely *within-speaker* and *across-speaker*. In both conditions, the segments $A$ and $B$ to be evaluated are generated by the same speaker. In the within-speaker condition, segment $X$ is generated by the same speaker as $A$ and $B$; In the across-speaker condition, $X$ is generated by a speaker different from $A$ and $B$.

## 3.6 Experiments

### 3.6.1 Experimental setup

Frame labeling by DPGMM clustering is implemented by an open-source tool developed by Chang et al. [50]. Speech frames for the three target languages are first processed by extracting 13-dimension MFCC features with cepstral mean normalization (CMN) and augmented with $\Delta$ and $\Delta\Delta$ to form a 39-dimension feature representation. The window length and step size for computing MFCCs are 25ms and 10ms respectively. Frame-level MFCC features are clustered by the DPGMM al-gorithm for each of three languages. The numbers of clustering iterations are $200, 240$ and $3000$ for English, French and Mandarin, respectively. The resulted numbers of DPGMM clusters are $1554, 1541$ and $381$. Each speech frame is assigned a DPGMM label.

The MTL-DNN model is trained with MFCCs+CMN spliced by $\pm 5$ contextual frames. There are 3 training tasks in MTL, each corresponding to a target language. The shared-hidden-layer (SHL) neural network structure is {1024, 1024, 1024, 1024, 40, 1024}. The dimensions of the language-specific output layers are $1554, 1541$ and $381$, as determined by the outcome of DPGMM clustering. All the layers are feed-forward. The activation function for SHLs is Sigmoid, except the 40-dimension linear BN layer. The weighted cross-entropy criterion is chosen as the objective function. Stochastic gradient descent (SGD) is used to optimize the objective function. The three tasks are assigned the equal weights in training. A 10% subset of training data is randomly chosen for cross-validation. The learning rate is 0.008 at the beginning of the training phase, and is halved when no improvement is observed on the cross-validation data. The mini-batch size is 256. After training, MTL-DNN is used to extract multilingual BNFs for test sets of target languages. The BNFs are further evaluated by the ABX subword discriminability task. The training of MTL-DNN and extraction of BNFs are implemented by Kaldi [88].

## 3.6.2 Results and analyses

The experimental results of ABX subword discriminability on multilingual BNFs are shown as in Tables 3.2 (across-speaker) and 3.3 (within-speaker). In addition, the ABX performance on raw MFCC features, supervised phone posteriorgram (provided by the Challenge organizers [14]), and multilingual BNFs proposed in [30], are also listed in the Tables as reference.

It can be observed from Tables 3.2 and 3.3 that the multilingual BNF representation obtained in our baseline system achieves significant improvements in both across- and within-speaker ABX discriminability tasks, as compared to the raw MFCC feature representation. The relative ABX error rate reduction is 40.3% in across-speaker condition, and 26.7% in within-speaker condition. It is also observed that multilingual BNFs perform consistently better on 10s and 120s test lengths than on 1s. On the contrary, MFCC features perform the same over different lengths. This can be explained by that MTL-DNN input features are MFCCs with CMN, while CMN is known to be ineffective for short utterances.

**Table 3.2:** Across-speaker ABX error rates (%) on the MFCC features, supervised phone posteriorgram and multilingual BNFs of ZeroSpeech 2017.

| | English | | | French | | | Mandarin | | | Avg. |
|---|---|---|---|---|---|---|---|---|---|---|
| | 1s | 10s | 120s | 1s | 10s | 120s | 1s | 10s | 120s | |
| MFCC [14] | 23.4 | 23.4 | 23.4 | 25.2 | 25.5 | 25.2 | 21.3 | 21.3 | 21.3 | 23.3 |
| Phone posteriorgram (supervised topline) [14] | 8.6 | 6.9 | 6.7 | 10.6 | 9.1 | 8.9 | 12.0 | 5.7 | 5.1 | 8.2 |
| Multilingual BNF [30] | 13.7 | 12.1 | 12.0 | 17.6 | 15.6 | 14.8 | 12.3 | 10.8 | 10.7 | 13.3 |
| Multilingual BNF | 13.5 | 12.4 | 12.4 | 17.8 | 16.4 | 16.1 | 12.6 | 11.9 | 12.0 | 13.9 |

**Table 3.3:** Within-speaker ABX error rates (%) on the MFCC features, supervised phone posteriorgram and multilingual BNFs of ZeroSpeech 2017.

| | English | | | French | | | Mandarin | | | Avg. |
|---|---|---|---|---|---|---|---|---|---|---|
| | 1s | 10s | 120s | 1s | 10s | 120s | 1s | 10s | 120s | |
| MFCC [14] | 12.0 | 12.1 | 12.1 | 12.5 | 12.6 | 12.6 | 11.5 | 11.5 | 11.5 | 12.0 |
| Phone posteriorgram (supervised topline) [14] | 6.5 | 5.3 | 5.1 | 8.0 | 6.8 | 6.8 | 9.5 | 4.2 | 4.0 | 6.2 |
| Multilingual BNF [30] | 8.5 | 7.3 | 7.2 | 11.1 | 9.5 | 9.4 | 10.5 | 8.5 | 8.4 | 8.9 |
| Multilingual BNF | 8.0 | 7.3 | 7.3 | 10.3 | 9.4 | 9.3 | 10.1 | 8.8 | 8.9 | 8.8 |

The multilingual BNF representation performs worse than phone posteriorgram over all the test utterance lengths and languages. This is reasonable, as the topline representation assumes transcribed training speech available during system development. The proposed multilingual BNFs are slightly better than that in [30] under the within-speaker condition, and slightly worse under the across-speaker condition. The system framework proposed in [30] is similar to our baseline. However, since the implementation details are publicly available, we are not able to reproduce the results.

### 3.6.3 Discussion

The baseline system with multilingual BNF representation has shown significant advancement over conventional spectral features, without requiring any in-domain or out-of-domain supervision information, e.g. transcription and phoneme inventory. Nevertheless, the present system design has a few shortcomings. For instance, MFCC feature as input features to DPGMM and MTL-DNN is not optimal, as it is known to depend significantly on non-linguistic factors such as speaker, emotion, channels. To address this issue, various approaches can be adopted, including speaker adaptation. Indeed MFCC features are affected significantly by speaker variation. This can be seen from the performance gap between across- and within-speaker ABX error rates in Tables 3.2 and 3.3. Speaker adaptation is an important and challenging problem particularly in acoustic modeling in the zero-resource scenario. While in the supervised scenario, reliable transcription can be used to ensure robustness of the learned subword units towards speaker change. In the unsupervised scenario, subword units can only be inferred from speech features.

On the other hand, DPGMM frame labels for DNN-BNF modeling in the baseline system could be improved in several aspects. DPGMM clustering assumes that neighboring speech frames are independent of each other. This is not in accordance with the nature of speech. To address this limitation, modeling of temporal dependency could be incorporated into the process of frame label acquisition. Improvement on frame labels could also be achieved by exploiting out-of-domain language-mismatched ASR systems.

# Chapter 4

# Speaker adaptation for unsupervised subword modeling

Speaker adaptation is important in unsupervised subword modeling. Speech data usually contains a diverse range of speaker variation. Removing speaker variation is beneficial to subword modeling. This chapter presents various speaker adaptation approaches that are applied to learning speaker-invariant feature representations for unsupervised subword modeling. These approaches include,

(a) fMLLR estimation by an out-of-domain ASR;

(b) disentangled speech representation learning, separating linguistic content and speaker characteristics into disjoint parts of speech representation;

(c) speaker adversarial training.

They can be illustrated by the system flow diagrams as shown in Figures 4.1a, 4.1b and 4.1c. These approaches can be directly applied to the baseline system mentioned in Chapter 3.

Generally speaking, all these approaches can be considered as learning to transform speech features from an original unadapted space to a speaker-adapted space. In terms of the type of transform, the approaches can be classified into two categories, namely, linear transform and nonlinear transform. FMLLR is a linear transform-

## 4. Speaker adaptation for unsupervised subword modeling

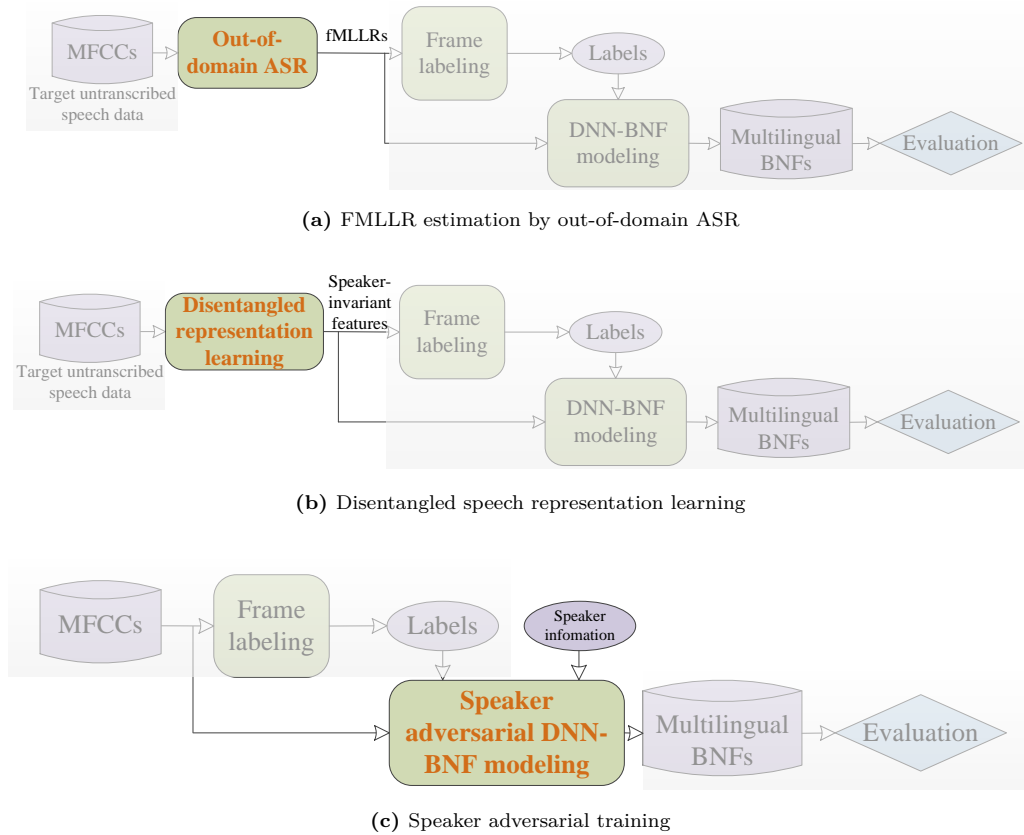

**(a)** FMLLR estimation by out-of-domain ASR

**(b)** Disentangled speech representation learning

**(c)** Speaker adversarial training

**Figure 4.1:** Three speaker adaptation approaches applied to improve subword modeling. In the above three sub-figures, shallow components are the same as in the baseline framework discussed in Chapter 3.

based approach, while disentangled representation learning and speaker adversarial training are both based on DNNs, which realizes nonlinear transforms of speech features. As shown in Figure 4.1, fMLLR and disentangled representation learning operate at front-end level, i.e., on MFCC or other types of unadapted features, while speaker adversarial training is applied to the back-end DNN-BNF modeling. From the perspective of resource utilization, fMLLR estimation exploits additional out-of-domain resources, i.e., transcribed speech data from resource-rich languages. On the contrary, disentangled representation learning and speaker adversarial training do not rely on any out-of-domain resources. A summary of properties of the approaches is given as in Table 4.1.

Section 4.1 describes fMLLR-based speaker adaptation based on an out-of-domain ASR system. Section 4.2 introduces disentangled speech representation learn-ing. Section 4.3 discusses speaker adversarial training. In Section 4.4, we investigate the combined use of speaker adaptation approaches. Section 4.5 describes the experiments to validate and evaluate the effectiveness of different approaches, as well as their combination. Section 4.6 summarizes the experimental results and comparison of all the three approaches.

**Table 4.1:** Summary of properties of the three speaker adaptation approaches. 'OOD' stands for out-of-domain.

| Approach ID | Transform type | Front/back-end | Require OOD resource |
|:-----------:|:--------------:|:--------------:|:--------------------:|
| (a) | Linear | Front | ✓ |
| (b) | Nonlinear | Front | ✗ |
| (c) | Nonlinear | Back | ✗ |

# 4.1 FMLLR by out-of-domain language mismatched ASR

We propose to apply fMLLR-based speaker adaptation method in the concerned task. This is motivated by the study achieving the best performance in ZeroSpeech 2017 [32]. Estimation of fMLLR features requires transcription. In [32], clustering-based pseudo transcription is generated beforehand to enable fMLLR estimation. In our thesis, an out-of-domain language-mismatched ASR system is exploited to estimate fMLLR features for in-domain low-resource speech. For major languages such as English and Chinese, large-scale speech corpora that include hundreds of speakers are available for training high-performance ASR systems [89, 90]. The richness of speaker diversity in these out-of-domain speech corpora could be leveraged to learn robust features representation of low-resource speech data.

FMLLR [91, 92] is a feature-based speaker adaptation method. It is effective in improving speaker invariance of speech features. FMLLR has been successfully applied in conventional acoustic modeling for large vocabulary ASR systems [93–95]. The general idea of fMLLR is to estimate speaker-specific linear transforms to reduce inter-speaker variability. By applying the transforms to MFCCs, the speaker-dependent features are mapped to a speaker-independent feature space. The trans-formed feature representation is referred to as fMLLR.

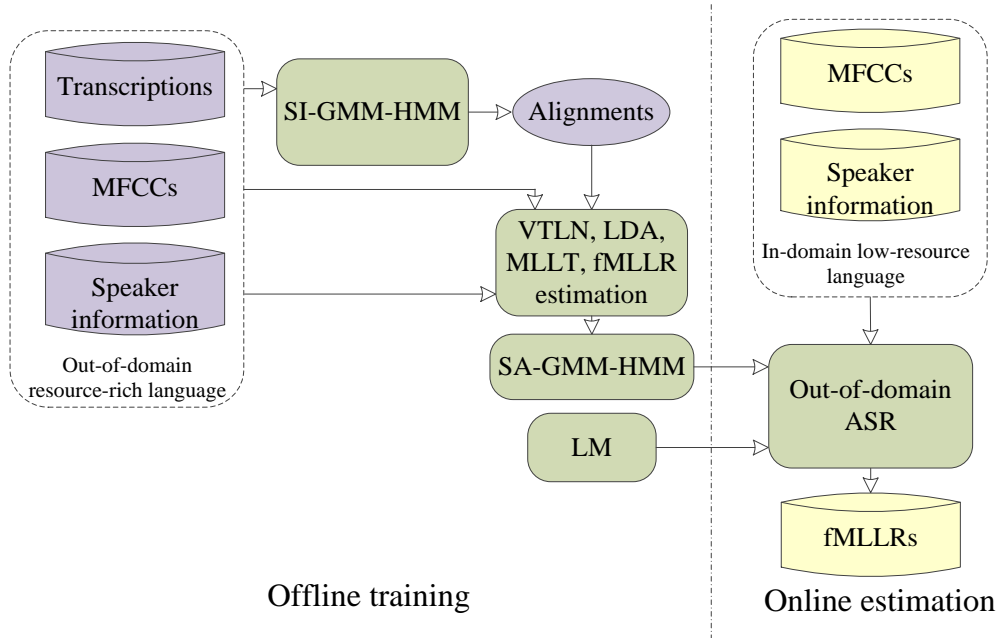

**Figure 4.2:** Process flow of utilizing an out-of-domain language-mismatched ASR to estimate in-domain fMLLR features.

Figure 4.2 depicts the process flow of utilizing an out-of-domain language-mismatched ASR to estimate in-domain fMLLR features. Given the out-of-domain training speech and transcriptions, speaker-independent GMM-HMM (SI-GMM-HMM) AMs are estimated with MFCCs as input features. The AMs are used to forced-align the out-of-domain training data to provide the supervision required for applying vocal tract length normalization (VTLN) [47], linear discriminant analysis (LDA) [96], maximum likelihood linear transforms (MLLT) [97] and fMLLR in a sequential manner. Subsequently, the fMLLR features are used to train speaker-adapted GMM-HMM AMs (SA-GMM-HMM). An LM is trained using transcriptions. The SA-GMM-HMM AMs and LM are used to build the out-of-domain ASR system. This ASR system is used to decode in-domain utterances and finds the best path for each utterance. Each path comprises a sequence of phone labels with time boundary information. The best paths are regarded as alignments and used in the estimation of fMLLR features for in-domain data.

## 4.2   Disentangled representation learning

Speaker-invariant features could be generated via disentangled representation learning and transformation. The main idea is to separate linguistically irrelevant factors, e.g. speaker, emotion etc. from linguistic information, e.g., phoneme, which are simultaneously encoded in the speech signal. Subsequently, a frame-level representation is constructed to mainly embed the linguistic part.

Separating linguistic content and speaker characteristics in speech is a non-trivial task in the zero-resource scenario [15]. In this study, we adopt the following assumption as proposed in [98]:

- Speaker characteristics tend to have a smaller amount of variation than linguistic content within a speech utterance, and linguistic content tends to have similar amounts of variation within and across utterances.

Intuitively and simplistically, this could be understood as the contrast that speaker identity affects the fundamental frequency (F0) at utterance level, and linguistic content affects spectral characteristics at segment level. Based on this assumption, the factorized hierarchical VAE (FHVAE) [99] is adopted for disentangling speaker and linguistic information in speech.

The FHVAE model was first proposed by Hsu et al. [99]. It is an unsupervised generative model extended from the VAE [100]. FHVAE learns to factorize sequence-level and segment-level attributes of sequential data (e.g. speech) into different latent variables, as compared to the VAE in which a single latent representation is learned. In the present study, the sequence- and segment-level attributes refer to speaker characteristics and linguistic content respectively. By discarding or unifying sequence-level latent representation, speaker-invariant features can be learned with FHVAE. This is a straightforward approach to learning speaker-invariant features. Its effectiveness has been proved in domain adaptation tasks, such as noise robust ASR [98], distant conversational ASR [101], and dialect identification [102]. This motivates us to apply this approach to learning speaker-invariant features in the unsupervised scenario.

### 4.2.1 FHVAE model

In FHVAE, the generation process of sequential data is formulated by imposing sequence-dependent priors and sequence-independent priors to different sets of variables. The overall structure of FHVAE is illustrated as in Figure 4.3a. Following notations and terminologies in [99], let $z_1$ and $z_2$ denote the latent segment variable and the latent sequence variable, respectively, $\mu_2$ be the sequence-dependent prior, known as *s-vector*. $\theta$ and $\phi$ denote the parameters of generative and inference models of FHVAEs. Let $\mathcal{D} = \{X^i\}_{i=1}^{M}$ denote a speech dataset containing $M$ sequences. This sequence $X^i$ contains $N^i$ speech segments $\{x^{(i,n)}\}_{n=1}^{N^i}$. Each segment $x^{(i,n)}$ contains a fixed number of frames. The FHVAE model generates $X$ from a random process as follows[1]:

(1) $\mu_2$ is drawn from a prior distribution $p_\theta(\mu_2)$ defined as

$$p_\theta(\mu_2) = \mathcal{N}(0, \sigma_{\mu_2}^2 I); \tag{4.1}$$

(2) $z_1^n$ and $z_2^n$ are drawn

$$p_\theta(z_1^n) = \mathcal{N}(0, \sigma_{z_1}^2 I), \tag{4.2}$$

$$p_\theta(z_2^n | \mu_2) = \mathcal{N}(\mu_2, \sigma_{z_2}^2 I); \tag{4.3}$$

(3) The segment $x^n$ is drawn from

$$p_\theta(x^n | z_1^n, z_2^n) = \mathcal{N}(f_{\mu_x}(z_1^n, z_2^n), diag(f_{\sigma_x^2}(z_1^n, z_2^n))). \tag{4.4}$$

Here $\mathcal{N}$ denotes standard normal distribution, $f_{\mu_x}(\cdot, \cdot)$ and $f_{\sigma_x^2}(\cdot, \cdot)$ are parameterized by DNN models.

A graphical illustration of FHVAE generative model $\theta$ is given as in Figure 4.3b. The joint probability for the generation of sequence $X$ is formulated as,

$$p_\theta(\mu_2) \prod_{n=1}^{N} p_\theta(z_1^n) p_\theta(z_2^n | \mu_2) p_\theta(x^n | z_1^n, z_2^n). \tag{4.5}$$

---

[1] Without causing any confusion, sequence index $i$ in $X^i$ is omitted for simplicity.

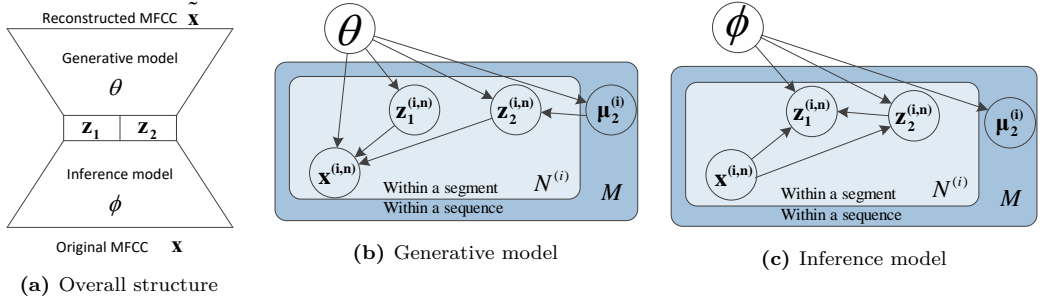

**(a)** Overall structure  **(b)** Generative model  **(c)** Inference model

**Figure 4.3:** Overall structure and graphic illustration of FHVAE generative and inference models.

In FHVAE, the exact posterior inference is intractable. Therefore an inference model $q_\phi(\cdot|\boldsymbol{X})$ is introduced to approximate the true posterior $p_\theta(\cdot|\boldsymbol{X})$ as follows,

$$q_\phi(\boldsymbol{\mu_2}) \prod_{n=1}^{N} q_\phi(\boldsymbol{z_2^n}|\boldsymbol{x^n}) q_\phi(\boldsymbol{z_1^n}|\boldsymbol{x^n}, \boldsymbol{z_2^n}), \tag{4.6}$$

where $q_\phi(\boldsymbol{\mu_2}), q_\phi(\boldsymbol{z_2^n}|\boldsymbol{x^n})$ and $q_\phi(\boldsymbol{z_1^n}|\boldsymbol{x^n}, \boldsymbol{z_2^n})$ are defined as,

$$q_\phi(\boldsymbol{\mu_2}) = \mathcal{N}(g_{\mu_{\mu_2}}(i), \sigma_{\tilde{\mu}_2}^2 \boldsymbol{I}), \tag{4.7}$$

$$q_\phi(\boldsymbol{z_2^n}|\boldsymbol{x^n}) = \mathcal{N}(g_{\mu_{z_2}}(\boldsymbol{x^n}), diag(g_{\sigma_{z_2}^2}(\boldsymbol{x^n})), \tag{4.8}$$

$$q_\phi(\boldsymbol{z_1^n}|\boldsymbol{x^n}, \boldsymbol{z_2^n}) = \mathcal{N}(g_{\mu_{z_1}}(\boldsymbol{x^n}, \boldsymbol{z_2^n}), diag(g_{\sigma_{z_1}^2}(\boldsymbol{x^n}, \boldsymbol{z_2^n})). \tag{4.9}$$

$g_{\mu_{z_2}}(\cdot), g_{\sigma_{z_2}^2}(\cdot), g_{\mu_{z_1}}(\cdot, \cdot)$ and $g_{\sigma_{z_1}^2}(\cdot, \cdot)$ are parameterized by two DNNs. For $q_\phi(\boldsymbol{\mu_2})$, during FHVAE training, a trainable lookup table containing posterior mean of $\boldsymbol{\mu_2}$ for each sequence is updated. For unseen test sequences, maximum a posteriori (MAP) estimation is used to infer $\boldsymbol{\mu_2}$. Details of $\boldsymbol{\mu_2}$ estimation can be found in [99]. A graphic illustration of FHVAE inference mode $\phi$ is given as in Figure 4.3c.

The training of FHVAE is done by optimizing **discriminative segmental**

**variational lower bound** $\mathcal{L}(\theta, \phi; \boldsymbol{x^{(i,n)}})$ [99], which is defined as,

$$
\begin{aligned}
& \mathbb{E}_{q_\phi(\boldsymbol{z_1^{(i,n)}}, \boldsymbol{z_2^{(i,n)}} | \boldsymbol{x^{(i,n)}})}[\log p_\theta(\boldsymbol{x^{(i,n)}} | \boldsymbol{z_1^{(i,n)}}, \boldsymbol{z_2^{(i,n)}})] \\
& - \mathbb{E}_{q_\phi(\boldsymbol{z_2^{(i,n)}} | \boldsymbol{x^{(i,n)}})}[\mathrm{KL}(q_\phi(\boldsymbol{z_1^{(i,n)}} | \boldsymbol{x^{(i,n)}}, \boldsymbol{z_2^{(i,n)}}) || p_\theta(\boldsymbol{z_1^{(i,n)}}))] \\
& - \mathrm{KL}(q_\phi(\boldsymbol{z_2^{(i,n)}} | \boldsymbol{x^{(i,n)}}) || p_\theta(\boldsymbol{z_2^{(i,n)}} | \boldsymbol{\tilde{\mu}_2^i})) \\
& + \frac{1}{N^i} \log p_\theta(\boldsymbol{\tilde{\mu}_2^i}) + \alpha \log p(i | \boldsymbol{z_2^{(i,n)}}),
\end{aligned}
$$

where $\boldsymbol{\tilde{\mu}_2^i}$ denotes posterior mean of $\boldsymbol{\mu_2}$ for the $i$-th sequence, $\alpha$ denotes the discriminative weight. The discriminative objective $\log p(i | \boldsymbol{z_2^{(i,n)}})$ is formulated as,

$$
\log p(i | \boldsymbol{z_2^{(i,n)}}) := \log p_\theta(\boldsymbol{z_2^{(i,n)}} | \boldsymbol{\tilde{\mu}_2^i}) - \log \sum_{j=1}^{M} p_\theta(\boldsymbol{z_2^{(j,n)}} | \boldsymbol{\tilde{\mu}_2^j}). \tag{4.10}
$$

After FHVAE training, $\boldsymbol{z_2}$ is expected to encode factors that are relatively consistent within a sequence. The discriminative objective encourage $\boldsymbol{z_2}$ to capture sequence-dependent information, instead of being collapsed into a trivial value [99]. $\boldsymbol{z_1}$ encodes the residual factors that are sequence-independent.

### 4.2.2 Speaker-invariant features by FHVAE

In this work, FHVAE is applied to learn speaker-invariant features in an unsupervised manner. The input data used to train the FHVAE are frame-level spectral features, e.g. MFCCs. Prior to training, feature sequences from the utterances of the same speaker are concatenated into a single sequence. With this arrangement, $\boldsymbol{z_2}$ is expected to encode largely speaker identity information and carry little phonetic information. $\boldsymbol{z_1}$ is expected to encode the residual information, primarily related to linguistic content.

Two different methods of FHVAE-based speaker-invariant feature learning can be performed. The first method in that latent segment variables $\{\boldsymbol{z_1^{(i,n)}}\}$ inferred from Equation 4.9 is treated as the learned representation.

In the second method, the feature representation is reconstructed based on a unified s-vector. After FHVAE training, a representative speaker is selected from the dataset. The s-vector of this speaker is denoted as $\boldsymbol{\mu_2^*}$. Then given an arbitrary speaker $i$, the respective latent sequence variable $\boldsymbol{z_2^{(i,n)}}$ inferred from Equation 4.8, is modified to $\hat{z}_2^{(i,n)}$ by applying a linear transformation as shown below,

$$\hat{z}_2^{(i,n)} = z_2^{(i,n)} - \mu_2^i + \mu_2^*, \tag{4.11}$$

where $\mu_2^i$ is the s-vector of speaker $i$. Finally, the FHVAE decoder (generative model) reconstructs speech segment $\hat{x}^{(i,n)}$ based on $p_\theta(\hat{x}^{(i,n)}|z_1^{(i,n)}, \hat{z}_2^{(i,n)})$ (Equation 4.4). The reconstructed features $\{\hat{x}^{(i,n)}\}$ are used as the learned feature representation. This method is named as *s-vector unification*. Compared to the original features, the reconstructed features are expected to retain linguistic content and capture the speaker's characteristics. In other words, speech synthesized from $\{\hat{x}^{(i,n)}\}$ would sound as if they were all spoken by the representative speaker.

In essence, Equation (4.11) is to add a universal bias $\mu_2^*$ towards $z_2^{(i,n)}$ of all the speech segments of all the speakers. One could also choose to set the universal bias to 0, or any consonant vector. However, as will be discussed in Section 4.5.2, the choice of $\mu_2^*$ affects the quality of reconstructed MFCC features.

## 4.3 Speaker adversarial training

Speaker adversarial training [103] is realized by the adversarial multi-task learning (AMTL) architecture as proposed in [104]. The main idea is to establish a speaker classification task on top of the hidden layers and reversely back-propagate the classification error, such that the lower layers are guided to extract representations irrelevant to speakers.

Figure 4.4 shows the architecture of a speaker adversarial training model. The model comprises a subword classification network ($M_p$), a speaker classification network ($M_s$) and a shared-hidden-layer feature extractor ($M_h$). This architecture is similar to the MTL-DNN as used in multilingual DNN ASR [86]. The difference of AMTL from MTL is on how learning error is propagated from $M_s$ to $M_h$. In AMTL, the speaker classification error is back-propagated by multiplying with a negative coefficient named the *adversarial weight*. By doing so, the output layer of $M_h$ is forced to learn speaker-invariant features so as to confuse $M_s$, while $M_s$ aims to correctly classify the outputs of $M_h$ into the corresponding speaker. Meanwhile, $M_p$ learns to predict subword-like labels, i.e. DPGMM frame labels, and back-propagate subword classification error to $M_h$ in a usual way. As a result, the output of $M_h$ gives a speaker-invariant and subword discriminative representation of input speech. The output dimension of $M_h$ is usually set to a small value, e.g. 40 to 80, to enable BNF extraction of $M_h$.

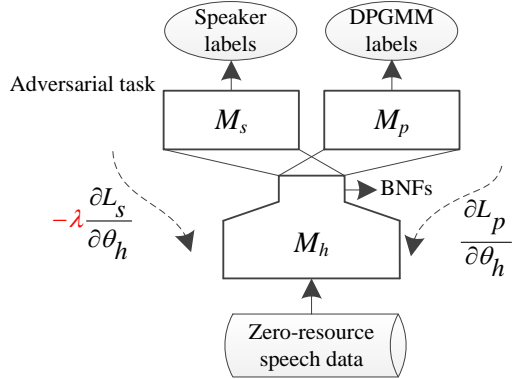

**Figure 4.4:** Speaker adversarial training based on AMTL.

Let $\theta_p, \theta_s$ and $\theta_h$ denote the network parameters of $M_p, M_s$ and $M_h$, respectively. Using the stochastic gradient descent (SGD) algorithm, these parameters are updated as,

$$\theta_p \leftarrow \theta_p - \delta \frac{\partial \mathcal{L}_p}{\partial \theta_p}, \tag{4.12}$$

$$\theta_s \leftarrow \theta_s - \delta \frac{\partial \mathcal{L}_s}{\partial \theta_s}, \tag{4.13}$$

$$\theta_h \leftarrow \theta_h - \delta \Big[ \frac{\partial \mathcal{L}_p}{\partial \theta_h} - \lambda \frac{\partial \mathcal{L}_s}{\partial \theta_h} \Big], \tag{4.14}$$

where $\delta$ is the learning rate, $\mathcal{L}_p$ and $\mathcal{L}_s$ are the loss values of subword and speaker classification tasks respectively, for both of which cross-entropy is adopted. A gradient reversal layer (GRL) [104] was designed and put in the middle of $M_h$ and $M_s$. The GRL acts as identity transform during forward-propagation and changes the sign of loss during back-propagation.

There are two main reasons why speaker adversarial training is adopted in this work. First, the idea of adversarial training is straightforward yet effective in various speech related tasks, such as domain-invariant ASR [103, 105], language recognition [106]. It only requires domain labels as supervision, such as speaker labels in speaker adversarial training. Second, speaker adversarial training is potentially com-plementary to previously mentioned two adaptation methods in this chapter. FM-LLR and disentangled representation learning are applied to the DNN-BNF framework at front-end, while speaker adversarial training is applied at back-end. It would be interesting to study the effectiveness of combining these methods.

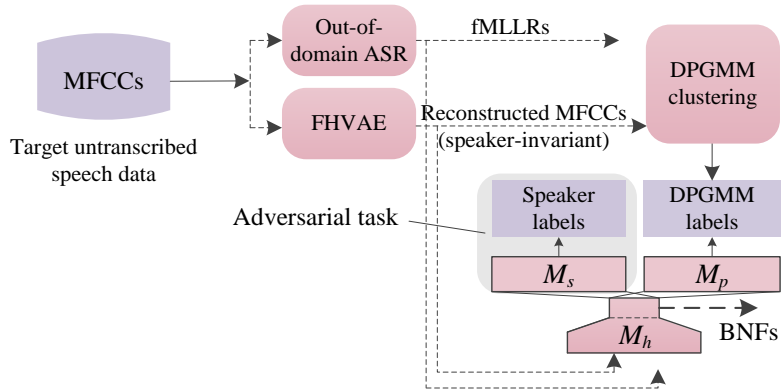

**Figure 4.5:** An illustration of combining speaker adaptation approaches.

## 4.4 Combination of adaptation approaches

In this section, combinations of speaker adaptation approaches described in previous sections are investigated. Specifically, speaker adversarial training is jointly used with either out-of-domain fMLLR estimation or disentangled speech representation learning. An illustration on combining these approaches is given in Figure 4.5. The input features to DPGMM and speaker AMTL are replaced by fMLLRs or reconstructed MFCCs. It is expected that with the replacement, speaker adversarial training could be further improved.

## 4.5 Experiments

Experiments that are carried out to validate the effectiveness of the speaker adaptation approaches in improving unsupervised subword modeling. The experiments are conducted using the ZeroSpeech 2017 development database, and on the ABX discriminability task.

### 4.5.1  FMLLR by out-of-domain ASR

#### Out-of-domain ASR system

A Cantonese ASR system is used as the out-of-domain language-mismatched ASR. Cantonese is a major Chinese dialect widely spoken in Hong Kong, Macau, southern part of mainland China and many overseas Chinese communities [107]. There is good reason to select Cantonese as the out-of-domain resource-rich language, despite that Mandarin, one of the target languages in ZeroSpeech 2017, and Cantonese are both Chinese languages. In fact, Cantonese and Mandarin are considered largely different in terms of acoustic-phonetic properties. These two languages are mutually unintelligible.

The Cantonese ASR system is trained with CUSENT, a read speech corpus developed by the Chinese University of Hong Kong [90]. CUSENT contains $20,378$ training utterances from 34 male and 34 female speakers, with a total of 19.3-hour speech. The ASR system is comprised of a context-dependent GMM-HMM (CD-GMM-HMM) AM with speaker adaptive training (CD-GMM-HMM-SAT), and a syllable trigram LM. For the AM, there are 33 phonemes (including a silence phoneme), each modeled by a 3-state HMM. The AM contains in total 2462 CD-HMM states. The input features to the AM are 40-dimension fMLLRs derived from sequentially applying VTLN, LDA, MLLT and fMLLR transforms towards 39-dimension MFCCs+$\Delta$+$\Delta\Delta$ with CMN. The AM is trained using Kaldi [88]. The LM is trained with CUSENT transcriptions using SRILM [108].

#### FMLLR estimation of target speech

The Cantonese ASR is used to perform fMLLR-based speaker adaptation of target zero-resource speech on the 39-dimension MFCCs in a two-pass procedure. In the first-pass, input speech utterances are decoded by the ASR in a speaker-independent manner, using unadapted features, from which initial fMLLR transforms are estimated. In the second-pass, input speech are decoded with initial fMLLRs in a speaker-adaptive manner. After the decoding, final fMLLR transforms for target speech utterances are estimated. The dimension of fMLLR features is 40. The ASR decoding result depends on the relative weighting of AM and LM. In our experi-ments, the LM carries a very small weight, such that the decoding result mainly reflects acoustic properties of target speech.

## Frame labeling and MTL-DNN training

After obtaining fMLLR features for the three target languages, DPGMM clustering is applied to generate frame-level labels. The numbers of clustering iterations for English, French and Mandarin corpora are 120, 200 and 3000. The resulted numbers of DPGMM clusters are 1118, 1345 and 596, respectively. Each frame is assigned with a DPGMM label.

The MTL-DNN model is trained with 40-dimension fMLLR features with context size $\pm 5$. There are 3 tasks involved in MTL, each corresponding to a target language. The DNN structure is the same as that used in the baseline system mentioned in Section 3.6.1, i.e., 1024-dimension 6-layer MLP except the 40-dimension linear BN layer located at the second topmost layer. The dimensions of the language-specific output layers are 1118, 1345 and 596. The objective function is weighted cross-entropy. SGD is used to optimize the objective function. After MTL-DNN training, the model is used to extract multilingual BNFs for test sets of target languages.

## Results and analyses

Experimental results of applying fMLLR features estimated by a Cantonese ASR are summarized in Tables 4.2 (across-speaker) and 4.3 (within-speaker). Each table contains two system groups marked as $\mathcal{A}$ and $\mathcal{B}$. Group $\mathcal{A}$ comprises systems without MTL-DNN training, thus are developed without using ZeroSpeech 2017 training data. The second row in $\mathcal{A}$ denotes fMLLR features obtained by our proposed methods. The third row in $\mathcal{A}$ was reported by Shibata et al. [1], which utilized a Japanese ASR for fMLLR estimation. Group $\mathcal{B}$ contains two systems generating multilingual BNFs but are trained with different input feature representations. For MUBNF0, the inputs to DPGMM frame labeling and MTL-DNN modeling are raw MFCCs. This system was reported in Section 3.6.2 as the baseline. For MUBNF, the inputs to DPGMM and MTL-DNN are fMLLRs estimated by the Cantonese ASR.

From Tables 4.2 and 4.3, several observations are made.

**Table 4.2:** Across-speaker ABX error rates (%) on raw MFCCs, fMLLRs by OOD ASRs and multilingual BNFs.

|   |   | English | | | French | | | Mandarin | | | Avg. |
|---|---|---|---|---|---|---|---|---|---|---|---|
|   |   | 1s | 10s | 120s | 1s | 10s | 120s | 1s | 10s | 120s | |
| $\mathcal{A}$ | MFCC [14] | 23.4 | 23.4 | 23.4 | 25.2 | 25.5 | 25.2 | 21.3 | 21.3 | 21.3 | 23.3 |
| | **FMLLR by OOD ASR** | 13.4 | 12.0 | 11.3 | 17.2 | 15.2 | 14.8 | 10.7 | 10.2 | 9.4 | **12.8** |
| | FMLLR by OOD ASR [1] | 14.2 | 11.9 | 11.3 | 17.6 | 15.2 | 14.4 | 12.7 | 13.6 | 10.0 | 13.4 |
| $\mathcal{B}$ | MUBNF0 (Section 3.6.2) | 13.5 | 12.4 | 12.4 | 17.8 | 16.4 | 16.1 | 12.6 | 11.9 | 12.0 | 13.9 |
| | **MUBNF** | 10.9 | 9.5 | 8.9 | 15.2 | 13.0 | 12.0 | 10.5 | 8.9 | 8.2 | **10.8** |

**Table 4.3:** Within-speaker ABX error rates (%) on raw MFCCs, fMLLRs by OOD ASRs and multilingual BNFs.

|   |   | English | | | French | | | Mandarin | | | Avg. |
|---|---|---|---|---|---|---|---|---|---|---|---|
|   |   | 1s | 10s | 120s | 1s | 10s | 120s | 1s | 10s | 120s | |
| $\mathcal{A}$ | MFCC [14] | 12.0 | 12.1 | 12.1 | 12.5 | 12.6 | 12.6 | 11.5 | 11.5 | 11.5 | 12.0 |
| | **FMLLR by OOD ASR** | 8.0 | 8.2 | 7.3 | 10.3 | 10.3 | 9.1 | 9.3 | 9.3 | 8.4 | **8.9** |
| | FMLLR by OOD ASR [1] | 7.8 | 7.7 | 7.0 | 10.4 | 10.5 | 9.2 | 9.2 | 11.4 | 8.8 | 9.1 |
| $\mathcal{B}$ | MUBNF0 (Section 3.6.2) | 8.0 | 7.3 | 7.3 | 10.3 | 9.4 | 9.3 | 10.1 | 8.8 | 8.9 | 8.8 |
| | **MUBNF** | 7.4 | 6.9 | 6.3 | 9.6 | 9.0 | 8.1 | 9.8 | 8.8 | 8.1 | **8.2** |

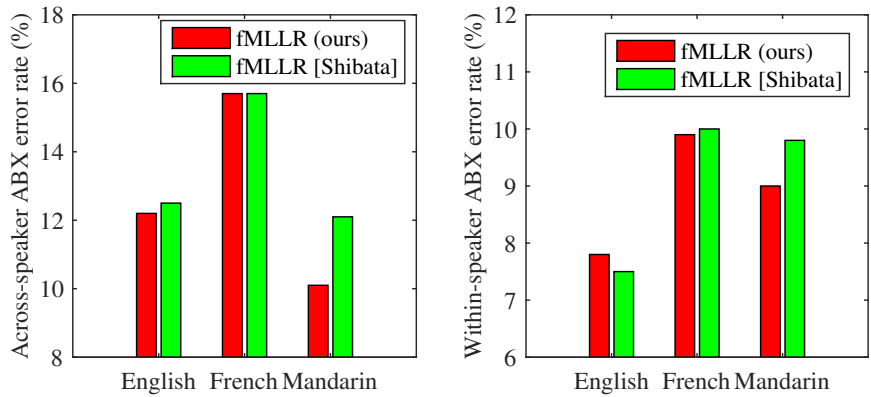

**Figure 4.6:** Comparison of fMLLRs obtained by our methods and by Shibata et al. [1]. The performance is computed by averaging over all utterance lengths within a target language.

(1). The fMLLR features consistently outperform MFCC features on all target languages. The relative performance improvement of our fMLLR features comparing to MFCCs is 45.1% in across-speaker condition and 25.8% in within-speaker condition. The results demonstrate that speaker adaptation based on an out-of-domain ASR system is both effective and efficient for unsupervised subword modeling. Unlike MFCCs which perform equally on different length conditions, fMLLRs achieve better performance on longer test utterances. This is reasonable, as fMLLR-based speaker adaptation is widely known to work less well on very short speech.

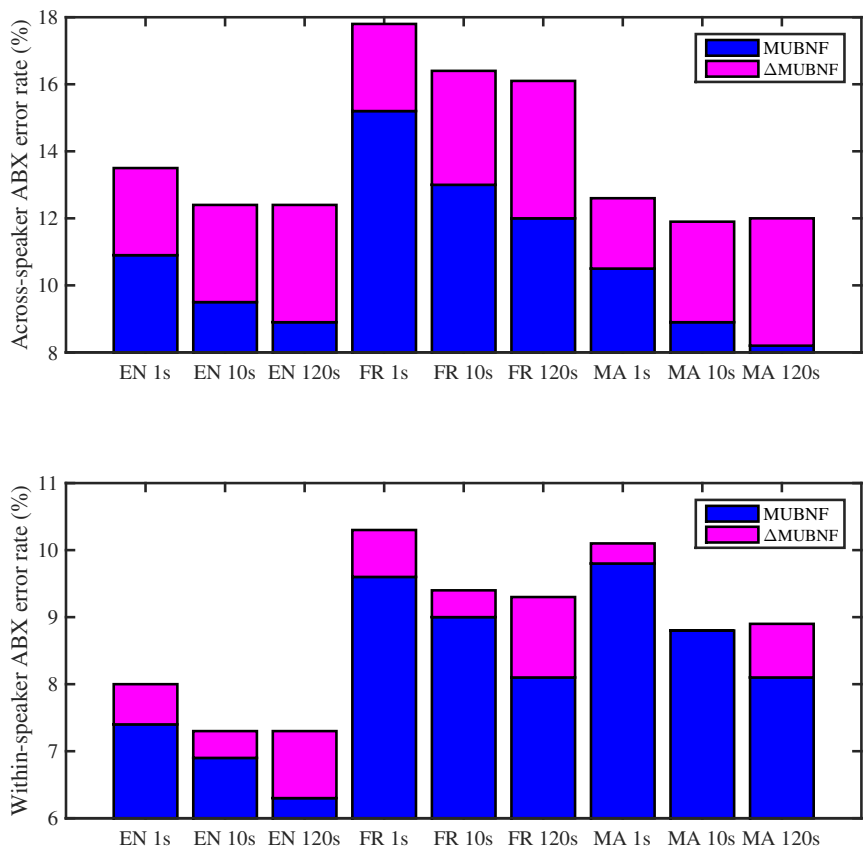

**Figure 4.7:** Comparison of MUBNF0 and MUBNF. 'ΔMUBNF' denotes difference of ABX error rate between MUBNF0 and MUBNF. 'EN, FR, MA' are abbreviations of English, French and Mandarin.

(2). Our fMLLR features perform slightly better than fMLLRs in [1]. Note that [1] used a 240-hour Japanese transcribed dataset to train an out-of-domain ASR, while our system used a 19.3-hour Cantonese dataset. It is interesting to see from Figure 4.6 that the two fMLLRs achieve similar ABX error rates on English and French, while our fMLLRs perform much better in Mandarin.

This can be partially explained that Cantonese and Mandarin are both Chinese languages, which may potentially provide additional benefit in modeling Mandarin speech, as compared to exploiting Japanese as the out-of-domain language resource.

(3). MUBNF outperforms MUBNF0 to a large extent, especially in the across-speaker test condition. The relative ABX error rate reduction of MUBNF over MUBNF0 is 22.3% in across-speaker condition and 6.8% in within-speaker condition. This suggests that speaker adaptation at input feature level is a critical step in obtaining speaker-invariant and subword-discriminative BNF representations. It is clearly seen from Figure 4.7 that MUBNF consistently outperform MUBNF0 over all the target languages and utterance lengths (except in Mandarin 10s). Moreover, Figure 4.7 also reveals different properties of MUBNF and MUBNF0 on 10s/120s test conditions. MUBNF0 performs almost the same on 10s and 120s conditions, whilst MUBNF performs significantly better on 120s. This indicates that fMLLR benefits from increasing the utterance from 10s to 120s, while CMN does not benefit from the increase longer than 10s.

## 4.5.2   Disentangled speech representation learning

### FHVAE setup and parameter tuning

FHVAE model parameters are determined by reference to [98]. The encoder and decoder networks of FHVAE are both 2-layer LSTMs with 256 neurons per layer. The dimensions of $z_1$ and $z_2$ are 32. Training data for the three target languages are merged to train the FHVAE. Input features are fixed-length speech segments randomly chosen from training utterances. The segment length $l$ is determined through a parameter tuning process, and will be discussed in the next paragraph. Each frame is represented by a 13-dimensional MFCC with CMN at speaker level. During the inference of latent segment variable $z_1$ and reconstructed MFCC features, the input segments are shifted by 1 frame. To match the length of extracted features with original MFCCs, the first and last frame are padded. Adam [109] with $\beta_1 = 0.95$ and $\beta_2 = 0.999$ is used to train the FHVAE. A 10% subset of training data is randomly selected for cross-validation. The rest part of training data is used for training. The training process is terminated if the lower bound on the cross-validation set does not improve for 20 epochs. The open-source tool [99] is used to train FHVAEs.

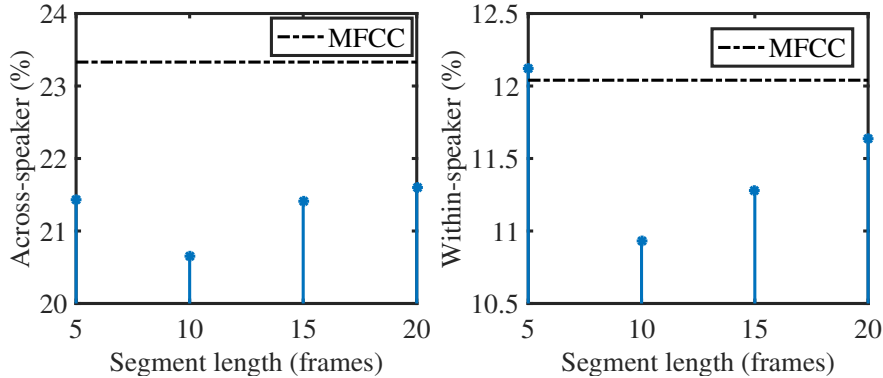

**Figure 4.8:** ABX error rates on $z_1$ with different segment lengths and raw MFCCs. The performance is computed by averaging over all languages.

In our preliminary experiments, the ABX performance of $z_1$ was found to be sensitive to the input segment length $l$. This could be explained as: a too large $l$ would reduce the capability of $z_1$ in modeling linguistic content at subword level; a too small $l$ would restrict the FHVAE from capturing sufficient temporal dependencies which are essential in modeling speech. ABX error rates on $z_1$ with different values of $l$ are shown in Figure 4.8. As a reference, ABX error rates on raw MFCCs given by Challenge organizers [14] are also shown in this Figure as dash-dotted lines. It can be seen that the optimal value of $l$ is 10. For the remaining experiments, $l$ is fixed to 10.

## Selecting representative speaker for reconstructed MFCC

As mentioned in Section 4.2.2, the extraction of reconstructed MFCCs $\{\hat{x}\}$ using s-vector unification assumes a pre-defined representative speaker. In order to validate the generalization ability of our proposed s-vector unification method and evaluate its sensitivity to the gender of the representative speaker, 6 English speakers {s0107, s3020, s4018, s0019, s1724, s2544}, 4 French speakers {M02R, M03R, F01R, F02R} and 2 Mandarin speakers {A08, C04}—namely, in total 12 speakers are randomly chosen from 'speaker-R' sets (defined in Table 3.1) of ZeroSpeech 2017 training data. The 12 speakers constitute candidates of the representative speaker in our experiments. The first half speakers inside each language set are male and the second half are female[2].

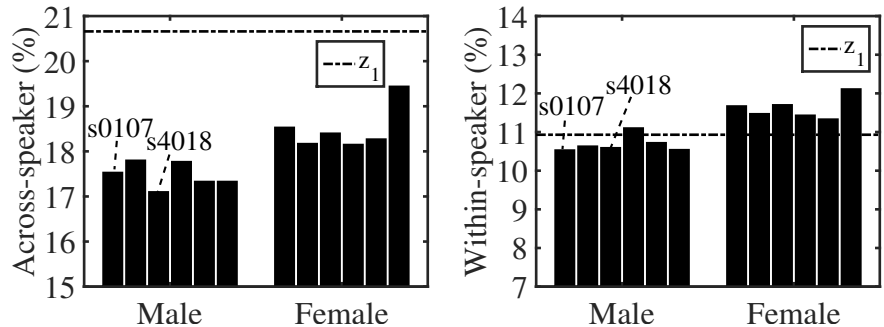

**Figure 4.9:** ABX error rates (%) on reconstructed MFCCs $\hat{x}$ using s-vector unification and latent segment variable $z_1$. Each bar corresponds to a representative speaker. The performance is computed by averaging over all languages.

Each time when extracting reconstructed MFCCs $\{\hat{x}\}$, a representative speaker is determined from the 12 speakers. The s-vector corresponding to this speaker is denoted as $\boldsymbol{\mu_2^*}$. During the extraction of $\{\hat{x}\}$, s-vectors $\{\boldsymbol{\mu_2^i}\}$ of all three target languages' utterances are modified to $\boldsymbol{\mu_2^*}$. After extraction, there are in total 12 groups of $\{\hat{x}\}$. They are evaluated by the ABX discriminability task. Their results are shown in Figure 4.9.

It can be observed from Figure 4.9 that in the across-speaker condition, $\{\hat{x}\}$ outperform $\{z_1\}$ regardless of choosing any of the 12 speakers as the representative. In the within-speaker condition, $\{\hat{x}\}$ perform slightly better than $\{z_1\}$ in most cases if the representative speaker is male, and are worse than $\{z_1\}$ in all female cases. It can be concluded that selecting a male representative speaker is generally better in extracting speaker-invariant reconstructed MFCC features. The male speaker 's4018' achieves the best overall performance. Further studies are required to explore why male speakers are more suitable than females for s-vector unification.

**Frame labeling and MTL-DNN training**

Figure 4.10 summarizes frame labeling and MTL-DNN training procedure in our experiments. Frame labels are generated by DPGMM clustering towards recon-structed MFCC features $\{\hat{x}\}$ appended by $\Delta$ and $\Delta\Delta$. The representative speaker for extracting $\{\hat{x}\}$ is selected from the 12 candidates. The numbers of clustering iterations for English, French and Mandarin are $80, 80$ and $1400$.

---

[2]Gender information is not released in ZeroSpeech database. We acquired this information via listening

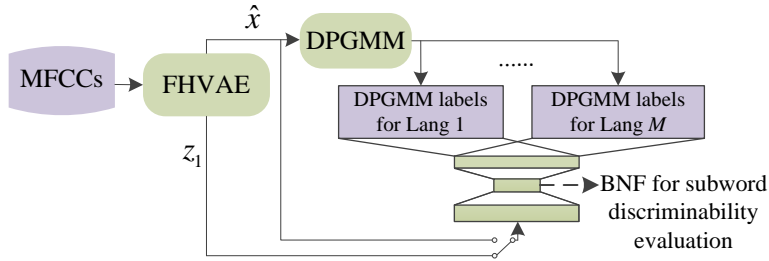

**Figure 4.10:** An illustration on applying speaker-invariant features extracted from an FHVAE in DPGMM and MTL-DNN.

The MTL-DNN model is trained with 3 equally-weighted tasks corresponding to the three target languages. The input features to MTL-DNN are either latent segment variables $\{z_1\}$ or reconstructed MFCCs $\{\tilde{x}\}$. Note that $\{\tilde{x}\}$ is slightly different from $\{\hat{x}\}$. During the inference of $\{\tilde{x}\}$ for training sets, s-vector unification is not applied; during the inference for test sets, s-vector unification is applied within every test subset with a subset-specific $\mu_2^*$. The reason is that MTL-DNN trained with $\{\tilde{x}\}$ was found to outperform that trained with $\{\hat{x}\}$. The layer-wise structure, objective function and training strategies of MTL-DNN are the same as those described in the baseline system in Section 3.6.1. After MTL-DNN training, the model is used to extract 40-dimension multilingual BNFs of test sets for ABX evaluation.

## Results and analyses

Experimental results on multilingual BNFs trained with and without FHVAE-based speaker-invariant features are summarized in Tables 4.4 (across-speaker) and 4.5 (within-speaker). In these Tables, each row represents a system, in which the input features to DPGMM clustering and MTL-DNN modeling are listed in the second and third column respectively. The baseline system using raw MFCC features as inputs to both DPGMM and MTL-DNN, and the system exploiting a Cantonese ASR for fMLLR estimation, are shown in the first two rows of the Tables. A similar work done by Chen et al. [30], which utilized VTLN-processed MFCCs as inputs to DPGMM is also listed for reference. The systems adopting FHVAE-based speaker-

# 4. Speaker adaptation for unsupervised subword modeling

**Table 4.4:** Across-speaker ABX error rates (%) on multilingual BNFs trained with-/without FHVAE-based speaker-invariant features.

| Ref. | Input feature | | English | | | French | | | Mandarin | | | Avg. |
|---|---|---|---|---|---|---|---|---|---|---|---|---|
| | DPGMM | DNN | 1s | 10s | 120s | 1s | 10s | 120s | 1s | 10s | 120s | |
| Sec. 3.6.2 | MFCC | MFCC | 13.5 | 12.4 | 12.4 | 17.8 | 16.4 | 16.1 | 12.6 | 11.9 | 12.0 | 13.9 |
| Sec. 4.5.1 | fMLLR | fMLLR | 10.9 | 9.5 | 8.9 | 15.2 | 13.0 | 12.0 | 10.5 | 8.9 | 8.2 | **10.8** |
| [30] | MFCC | FB+F0 | 13.7 | 12.1 | 12.0 | 17.6 | 15.6 | 14.8 | 12.3 | 10.8 | 10.7 | 13.3 |
| | +VTLN | FB+F0 | 12.7 | 11.0 | 10.8 | 17.0 | 14.5 | 14.1 | 11.9 | 10.3 | 10.1 | 12.5 |
| ① | MFCC | $z_1$ | 12.9 | 11.7 | 11.7 | 17.2 | 15.5 | 15.2 | 12.5 | 11.4 | 11.5 | 13.3 |
| | MFCC | $\tilde{x}$ | 12.8 | 11.7 | 11.5 | 17.8 | 15.5 | 15.1 | 12.3 | 10.9 | 10.7 | 13.1 |
| ② | $\hat{x}$-s0107 | $z_1$ | 11.2 | 10.1 | 10.1 | 15.5 | 13.8 | 13.7 | 11.5 | 10.2 | 10.0 | 11.8 |
| | $\hat{x}$-s0107 | $\tilde{x}$ | 11.6 | 10.4 | 10.1 | 16.1 | 13.9 | 13.7 | 11.9 | 10.2 | 10.4 | 12.0 |
| ③ | $\hat{x}$-s4018 | $z_1$ | 11.0 | 9.8 | 9.8 | 14.9 | 13.4 | 13.0 | 11.4 | 10.1 | 10.0 | **11.5** |
| | $\hat{x}$-s4018 | $\tilde{x}$ | 11.3 | 10.0 | 9.8 | 15.7 | 13.6 | 13.3 | 11.8 | 10.0 | 10.4 | 11.8 |

**Table 4.5:** Within-speaker ABX error rates (%) on multilingual BNFs trained with-/without FHVAE-based speaker-invariant features.

| Ref. | Input feature | | English | | | French | | | Mandarin | | | Avg. |
|---|---|---|---|---|---|---|---|---|---|---|---|---|
| | DPGMM | DNN | 1s | 10s | 120s | 1s | 10s | 120s | 1s | 10s | 120s | |
| Sec. 3.6.2 | MFCC | MFCC | 8.0 | 7.3 | 7.3 | 10.3 | 9.4 | 9.3 | 10.1 | 8.8 | 8.9 | 8.8 |
| Sec. 4.5.1 | fMLLR | fMLLR | 7.4 | 6.9 | 6.3 | 9.6 | 9.0 | 8.1 | 9.8 | 8.8 | 8.1 | **8.2** |
| [30] | MFCC | FB+F0 | 8.5 | 7.3 | 7.2 | 11.1 | 9.5 | 9.4 | 10.5 | 8.5 | 8.4 | 8.9 |
| | +VTLN | FB+F0 | 8.5 | 7.3 | 7.2 | 11.2 | 9.4 | 9.4 | 10.5 | 8.7 | 8.5 | 9.0 |
| ① | MFCC | $z_1$ | 8.2 | 7.0 | 7.0 | 10.7 | 9.2 | 9.1 | 10.4 | 8.8 | 8.7 | 8.8 |
| | MFCC | $\tilde{x}$ | 8.2 | 7.3 | 7.0 | 10.6 | 9.3 | 8.9 | 10.5 | 8.8 | 8.7 | 8.8 |
| ② | $\hat{x}$-s0107 | $z_1$ | 7.3 | 6.4 | 6.6 | 10.1 | 8.9 | 8.8 | 10.4 | 8.5 | 8.4 | 8.4 |
| | $\hat{x}$-s0107 | $\tilde{x}$ | 7.8 | 6.7 | 6.5 | 10.5 | 9.6 | 9.3 | 10.8 | 8.6 | 8.7 | 8.7 |
| ③ | $\hat{x}$-s4018 | $z_1$ | 7.3 | 6.3 | 6.3 | 9.7 | 8.6 | 8.4 | 10.1 | 8.5 | 8.4 | **8.2** |
| | $\hat{x}$-s4018 | $\tilde{x}$ | 7.8 | 6.5 | 6.5 | 10.1 | 9.1 | 8.8 | 10.6 | 8.7 | 8.7 | 8.5 |

invariant features are indexed as ①, ② and ③. '$\hat{x}$-s0107' and '$\hat{x}$-s4018' denote reconstructed MFCCs with speaker s0107 and speaker s4018 as the representative speaker respectively. Here, $\hat{x}$-s4018 is used to represent the ideal scenario, as s4018 performs the best among all the representative candidates (shown in 4.9). $\hat{x}$-s0107 is used to represent the general scenario, as s0107 performs moderately among all the male candidates.

From Tables 4.4 and 4.5, several observations can be made.

(1). By comparing Sec. 3.6.2 with ①, it is shown that without improving input feature representation to DPGMM clustering, the MTL-DNN trained with $z_1$ or $\tilde{x}$ outperforms that trained with raw MFCCs in the across-speaker test condition. The relative ABX error rate reduction is 4.3% for the first line in ① and 5.8% for the second line in ①.

(2). Reconstructed MFCC features $\hat{x}$ significantly outperform original MFCC features in DPGMM clustering. In the ideal scenario where s4018 is selected as the representative speaker, by comparing the first lines in ③ and ①, the across- and within-speaker relative ABX error rate reduction are 13.5% and 6.9% respectively. Even in the general scenario where s0107 is selected as the representative speaker, by comparing the first lines in ② and ①, the across- and within-speaker relative ABX error rate reduction are 11.3% and 4.7%. These results clearly demonstrate that improving speaker invariance of input features to DPGMM clustering is crucial in achieving better BNFs for unsupervised subword modeling, especially in the across-speaker scenario.

(3). The advancement of ABX task performance contributed from improving DPGMM inputs is more prominent than that from improving DNN inputs. This observation is true for both across- and within-speaker conditions. In fact, Figure 4.5 tells that replacing DNN input features from original MFCCs to either $z_1$ or $\tilde{x}$ (system Sec. 3.6.2 $\longrightarrow$ ①) does not achieve within-speaker performance improvement, if DPGMM labels keep unchanged.

(4). Our best system (first line in ③) achieves 17.3% and 7.3% relative ABX error rate reduction compared to the baseline system reported in Section 3.6.2. This improvement is achieved without exploiting any out-of-domain resources. It is attributed to better speaker-invariant features learned from FHVAE-based disentangled representation learning. Compared to system Sec. 4.5.1, which utilized Cantonese transcribed speech, the best system is slightly better in the within-speaker condition, while slightly worse in the across-speaker condition.

(5). Speaker-invariant features generated by FHVAEs are better than VTLN in improving unsupervised subword modeling. While our baseline system is inferior to the baseline MFCC system in [30] in the across-speaker condition, our best system consistently outperforms MFCC+VTLN in all target languages and utterance lengths. In the within-speaker condition, our proposed methods also achieve better performance.

**Visualization of speaker-invariant features**

Apart from evaluating the effectiveness of our proposed FHVAE-based speaker-invariant features by ABX discriminability, we also show 2-dimension visualization of the learned features to demonstrate their robustness towards speaker variation. To this end, two Mandarin speakers from ZeroSpeech 2017 training data are randomly selected. For each speaker, frame-level speech features from 10-second speech utterances (i.e. 1000 frames) are used for t-SNE based 2-dimension visualization, using open-source tools developed by [110].

The visualization results are shown in Figure 4.11. This Figure contains five sub-Figures corresponding to different feature representations, namely, original MFCC, original MFCC+CMN, $z_1$, $z_2$ and reconstructed MFCC $\hat{x}$ with representative speaker s4018. Each data point inside the sub-Figure denotes a speech frame. The two speakers are marked by different colors. Figure 4.11e clearly shows the disentanglement of speech features towards speaker variation, which is in agreement of our expectation. Figures 4.11a and 4.11b indicate that CMN alleviates speaker variation encoded in MFCC features while this variation is still perceptible. In comparison, reconstructed MFCCs, as shown in 4.11d, demonstrate much higher robustness towards speaker variation.

## 4.5.3 Speaker adversarial training

**Experimental setup**

The AMTL-DNN architecture [104] is adopted to train multilingual BNFs, as shown in Figure 4.4. There are two types of labels involved during training, namely, DPGMM labels and speaker labels. DPGMM labels are obtained following the baseline experimental settings in Section 3.6.1. Speaker labels are released in the database. DPGMM labels support the training of $M_p$, while speaker labels support $M_s$. The input features to AMTL-DNN are 39-dimension MFCCs with $\Delta$ and $\Delta\Delta$ with context size $\pm 5$. The layer-wise structure of $M_h$ is $\{1024 \times 5, 40\}$. $M_s$ and $M_p$ both have 3 sub-networks, each corresponding to a target language. Each sub-network contains a 1024-dimension feed-forward layer followed by a soft-max layer. During AMTL-DNN training, the learning rate starts from $8 \cdot 10^{-3}$ to $8 \cdot 10^{-4}$ with

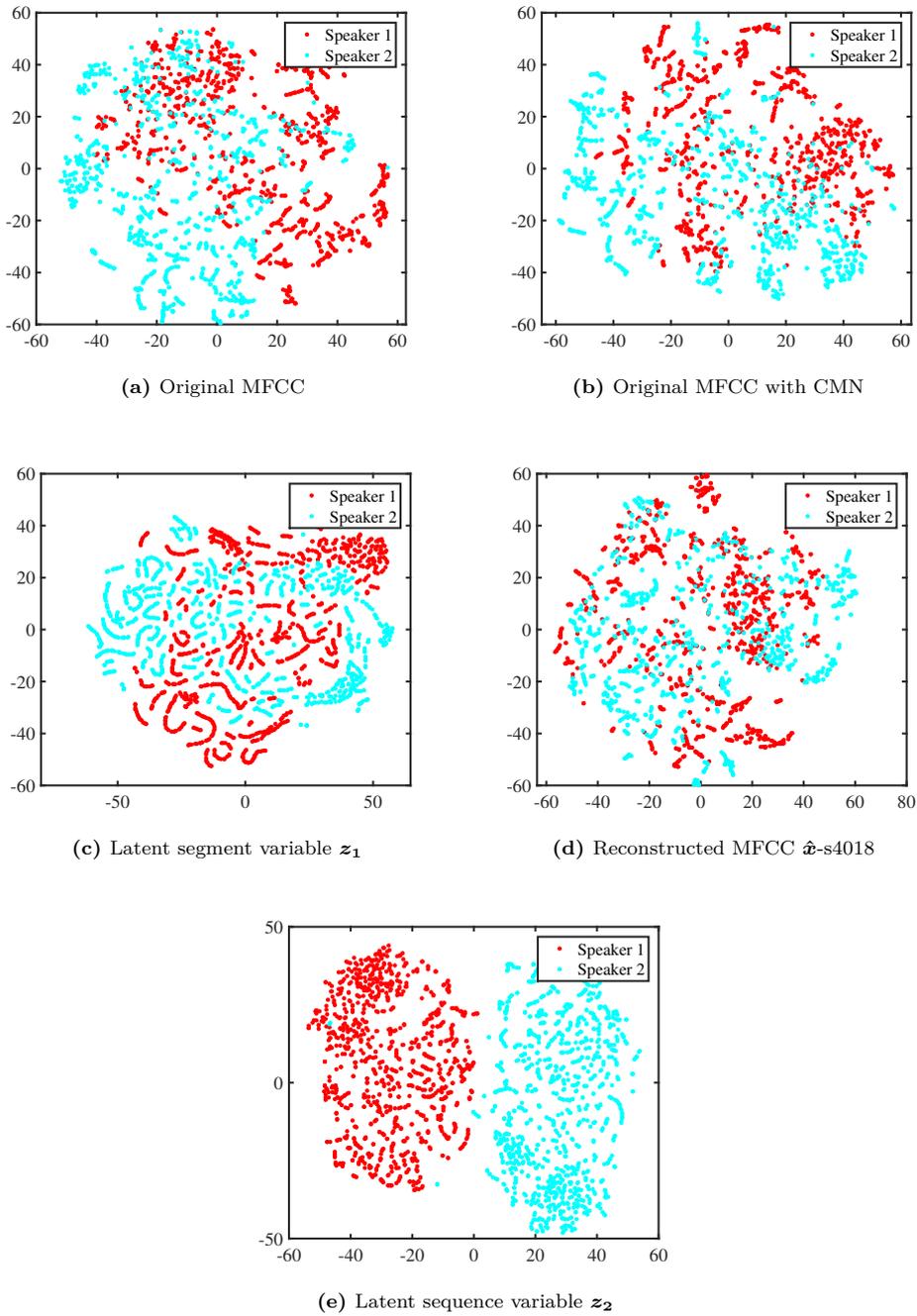

**(a)** Original MFCC

**(b)** Original MFCC with CMN

**(c)** Latent segment variable $z_1$

**(d)** Reconstructed MFCC $\hat{x}$-s4018

**(e)** Latent sequence variable $z_2$

**Figure 4.11:** T-SNE visualization of frame-level feature representations from two speakers in ZeroSpeech 2017 Mandarin subset.

exponential decay. The number of epochs is 5. Speaker adversarial weight $\lambda$ ranges from 0 to 0.1, with an interval of 0.02. After training, multilingual BNFs extracted

**Table 4.6:** Across-speaker ABX error rates (%) on multilingual BNFs extracted from speaker AMTL-DNN with different adversarial weights $\lambda$.

| Ref. | $\lambda$ | English | | | French | | | Mandarin | | | Avg. |
|------|-----------|------|------|------|------|------|------|------|------|------|------|
| | | 1s | 10s | 120s | 1s | 10s | 120s | 1s | 10s | 120s | |
| | 0 | 13.1 | 12.0 | 12.0 | 17.9 | 15.7 | 15.6 | 12.2 | 11.5 | 11.5 | 13.50 |
| | 0.02 | 13.0 | 11.9 | 12.0 | 17.5 | 15.6 | 15.4 | 12.3 | 11.3 | 11.3 | 13.37 |
| | 0.04 | 12.8 | 11.8 | 11.8 | 17.3 | 15.4 | 15.3 | 12.1 | 11.3 | 11.2 | **13.22** |
| | 0.06 | 13.1 | 12.0 | 11.9 | 17.5 | 15.5 | 15.5 | 12.1 | 11.3 | 11.3 | 13.36 |
| | 0.08 | 13.1 | 11.8 | 11.8 | 17.5 | 15.4 | 15.3 | 12.0 | 11.3 | 11.3 | 13.28 |
| | 0.1 | 13.0 | 11.8 | 11.8 | 17.5 | 15.3 | 15.4 | 12.2 | 11.3 | 11.3 | 13.29 |
| R0 | N/A | 13.5 | 12.4 | 12.4 | 17.8 | 16.4 | 16.1 | 12.6 | 11.9 | 12.0 | 13.90 |
| R-F | N/A | 11.0 | 9.8 | 9.8 | 14.9 | 13.4 | 13.0 | 11.4 | 10.1 | 10.0 | 11.49 |

from the output of $M_h$ are evaluated by ABX discriminability.

It must be noted that experiments applying speaker AMTL-DNN are conducted using a different Kaldi implementation[3] from those applying MTL-DNN in Sections 4.5.1 and 4.5.2.

## Results and analyses

Experimental results on multilingual BNFs extracted from speaker AMTL-DNN are listed in Tables 4.6 (across-speaker) and 4.7 (within-speaker). The Tables also contain two reference systems. 'R0' denotes our baseline system which is introduced in Section 3.6.2. 'R-F' denotes the best system adopting FHVAE-based disentangled speech representation learning, which is discussed in Section 4.5.2. Note that when $\lambda = 0$, speaker adversarial training is not adopted, and the AMTL-DNN model is collapsed to an MTL-DNN.

From the Tables, it can be observed that speaker adversarial training could reduce ABX error rates when $\lambda$ is set between 0 and 0.1. The absolute error rate reduction is 0.28% in the across-speaker condition, and 0.23% in the within-speaker condition. The amount of performance improvement is in accordance with the findings in a relevant study [34], despite that [34] exploited transcription resources for in-domain English speech utterances during AMTL-DNN training.

In principle, the system with $\lambda = 0$ is the same as our baseline system R0. The performance difference is caused by different implementation in MTL-DNN training.

---

[3]AMTL-DNN is implemented as Kaldi `nnet3`, while MTL-DNN in Sections 4.5.1 and 4.5.2 is implemented as Kaldi `nnet1`.

**Table 4.7:** Within-speaker ABX error rates (%) on multilingual BNFs extracted from speaker AMTL-DNN with different adversarial weights $\lambda$.

| Ref. | $\lambda$ | English | | | French | | | Mandarin | | | Avg. |
|------|-----------|---------|-----|------|--------|-----|------|----------|-----|------|------|
| | | 1s | 10s | 120s | 1s | 10s | 120s | 1s | 10s | 120s | |
| | 0 | 7.7 | 6.9 | 6.9 | 10.6 | 9.0 | 8.9 | 10.0 | 8.7 | 8.6 | 8.59 |
| | 0.02 | 7.5 | 6.8 | 6.9 | 10.5 | 9.1 | 8.9 | 9.9 | 8.6 | 8.4 | 8.51 |
| | 0.04 | 7.5 | 6.7 | 6.8 | 10.2 | 8.9 | 8.8 | 9.9 | 8.6 | 8.4 | 8.42 |
| | 0.06 | 7.4 | 6.7 | 6.8 | 10.1 | 8.8 | 8.8 | 9.8 | 8.5 | 8.3 | **8.36** |
| | 0.08 | 7.4 | 6.7 | 6.8 | 9.8 | 8.9 | 8.8 | 9.9 | 8.6 | 8.5 | 8.38 |
| | 0.1 | 7.3 | 6.7 | 6.8 | 9.9 | 8.9 | 8.7 | 9.9 | 8.6 | 8.6 | 8.38 |
| R0 | N/A | 8.0 | 7.3 | 7.3 | 10.3 | 9.4 | 9.3 | 10.1 | 8.8 | 8.9 | 8.82 |
| R-F | N/A | 7.3 | 6.3 | 6.3 | 9.7 | 8.6 | 8.4 | 10.1 | 8.5 | 8.4 | 8.18 |

By comparing R-F with the best system adopting speaker adversarial training, it can be observed that although the baseline is inferior, R-F achieves better performance in both across- and within-speaker conditions.

### 4.5.4 Combination of adaptation approaches

**Experimental setup**

For experiments combining approaches of out-of-domain ASR based fMLLR and speaker adversarial training, the input features to DPGMM are 40-dimension fMLLR features estimated by a Cantonese ASR. The input features to AMTL-DNN are either MFCCs or fMLLRs. For experiments combining approaches of disentangled representation learning and speaker adversarial training, the input features to DPGMM reconstructed MFCCs with s-vector unification. The input features to AMTL-DNN are either original or reconstructed MFCCs. The layer-wise structure of AMTL-DNN model keeps the same as mentioned in Section 4.5.3.

**Results and analyses**

Experimental results on multilingual BNFs by combining speaker adaptation approaches are summarized in Tables 4.8 (across-speaker) and 4.9 (within-speaker) respectively. The Tables comprise two groups marked with $\mathcal{A}$ and $\mathcal{B}$. Group $\mathcal{A}$ contains systems using fMLLR features as inputs to DPGMM based frame labeling, while $\mathcal{B}$ contains systems using reconstructed MFCCs as inputs to frame labeling.

**Table 4.8:** Across-speaker ABX error rates (%) on systems combining different speaker adaptation approaches.

| Ref. | Input feature | | $\lambda$ | English | | | French | | | Mandarin | | | Avg. |
|---|---|---|---|---|---|---|---|---|---|---|---|---|---|
| | DPGMM | DNN | | 1s | 10s | 120s | 1s | 10s | 120s | 1s | 10s | 120s | |
| $\mathcal{A}$ | fMLLR | MFCC | 0 | 10.3 | 9.3 | 9.2 | 14.5 | 12.9 | 12.8 | 10.3 | 9.2 | 9.1 | <u>10.84</u> |
| | | | 0.02 | 10.2 | 9.1 | 9.1 | 14.4 | 12.7 | 12.6 | 10.2 | 9.2 | 9.1 | 10.73 |
| | | | 0.04 | 10.2 | 9.1 | 9.0 | 14.2 | 12.5 | 12.3 | 10.4 | 9.2 | 9.1 | **10.67** |
| | | | 0.06 | 10.2 | 9.0 | 8.9 | 14.3 | 12.6 | 12.5 | 10.4 | 9.2 | 9.1 | 10.69 |
| | | | 0.08 | 10.1 | 8.9 | 8.9 | 14.7 | 12.6 | 12.5 | 10.3 | 9.2 | 9.1 | 10.70 |
| | | | 0.1 | 10.1 | 8.9 | 8.9 | 14.6 | 12.7 | 12.5 | 10.3 | 9.2 | 9.1 | 10.70 |
| | | fMLLR | 0 | 10.6 | 9.4 | 8.9 | 14.5 | 13.0 | 11.9 | 10.0 | 8.5 | 7.8 | <u>10.51</u> |
| | | | 0.02 | 10.5 | 9.3 | 8.7 | 14.5 | 12.9 | 11.9 | 10.0 | 8.4 | 7.9 | 10.46 |
| | | | 0.04 | 10.5 | 9.2 | 8.7 | 14.5 | 12.8 | 11.7 | 9.9 | 8.4 | 7.9 | 10.40 |
| | | | 0.06 | 10.4 | 9.1 | 8.6 | 14.5 | 12.7 | 11.7 | 9.9 | 8.4 | 7.8 | 10.34 |
| | | | 0.08 | 10.4 | 9.1 | 8.6 | 14.5 | 12.7 | 11.7 | 10.0 | 8.3 | 7.7 | <span style="color:red">**10.33**</span> |
| | | | 0.1 | 10.5 | 9.2 | 8.7 | 14.7 | 12.8 | 11.7 | 10.0 | 8.4 | 7.8 | 10.42 |
| $\mathcal{B}$ | $\hat{x}$-s4018 | MFCC | 0 | 11.4 | 9.8 | 9.8 | 15.6 | 13.3 | 13.1 | 11.4 | 9.9 | 9.9 | <u>11.58</u> |
| | | | 0.02 | 11.4 | 9.7 | 9.6 | 15.3 | 13.2 | 12.9 | 11.6 | 10.1 | 10.0 | **11.53** |
| | | | 0.04 | 11.5 | 9.8 | 9.7 | 15.6 | 13.3 | 13.1 | 11.9 | 10.3 | 10.2 | 11.71 |
| | | | 0.06 | 11.4 | 9.7 | 9.6 | 15.4 | 13.1 | 12.9 | 11.9 | 10.2 | 10.1 | 11.59 |
| | | | 0.08 | 11.4 | 9.7 | 9.6 | 15.7 | 13.2 | 13.0 | 11.7 | 10.0 | 9.9 | 11.58 |
| | | | 0.1 | 11.4 | 9.6 | 9.6 | 15.6 | 13.2 | 13.0 | 11.8 | 10.1 | 9.9 | 11.58 |
| | | $\tilde{x}$ | 0 | 10.7 | 9.9 | 9.9 | 14.7 | 13.5 | 13.0 | 10.7 | 9.5 | 9.9 | <u>11.31</u> |
| | | | 0.02 | 10.6 | 9.9 | 9.7 | 14.6 | 13.2 | 13.0 | 11.1 | 9.9 | 10.3 | 11.37 |
| | | | 0.04 | 10.7 | 9.9 | 9.8 | 14.7 | 13.1 | 12.9 | 11.2 | 9.9 | 10.3 | 11.39 |
| | | | 0.06 | 10.6 | 9.7 | 9.6 | 14.6 | 13.1 | 12.9 | 11.4 | 9.9 | 10.3 | 11.34 |
| | | | 0.08 | 10.7 | 9.9 | 9.8 | 14.7 | 13.1 | 13.0 | 11.2 | 9.9 | 10.3 | 11.40 |
| | | | 0.1 | 10.6 | 9.9 | 9.8 | 14.8 | 13.2 | 12.9 | 11.3 | 10.0 | 10.3 | 11.42 |
| Heck et al. [32] (1-st in ZRSC17) | | | | 10.1 | 8.7 | 8.5 | 13.6 | 11.7 | 11.3 | 8.8 | 7.4 | 7.3 | 9.71 |

The best submitted system to ZeroSpeech 2017 proposed by Heck et al. [32] is also listed in the Tables for reference.

It can be observed from Tables 4.8 & 4.9 that,

(1). Speaker adversarial training is complementary to out-of-domain ASR based speaker adaptation, especially in the across-speaker condition. The absolute error rate reduction in the two conditions achieved by adopting fMLLRs as inputs to DPGMM and AMTL-DNN are 0.18% and 0.15%.

(2). Speaker adversarial training has little effectiveness on systems adopting FHVAE-based disentangled representation. In the within-speaker condition, systems using reconstructed MFCCs in DPGMM frame labeling witness a severe degradation.

(3). The best performance is achieved with fMLLR features as inputs to DPGMM and AMTL-DNN. The across-/within-speaker ABX error rate is 10.33%/7.81%.

**Table 4.9:** Within-speaker ABX error rates (%) on systems combining different speaker adaptation approaches.

| Ref. | Input feature DPGMM | DNN | $\lambda$ | English | | | French | | | Mandarin | | | Avg. |
|---|---|---|---|---|---|---|---|---|---|---|---|---|---|
| | | | | 1s | 10s | 120s | 1s | 10s | 120s | 1s | 10s | 120s | |
| $\mathcal{A}$ | fMLLR | MFCC | 0 | 6.9 | 6.1 | 6.1 | 9.5 | 8.2 | 8.1 | 9.5 | 8.1 | 8.1 | <u>7.84</u> |
| | | | 0.02 | 7.0 | 6.3 | 6.2 | 9.7 | 8.1 | 7.9 | 9.4 | 8.3 | 8.1 | 7.89 |
| | | | 0.04 | 7.0 | 6.1 | 6.2 | 9.5 | 8.3 | 8.0 | 9.4 | 8.3 | 8.2 | 7.89 |
| | | | 0.06 | 7.0 | 6.2 | 6.2 | 9.5 | 8.3 | 7.8 | 9.7 | 8.3 | 8.3 | 7.92 |
| | | | 0.08 | 7.1 | 6.1 | 6.1 | 9.5 | 8.6 | 8.0 | 9.7 | 8.2 | 8.1 | 7.93 |
| | | | 0.1 | 7.0 | 6.1 | 6.2 | 9.6 | 8.5 | 8.2 | 9.7 | 8.2 | 8.1 | 7.96 |
| | | fMLLR | 0 | 7.1 | 6.5 | 6.2 | 9.4 | 9.2 | 7.9 | 9.4 | 8.3 | 7.6 | <u>7.96</u> |
| | | | 0.02 | 7.0 | 6.5 | 6.1 | 9.4 | 9.0 | 7.9 | 9.2 | 8.2 | 7.6 | 7.88 |
| | | | 0.04 | 7.0 | 6.5 | 6.1 | 9.2 | 8.9 | 7.9 | 9.2 | 8.3 | 7.6 | 7.86 |
| | | | 0.06 | 7.0 | 6.4 | 6.1 | 9.3 | 8.8 | 8.0 | 9.1 | 8.3 | 7.6 | 7.84 |
| | | | 0.08 | 6.9 | 6.4 | 6.0 | 9.3 | 8.8 | 7.7 | 9.5 | 8.1 | 7.6 | <span style="color:red">7.81</span> |
| | | | 0.1 | 7.1 | 6.5 | 6.1 | 9.4 | 8.7 | 7.7 | 9.2 | 8.0 | 7.6 | **7.81** |
| $\mathcal{B}$ | $\hat{x}$-s4018 | MFCC | 0 | 7.4 | 6.2 | 6.2 | 9.8 | 8.4 | 8.1 | 10.3 | 8.4 | 8.4 | <u>8.13</u> |
| | | | 0.02 | 7.6 | 6.2 | 6.2 | 10.1 | 8.4 | 8.2 | 10.6 | 8.6 | 8.6 | 8.28 |
| | | | 0.04 | 7.6 | 6.3 | 6.2 | 10.1 | 8.5 | 8.2 | 10.8 | 8.8 | 8.8 | 8.37 |
| | | | 0.06 | 7.7 | 6.3 | 6.2 | 10.0 | 8.3 | 8.2 | 10.7 | 8.8 | 8.7 | 8.32 |
| | | | 0.08 | 7.7 | 6.2 | 6.3 | 10.1 | 8.5 | 8.3 | 10.8 | 8.8 | 8.6 | 8.37 |
| | | | 0.1 | 7.8 | 6.4 | 6.3 | 10.2 | 8.5 | 8.3 | 11.0 | 8.6 | 8.7 | 8.42 |
| | | $\tilde{x}$ | 0 | 6.9 | 6.3 | 6.3 | 9.2 | 8.5 | 8.2 | 9.4 | 8.4 | 8.4 | <u>7.96</u> |
| | | | 0.02 | 7.0 | 6.3 | 6.2 | 9.5 | 8.7 | 8.4 | 9.9 | 8.8 | 9.0 | 8.20 |
| | | | 0.04 | 7.0 | 6.5 | 6.3 | 9.5 | 8.6 | 8.3 | 10.1 | 8.8 | 9.0 | 8.23 |
| | | | 0.06 | 7.0 | 6.4 | 6.3 | 9.5 | 8.7 | 8.4 | 10.4 | 9.0 | 9.0 | 8.30 |
| | | | 0.08 | 7.1 | 6.5 | 6.3 | 9.4 | 8.6 | 8.5 | 10.2 | 8.9 | 9.0 | 8.28 |
| | | | 0.1 | 7.1 | 6.4 | 6.3 | 9.5 | 8.6 | 8.4 | 10.3 | 8.9 | 9.0 | 8.28 |
| Heck et al. [32] (1-st in ZRSC17) | | | | 6.9 | 6.2 | 6.0 | 9.7 | 8.7 | 8.4 | 8.8 | 7.9 | 7.8 | 7.82 |

Our best system is competitive with the Challenge winner [32] in the within-speaker condition, while slightly worse than theirs in the across-speaker condition.

## 4.5.5 Discussion

Three different approaches to speaker adaptation have been investigated toward improving the robustness of DNN-BNF features for unsupervised subword modeling. They serve to achieve different goals and at the same time, they have different requirements on the adaptation data. fMLLR and disentangled representation learning are applied at the front end to make the input features speaker-invariant for DNN-BNF training and DPGMM frame labeling, whilst speaker adversarial training is applied at the bank end to suppress speaker variation of the BNF representation.

The experimental results show that techniques applied at the front end are more effective than at the back end.

The use of fMLLR requires transcribed out-of-domain speech data. As out-of-domain data are literally unlimited in amount and diversity, the benefit of this approach could be further exploited. Indeed the experimental results show that fMLLR with out-of-domain data give better performance than disentangled representation learning and speaker adversarial training without out-of-domain data. It must be noted that adversarial training can also be applied with out-of-domain data (either transcribed or untranscribed). Its effectiveness is worth further investigation.

## 4.6   Summary

In this Section, three speaker adaptation approaches that can be directly applied to our baseline unsupervised subword modeling system are introduced and extensively studied. The three approaches, namely, fMLLR estimation by an out-of-domain ASR, disentangled speech representation learning and speaker adversarial training, are investigated individually in our concerned task. Their combination is further studied.

The experiments are conducted using official database and evaluation metrics in ZeroSpeech 2017. Experimental results demonstrate the effectiveness of all the three approaches. The out-of-domain ASR based adaptation achieves the most significant performance improvement compared to our baseline. Speaker adversarial training achieves the least amount of improvement among the three approaches. Combining out-of-domain ASR based adaptation and adversarial training contributes to further improvement, in which our best performance (10.3%/7.8%) is achieved. In contrast, adversarial training is not complementary to disentangled representation learning.

# Chapter 5

# Frame labeling in unsupervised subword modeling

In the baseline DNN-BNF system, frame labeling is realized by applying DPGMM clustering on speech frames. DPGMM clustering does not require a pre-defined cluster number, making it suitable for zero-resource speech modeling. Previous studies showed its effectiveness in unsupervised subword modeling [27, 32]. Nevertheless, DPGMM-based frame labeling has two major limitations. First, as neighboring frames are assumed to be independent, contextual information in speech is not taken into account in determining the labels. Second, DPGMM is prone to producing over-fragmented speech units [111, 112].

Towards both limitations, methods of improving DPGMM frame labels are developed in this chapter. On one hand, a full-fledged GMM-HMM is trained to facilitate better modeling of contextual information. The transcriptions required for the training are initialized via DPGMM clustering. Following the terminologies used in [46], the resulted model is referred to as *DPGMM-HMM*.

To alleviate the over-fragment problem, a new algorithm is proposed to filter out infrequent labels in DPGMM clustering results, so as to control the number of inferred clusters. Experimental results reveal that these infrequent labels adversely affect the performance of unsupervised subword modeling, and that the proposed label filtering algorithm is effective.

This chapter also presents our attempt of leveraging benefits of different types of frame labels. Out-of-domain speech and language resources are exploited to enable language-mismatched frame labeling. Such frame labels are obtained from decoding results of one or multiple out-of-domain ASR systems. It is expected that labels generated by out-of-domain ASR decoding and those via DPGMM-HMM acoustic modeling are complementary and can be jointly used in unsupervised subword modeling.

Section 5.1 describes the DPGMM-HMM frame labeling approach, and the proposed label filtering algorithm. The out-of-domain ASR based frame labeling approach, as well as the combined use of different frame label types, are introduced in Section 5.2. Experiments are reported in Section 5.3, followed by the chapter summary in Section 5.4.

## 5.1 DPGMM-HMM frame labeling

The proposed DPGMM-HMM frame labeling approach consists of three stages, i.e., DPGMM clustering, frame label filtering and supervised GMM-HMM acoustic modeling. DPGMM clustering is performed to provide initial labels. These initial labels are processed by a label filtering algorithm to discard infrequent frame clusters. The filtered labels are subsequently used in supervised context-dependent GMM-HMM (CD-GMM-HMM) acoustic modeling. Finally, the CD-GMM-HMM is applied to force-align target zero-resource speech to generate the desired frame labels.

DPGMM clustering is implemented in the same way as discussed in Section 3.2. The label filtering algorithm and DPGMM-HMM acoustic modeling are explained in the following sections.

### 5.1.1 Label filtering

For a specific target language, let us assume that $K$ Gaussian components (clusters) are obtained by DPGMM clustering. The labels are denoted as $l_1, l_2, \ldots, l_N$ for an utterance with $N$ frames. Let $c_k$ be the number of frames labeled as cluster $k$,

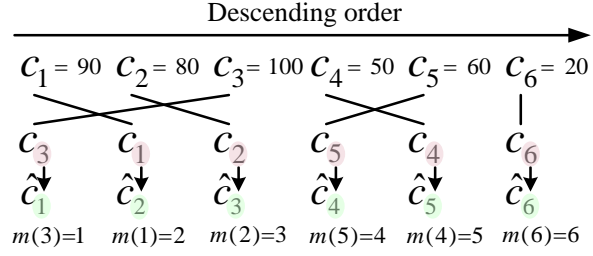

**Figure 5.1:** Example of cluster-size sorting.

i.e.,

$$c_k = \sum_{i=1}^{N} \mathbb{1}(l_i = k), k \in \{1, 2, \ldots, K\}, \tag{5.1}$$

where $\mathbb{1}(\cdot)$ denotes the indicator function.

The elements in $\{c_1, c_2, \ldots, c_K\}$ are sorted in descending order

$$\{\hat{c}_1, \hat{c}_2, \ldots, \hat{c}_K | \hat{c}_1 \geq \hat{c}_2 \geq \ldots \geq \hat{c}_K\}. \tag{5.2}$$

$m(\cdot)$ denotes the index mapping function, i.e.,

$$\hat{c}_k = c_{m(k)}. \tag{5.3}$$

Figure 5.1 shows an example of cluster-size sorting.

Let $P$ be the designated percentage of frame labels to be retained. The relevant frames are from $K_{cut}$ "dominant" clusters, where

$$K_{\text{cut}} = \arg\min_{K'} \frac{\sum_{k=1}^{K'} \hat{c}_k}{N} \geq P. \tag{5.4}$$

Let $\mathcal{O}$ denote the collection of all removed frame labels, i.e.,

$$\mathcal{O} = \left\{ l_i : l_i \in \mathcal{F}, i \in \{1, 2, \ldots, N\} \right\}, \tag{5.5}$$

where

$$\mathcal{F} = \left\{ m(K_{\text{cut}} + 1), \ldots, m(K) \right\}. \tag{5.6}$$

$\mathcal{F}$ contains the indices of $K - K_{cut}$ clusters that are the least frequent to occur.

Speech frames being assigned to these clusters are considered as outliers.

In the extreme case when $P$ is set to 1, $\mathcal{F}$ and $\mathcal{O}$ would be empty sets. The smaller the value of $P$, the larger the proportion of removed labels. The label filtering algorithm is summarized as in Algorithm 5.1.

---

**Algorithm 5.1:** DPGMM label filtering algorithm

---

   **Input:** $l_1, l_2, \ldots, l_N$, $P$
   **Output:** $\mathcal{O}$

1: Calculate $c_k$ by Equation (5.1).
2: Sort $\{c_1, c_2, \ldots, c_K\}$ in descending order.
3: Calculate $m(k)$ by Equation (5.3).
4: Calculate $K_{\text{cut}}$ by Equation (5.4) and $P$.
5: Select a subset of $l_1, l_2, \ldots, l_N$ as $\mathcal{O}$, by Equation (5.5)&(5.6). {Frame labels that are removed.}

---

It is worth noting that filtering out small-sized clusters produced by DPGMM is mainly for the consideration of practical implementation. The DPGMM algorithm automatically generates variable-sized clusters according to the underlying structure of input data. Small-sized clusters are meaningful from the perspective of the DPGMM algorithm itself. Nevertheless, in the concerned task, filtering out small-sized clusters is a simple yet effective (as will be shown in experiments) method to discard less useful labels and improve the overall quality of frame labels.

## 5.1.2 DPGMM-HMM acoustic modeling

Each of the DPGMM clusters can be regarded as a phone-like speech unit, or pseudo phone. The sequence of frame labels (after label filtering) can be converted into a pseudo phone transcription by collapsing neighboring identical labels. For example, a sequence of frame labels "1,3,3,3,7,10,10" would lead to transcription "1,3,7,10". Based on pseudo transcriptions, DPGMM-HMM acoustic modeling is done by following the standard supervised training pipeline, i.e., proceeding from monophone model training with uniform time alignment to CD-GMM-HMM. The CD-GMM-HMM AM is further refined by performing speaker adaptive training. The resulted AM is referred to as the DPGMM-HMM AM.

The DPGMM-HMM AM is used to produce time alignment information for subsequent MTL-DNN modeling. To be distinguishable from DPGMM labels, the labels obtained form DPGMM-HMM forced alignment are referred to as DPGMM-HMM labels.

## 5.2 MTL-DNN training with multiple types of labels

### 5.2.1 Out-of-domain ASR decoding labels

Frame labeling could be done with an out-of-domain ASR system, which is typically trained with a large amount of transcribed speech in a resource-rich language. The AM in such an ASR system provides fine-grained speech representation of the original language. Given a speech utterance in a different language, the ASR system can be applied to assign a language-mismatched state-level or phone-level label to each frame in the utterance. The idea can be naturally extended to using multiple ASR systems that desirably provide a wide coverage of phonetic diversity.

The ASR decoding output depends on the relative weighting of AM and LM. In this study, the LM is assigned a very small weight, such that the acquired frame labels mainly reflect acoustic properties of the target speech.

### 5.2.2 Multi-task learning with multiple label prediction tasks

Out-of-domain ASR decoding and DPGMM-HMM acoustic modeling provide two different types of frame labels for target speech utterances. The DPGMM-HMM labels incorporate statistical information of the acoustic properties of target speech. The ASR-decoded labels leverage the phonetic information acquired from other languages. It is expected that they would contribute complementarily in subword-discriminative feature learning. In this study, their complementarity is investigated in the multi-task learning (MTL) framework.

The proposed MTL-DNN system is depicted as in Fig 5.2. The training involves a total of $M + N$ tasks, which includes $M$ zero-resource target languages and $N$ out-of-domain ASR systems. Each of the tasks is represented by a task-specific softmax output layer in the DNN. For the zero-resource language tasks, state-level or phone-level DPGMM-HMM labels are used as target labels. The decoding output from each of the out-of-domain ASR systems provides one set of labels.

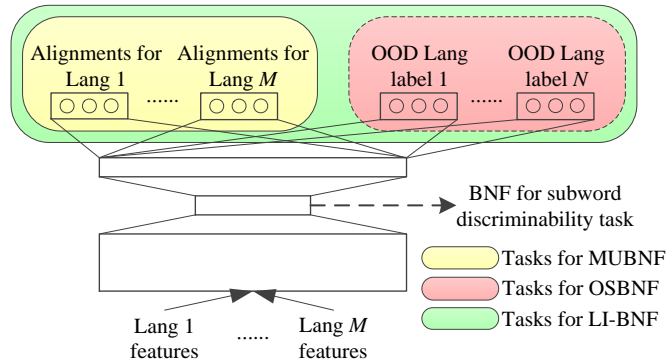

**Figure 5.2:** MTL-DNN for extracting LI-BNF, MUBNF and OSBNF. "OOD" stands for out-of-domain.

If the MTL-DNN trained only on the $M$ target language tasks, the extracted BNFs are referred to as multilingual unsupervised BNFs (MUBNFs). When out-of-domain ASR tasks are added, the BNFs are named language-independent BNFs (LI-BNFs). In the case that only the out-of-domain ASR tasks are involved, the extracted BNFs are referred to as out-of-domain supervised BNFs (OSBNFs).

For the shared-hidden-layer structure in the MTL-DNN, multi-layer perceptron (MLP) has been commonly used [30, 31, 34, 113]. In this study, in addition to MLP, we investigate the use of long short-term memory (LSTM) [114] and bi-directional LSTM (BLSTM) [115], which were shown to perform better than MLP in conventional supervised acoustic modeling.

On the other hand, BNF representation can also be obtained from the DNN AM pre-trained on a resource-rich language [1]. This is considered as a transfer learning approach [116]. The resulted BNF, denoted as transfer learning BNF (TLBNF), is expected to further enrich the feature representation for subword modeling.

## 5.2.3 Complete system

The complete system framework of multi-label assisted unsupervised subword modeling is shown as in Figure 5.3. As can be seen, the input features to DPGMM-

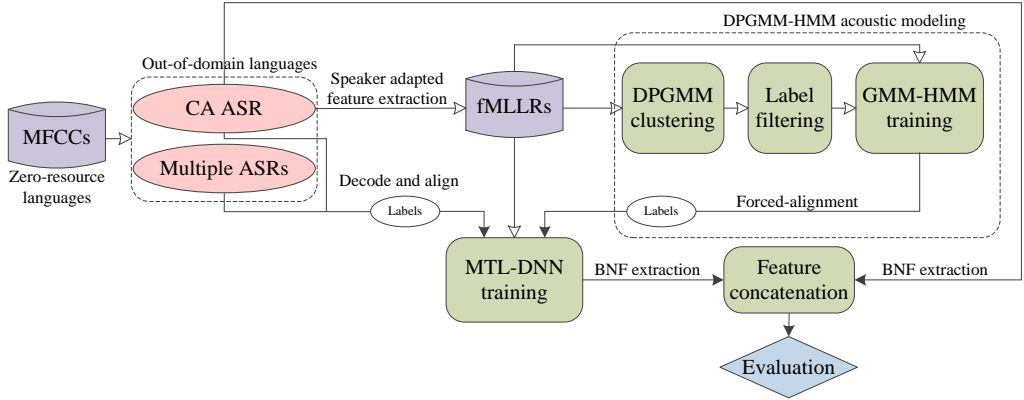

**Figure 5.3:** Complete system framework of multi-label assisted unsupervised subword modeling.

HMM acoustic modeling and MTL-DNN are both fMLLR features estimated by an out-of-domain ASR. Frame labels required for MTL-DNN training is performed with multiple types of labels as described in previous sections. BNFs are extracted from the trained MTL-DNN and evaluated on the ABX discriminability task. The TLBNF representation is optionally concatenated to the BNFs by the MTL-DNN, and evaluated on the same task.

## 5.3 Experiments

### 5.3.1 Experimental setup

#### Out-of-domain ASR systems

Four out-of-domain ASR systems are utilized and investigated in our experiments. They cover the languages of Cantonese (CA), Czech (CZ), Hungarian (HU) and Russian (RU). The Cantonese ASR system is trained with the settings as given in Section 4.5.1. Two sets of AMs, i.e., CD-GMM-HMM-SAT and DNN-HMM models, are investigated. The CD-GMM-HMM-SAT is the same as that in the baseline. The DNN-HMM is trained using Kaldi [88]. The training procedure follows the settings used in training the baseline DNN as described in Section 3.6.1. The input features for DNN-HMM are fMLLRs with contextual splicing $\{0, \pm1, \pm2, \pm3, \pm4, \pm5\}$, with the fMLLRs estimated by the CD-GMM-HMM-SAT. The DNN-HMM training labels are acquired from CD-GMM-HMM-SAT time alignment. The DNN-HMM model is a 7-layer MLP, with the layer configuration (from bottom to top) 440-1024 × 5-40-1024-2462. The dimension of the output layer is determined by the number of states in CD-GMM-HMM-SAT. The layers are activated with sigmoid function, except for the 40-dimension linear BN layer. The network is trained to optimize the cross-entropy criterion.

The CZ, HU and RU phone recognizers are developed by the Brno University of Technology (BUT) [117]. They adopt a 3-layer MLP structure, in which the first two are sigmoid layers and the third is a softmax layer. The MLP were trained with the SpeechDat-E databases [118]. The numbers of modeled phones for CZ, HU and RU are 45, 61 and 52, respectively. The amount of training data are 9.7, 7.9 and 14.0 hours, respectively. The cross-entropy criterion was used for MLP training.

## DPGMM frame clustering and label filtering

For each target language, training speech frames for different languages are clustered by applying the DPGMM algorithm to 40-dimension fMLLR features estimated by the Cantonese ASR, (refer to Section 4.5.1). Hyper-parameters of DPGMM clustering is not tuned. The numbers of iterations for English and French are determined by preliminary experiments. More specifically, the number of iterations tested for English is in the range $\{40, 80, \ldots, 680\}$ and that for French in the range $\{40, 80, \ldots, 400\}$. The optimal numbers of iterations were 120 and 200 respectively. For Mandarin, the number of iterations was empirically determined to be 3000. The resulted numbers of DPGMM clusters for English, French and Mandarin are 1118, 1345 and 596, respectively. After frame clustering, each frame is assigned a cluster label. Figure 5.4 shows the frame labeling results in the form of cumulative distribution function (CDF) for the three target languages. The clusters are sorted according to their cluster sizes in descending order. The horizontal axis denotes the number of DPGMM clusters and the vertical axis denotes the proportion of frames. Each point $(K_i, Q_i)$ on the CDF curve represents the proportion of frame labels $Q_i$ that the largest $K_i$ clusters can cover.

For label filtering, we evaluated the value of $P$ in the range from 0.6 to 0.95

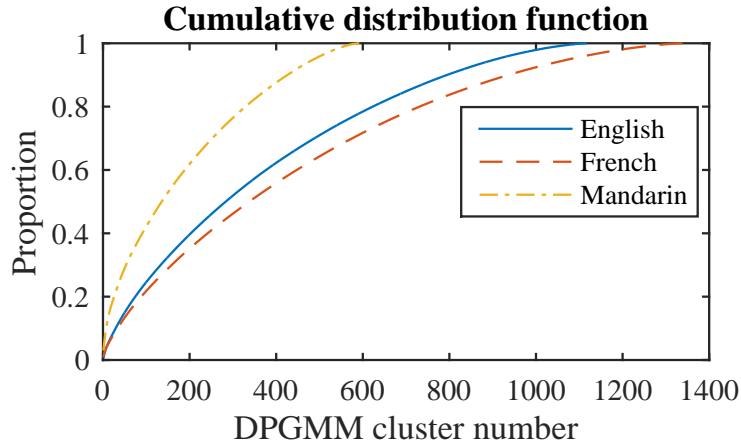

**Figure 5.4:** Clustering results in the form of cumulative distribution function for the three target languages. Clusters are sorted according to cluster size in descending order.

(step size of 0.05). After filtering, the frame-level label sequences are converted into pseudo transcriptions, for the training of DPGMM-HMM AMs.

## DPGMM-HMM acoustic modeling

Supervised training of the DPGMM-HMM AMs is carried out with the pseudo transcriptions. Different from the conventional 3-state HMM topology, during DPGMM-HMM training a 1-state HMM is used to model each pseudo phone. This alleviates the problem of unsuccessful forced alignments, as the numbers of pseudo phones for target languages are significantly larger than the number of linguistically-defined phones in a typical language. The input features for DPGMM-HMM are 40-dimension fMLLRs estimated by the Cantonese ASR. The training starts from CI-GMM-HMM to CD-GMM-HMM, followed by VTLN and fMLLR-based SAT[1]. After training, the numbers of CD-HMM states for English, French and Mandarin are 2818, 2856 and 2688, respectively. The training of DPGMM-HMM AMs are implemented by Kaldi.

## MTL-DNN training for BNF generation

The MTL-DNN model is trained with all the three target zero-resource languages, from which BNFs are extracted and evaluated by the ABX subword dis-criminability task. There are two types of tasks for MTL, namely, DPGMM-HMM alignment prediction task and out-of-domain ASR label prediction task. In the first case, three tasks are included, i.e., frame alignments generated by DPGMM-HMM AMs, one for each target zero-resource language. In the second case, four tasks corresponding to Cantonese, Czech, Hungarian and Russian recognizers' senone labels are included. The senone labels are generated by decoding with LM to AM weight ratio set to 0.001. After MTL-DNN training, 40-dimension HMM-LI-BNFs[2] are extracted for the ABX task evaluation. Similarly, HMM-MUBNFs[3], extracted by MTL-DNN with DPGMM-HMM alignment tasks, and OSBNFs, extracted by MTL-DNN with one or more out-of-domain phone recognizers' senone labels, are also evaluated by the ABX task. The dimensions of both HMM-MUBNFs and OSBNFs are 40. As illustrated in Figure 5.2, we defined several BNF representations according to the tasks included in MTL-DNN training. The configurations for (HMM-)MUBNF, OSBNF and (HMM-)LI-BNF are listed in Table 5.1.

---

[1]LDA and MLLT are not estimated, as no improvement was found.

**Table 5.1:** Configurations for (HMM-)MUBNF, OSBNF and (HMM-)LI-BNF representations. 'EN', 'FR' and 'MA' are abbreviations for English, French and Mandarin.

| Task label from | DPGMM | | | DPGMM-HMM | | | CA | CZ | HU | RU |
|---|---|---|---|---|---|---|---|---|---|---|
| Train set | EN | FR | MA | EN | FR | MA | Pooling EN, FR and MA | | | |
| MUBNF | ✓ | ✓ | ✓ | | | | | | | |
| OSBNF1 | | | | | | | ✓ | | | |
| OSBNF2 | | | | | | | ✓ | ✓ | ✓ | ✓ |
| LI-BNF1 | ✓ | ✓ | ✓ | | | | ✓ | | | |
| LI-BNF2 | ✓ | ✓ | ✓ | | | | ✓ | ✓ | ✓ | ✓ |
| HMM-MUBNF | | | | ✓ | ✓ | ✓ | | | | |
| HMM-LI-BNF1 | | | | ✓ | ✓ | ✓ | ✓ | | | |
| HMM-LI-BNF2 | | | | ✓ | ✓ | ✓ | ✓ | ✓ | ✓ | ✓ |

## DNN structures

The MTL-DNN is implemented in three different model structures: MLP, LSTM and BLSTM. The input features are 40-dimension fMLLRs spliced with context size ±5. For MLP, we follow the structure in our baseline system, i.e., the dimensions of shared hidden layers in the MLP are 440-1024×5-40-1024. Sigmoid activation is used in all hidden layers except that the 40 neurons in the BN layer use linear activation functions. The learning rate for MLP training is set at 0.008 at the beginning, and halved when no improvement is observed on a cross-validation set. The mini-batch size is 256.

---

[2]The prefix 'HMM-' emphasizes the use of DPGMM-HMM alignments, rather than DPGMM cluster labels.

The LSTM model comprises 2 LSTM layers with 320-dimension cell activation vectors, and 1024-dimension outputs. A 40-dimension BN layer followed by a 1024-dimension fully-connected (FC) layer is set on top of LSTMs. For the BLSTM model, there are 2 pairs of forward and backward LSTM layers. Each bi-directional layer has 320-dimension cell activation vectors and 512-dimension outputs. A BN layer followed by an FC layer is set on top of BLSTMs, with the same configuration as in the LSTM. The activation function in (B)LSTMs is tanh. The learning rate is $2e-4$ initially, and halved under the same criteria as for MLP. The truncated back-propagation through time (BPTT) algorithm [119] is used to train (B)LSTM, with a fixed time step $T_{bptt} = 20$. Note that the model parameters of LSTM and BLSTM structures were tuned in preliminary studies.

**TLBNF generation**

The TLBNFs for target zero-resource languages are generated by applying the Cantonese DNN-HMM AM as the feature extractor. During TLBNF extraction, all the parameters of the DNN-HMM are fixed. The fMLLR features for target languages are fed as inputs to the DNN-HMM till its BN layer to generate TLBNFs.

## 5.3.2   Results and discussion

Tables 5.2 and 5.3 provide a master summary to facilitate performance comparison among different BNF representations, in the across- and within-speaker conditions respectively. The feature representations being compared are organized in three groups, marked by $\mathcal{A}, \mathcal{B}$ and $\mathcal{C}$ in the Tables.

Systems in groups $\mathcal{A}$ and $\mathcal{B}$ all use multilingual BNF representations, which are learned with different combinations of frame labels. The DPGMM labels and DPGMM-HMM labels are used in group $\mathcal{A}$ and group $\mathcal{B}$ respectively. As described in Section 5.3.1 and Table 5.1, OSBNF1 and OSBNF2 are trained with out-of-domain ASR senone labels, and LI-BNF1 and LI-BNF2 are trained with both DPGMM labels and out-of-domain ASR senone labels. In the third group, "HMM(S)" and "HMM(P)" denote the use of state-level and phone-level HMM alignments respectively for label generation. Systems in group $\mathcal{C}$ are built on different combination of BNF features. The "+" sign represents the concatenation of two frame-level feature representations. Our experimental results shown in Tables 5.2 and 5.3 are obtained by using the MLP structure in MTL-DNN. Label filtering is not applied at this stage. In addition, two representative systems that achieved very good performances in ZeroSpeech 2017 [32, 37] are also listed in the Tables.

**Table 5.2:** Across-speaker ABX error rates (%) on BNFs learned by our proposed multiple frame labeling approaches and state of the art of ZeroSpeech 2017. MLP is adopted as the DNN structure. Label filtering is not applied.

| | | English | | | French | | | Mandarin | | | Avg. |
|---|---|---|---|---|---|---|---|---|---|---|---|
| | | 1s | 10s | 120s | 1s | 10s | 120s | 1s | 10s | 120s | |
| | MUBNF (in Section 4.5.1) | 10.9 | 9.5 | 8.9 | 15.2 | 13.0 | 12.0 | 10.5 | 8.9 | 8.2 | 10.8 |
| $\mathcal{A}$ | OSBNF1 | 10.0 | 9.7 | 8.6 | 13.9 | 13.4 | 11.6 | 9.0 | 8.4 | 7.5 | 10.2 |
| | OSBNF2 | 9.5 | 9.2 | 7.9 | 13.1 | 13.0 | 11.3 | 9.4 | 8.7 | 7.9 | 10.0 |
| | LI-BNF1 | 10.0 | 8.9 | 8.2 | 14.3 | 12.9 | 11.5 | 9.5 | 8.5 | 7.7 | 10.2 |
| | LI-BNF2 | 9.4 | 8.7 | 7.8 | 13.4 | 12.7 | 11.0 | 9.3 | 8.6 | 7.7 | 9.8 |
| $\mathcal{B}$ | HMM(S)-MUBNF | 10.2 | 9.3 | 8.6 | 14.5 | 13.0 | 11.9 | 10.7 | 9.2 | 8.4 | 10.6 |
| | HMM(P)-MUBNF | 10.4 | 9.2 | 8.7 | 14.5 | 12.7 | 11.7 | 10.4 | 8.9 | 8.2 | 10.5 |
| | HMM(P)-LI-BNF1 | 9.7 | 8.7 | 8.0 | 13.7 | 12.3 | 11.1 | 9.7 | 8.4 | 7.6 | 9.9 |
| | HMM(P)-LI-BNF2 | 9.3 | 8.7 | 7.8 | 13.0 | 12.4 | 11.0 | 9.5 | 8.5 | 7.7 | 9.8 |
| $\mathcal{C}$ | TLBNF | 10.6 | 9.6 | 8.7 | 14.2 | 13.2 | 11.5 | 8.5 | 7.6 | 6.7 | 10.1 |
| | TLBNF+LI-BNF1 | 10.3 | 9.3 | 8.4 | 13.9 | 12.9 | 11.4 | 8.5 | 7.6 | 6.7 | 9.9 |
| | TLBNF+LI-BNF2 | 10.4 | 9.4 | 8.5 | 14.0 | 13.0 | 11.3 | 8.5 | 7.6 | 6.6 | 9.9 |
| | TLBNF+HMM(P)-LI-BNF1 | 10.3 | 9.4 | 8.4 | 13.9 | 12.9 | 11.3 | 8.5 | 7.6 | 6.6 | 9.9 |
| | TLBNF+MUBNF+OSBNF1 | 9.9 | 9.0 | 8.2 | 13.6 | 12.6 | 11.1 | 8.4 | 7.7 | 6.7 | 9.7 |
| | TLBNF+HMM(P)-MUBNF+OSBNF1 | 10.0 | 9.0 | 8.2 | 13.6 | 12.6 | 11.1 | 8.4 | 7.6 | 6.7 | 9.7 |
| | TLBNF+HMM(P)-MUBNF+OSBNF2 | 10.0 | 9.0 | 8.2 | 13.6 | 12.6 | 11.1 | 8.4 | 7.6 | 6.7 | 9.7 |
| | Heck et al. [32] (1-st in ZRSC17) | 10.1 | 8.7 | 8.5 | 13.6 | 11.7 | 11.3 | 8.8 | 7.4 | 7.3 | 9.7 |
| | Chorowski et al. [37] | 9.3 | 9.3 | 9.3 | 11.9 | 11.4 | 11.6 | 8.6 | 8.5 | 8.5 | 9.8 |
| | Supervised topline [14] | 8.6 | 6.9 | 6.7 | 10.6 | 9.1 | 8.9 | 12.0 | 5.7 | 5.1 | 8.2 |

**Effectiveness of multilingual BNFs**

The following observations can be made on the performances of the learned multilingual BNF representations:

(1). The effectiveness of MUBNF, which is reported in Section 4.5.1, can be improved by training the MTL-DNN with additional out-of-domain ASRs' senone labels. With the Cantonese ASR's senone labels included as one of the training tasks, the LI-BNF1 representation reduces within-/across-speaker ABX error rates by absolute 0.2%/0.6% as compared to MUBNF. When the senone labels of Czech, Hungarian and Russian are added, the resulted LI-BNF2 representation shows a further improvement of absolute 0.4% under the across-speaker condition. This shows that out-of-domain acoustic-phonetic knowledge provides complementary information to the in-domain clustering labels for feature learning. The performance gain of OSBNF2 over OSBNF1, as well as that of LI-BNF2 over LI-BNF1, confirm the benefit of exploiting a wider coverage of language resources.

**Table 5.3:** Within-speaker ABX error rates (%) on BNFs learned by our proposed multiple frame labeling approaches and state of the art of ZeroSpeech 2017. MLP is adopted as the DNN structure. Label filtering is not applied.

|  |  | English | | | French | | | Mandarin | | | Avg. |
|---|---|---|---|---|---|---|---|---|---|---|---|
|  |  | 1s | 10s | 120s | 1s | 10s | 120s | 1s | 10s | 120s | |
|  | MUBNF (in Section 4.5.1) | 7.4 | 6.9 | 6.3 | 9.6 | 9.0 | 8.1 | 9.8 | 8.8 | 8.1 | 8.2 |
| $\mathcal{A}$ | OSBNF1 | 7.2 | 7.1 | 6.3 | 10.2 | 9.7 | 8.7 | 9.1 | 8.6 | 7.6 | 8.3 |
|  | OSBNF2 | 6.8 | 6.7 | 5.9 | 9.5 | 9.2 | 8.3 | 9.7 | 8.9 | 8.0 | 8.1 |
|  | LI-BNF1 | 6.9 | 6.6 | 6.1 | 9.5 | 9.2 | 8.4 | 9.2 | 8.5 | 7.9 | 8.0 |
|  | LI-BNF2 | 6.6 | 6.4 | 5.7 | 9.1 | 9.3 | 8.2 | 9.5 | 8.7 | 8.1 | 8.0 |
| $\mathcal{B}$ | HMM(S)-MUBNF | 7.2 | 6.7 | 6.3 | 9.7 | 9.2 | 8.3 | 10.4 | 9.2 | 8.5 | 8.4 |
|  | HMM(P)-MUBNF | 7.1 | 6.6 | 6.2 | 9.4 | 9.1 | 7.8 | 9.9 | 8.8 | 8.2 | 8.1 |
|  | HMM(P)-LI-BNF1 | 6.8 | 6.3 | 5.8 | 9.1 | 8.7 | 7.8 | 9.1 | 8.5 | 7.6 | 7.7 |
|  | HMM(P)-LI-BNF2 | 6.6 | 6.4 | 5.7 | 9.2 | 8.8 | 8.1 | 9.2 | 8.6 | 7.9 | 7.8 |
| $\mathcal{C}$ | TLBNF | 7.2 | 6.8 | 6.1 | 9.6 | 9.0 | 8.0 | 8.7 | 7.6 | 6.8 | 7.8 |
|  | TLBNF+LI-BNF1 | 7.0 | 6.6 | 6.0 | 9.3 | 8.8 | 7.9 | 8.6 | 7.5 | 6.7 | 7.6 |
|  | TLBNF+LI-BNF2 | 7.1 | 6.6 | 6.0 | 9.4 | 8.9 | 7.8 | 8.7 | 7.5 | 6.8 | 7.6 |
|  | TLBNF+HMM(P)-LI-BNF1 | 7.0 | 6.6 | 6.0 | 9.4 | 8.8 | 7.8 | 8.6 | 7.5 | 6.7 | 7.6 |
|  | TLBNF+MUBNF+OSBNF1 | 6.8 | 6.4 | 5.8 | 9.0 | 8.8 | 7.8 | 8.5 | 7.7 | 6.8 | 7.5 |
|  | TLBNF+HMM(P)-MUBNF+OSBNF1 | 6.8 | 6.4 | 5.7 | 8.8 | 8.7 | 7.5 | 8.4 | 7.5 | 6.8 | 7.4 |
|  | TLBNF+HMM(P)-MUBNF+OSBNF2 | 6.7 | 6.4 | 5.8 | 9.0 | 8.8 | 7.5 | 8.3 | 7.5 | 6.8 | 7.4 |
|  | Heck et al. [32] (1-st in ZRSC17) | 6.9 | 6.2 | 6.0 | 9.7 | 8.7 | 8.4 | 8.8 | 7.9 | 7.8 | 7.8 |
|  | Chorowski et al. [37] | 5.8 | 5.7 | 5.8 | 7.1 | 7.0 | 6.9 | 7.4 | 7.2 | 7.1 | 6.7 |
|  | Supervised topline [14] | 6.5 | 5.3 | 5.1 | 8.0 | 6.8 | 6.8 | 9.5 | 4.2 | 4.0 | 6.2 |

The performance of OSBNF2 is inferior to OSBNF1 on Mandarin test set, but not on English and French. It is noted that OSBNF1 is learned by using the Cantonese ASR senone labels while OSBNF2 is learned by involving Cantonese and the other three European languages. Cantonese, being a Chinese dialect, is apparently closer to Mandarin than Czech, Hungarian and Russian in terms of acoustic-phonetic properties. The experimental results imply that the frame labels generated by involving highly-mismatched out-of-domain languages may be of low quality and not suitable for feature learning.

(2). As discussed in Sections 5.1.2 and 3.2, DPGMM-HMM labels are obtained by modeling temporal dependency of speech and DPGMM labels are determined with the assumption that neighboring speech frames are independent. Comparing the corresponding systems in groups $\mathcal{A}$ and $\mathcal{B}$ in Tables 5.2 and 5.3, it is noted that DPGMM-HMM labels perform slightly better than DPGMM labels. The ABX error rates attained with HMM(P)-MUBNF, HMM(P)-LI-BNF1 and HMM(P)-LI-BNF2 are about absolute 0.2% - 0.3% lower than those with MUBNF, LI-BNF1 and LI-BNF2 respectively, except for HMM(P)-LI-BNF2 under the across-speaker condition. This demonstrates that capturing temporal dependency in speech is beneficial to feature learning for subword modeling, as was found in [46]. It is also noted that phone-level HMM alignments are better than state-level ones.

(3). Combining different types of BNF feature representations leads to further improvement of performance. Specifically, by concatenating HMM(P)-MUBNF, OSBNF1 and TLBNF, the best ABX error rates under both within-speaker and across-speaker conditions are achieved (7.4% and 9.7%). It is found that BNFs learned from in-domain unsupervised data (HMM(P)-MUBNF, OS-BNF1) and learned via transfer learning (TLBNF) can be jointly used to compose an optimal feature representation that is better than any individual BNF.

The best performance attained in this study is competitive to the best submitted system for the ZeroSpeech 2017 challenge, which is based on the combination of multiple DPGMM posteriorgrams [32]. These posteriograms were generated with unsupervisedly estimated fMLLRs based on different implementation parameters. The combination of posteriorgrams led to 3.0% and 3.3% relative error rate reduction under the within-speaker and across-speaker conditions, compared to the use of single posteriorgram representation. In our work, concatenating the three aforementioned BNF representations results in 5.1% and 4.0% relative error rate reduction, as compared with the best system with single BNF. It must be noted that no out-of-domain transcribed speech was involved in the system of [32].

In a very recent work [37], vector quantized VAE (VQ-VAE) was applied to develop a system of unsupervised subword modeling. The reported average ABX error rate was 6.7% for within-speaker condition, which is the best among all reported systems so far. For the across-speaker condition, our proposed systems with combined BNF features have slightly better performance than VQ-VAE (9.8%). Our systems are found to be more effective on long utterances than VQ-VAE. In Tables 5.2 and 5.3, it is noted that the performance of VQ-VAE does not depend on utterance duration. For English and Mandarin, the ABX error rates are almost exactly the same between the cases of 1s and 120s. One possible reason is that the VQ-VAE system does not perform explicit utterance-level speaker normalization on input features. On the contrary, the BNF representations investigated in the study perform significantly better on longer utterances (10s & 120s) than on 1s ones. It is also noted that our systems are more effective for Mandarin in the across-speaker condition. This may be due to the use of Cantonese speech in feature learning. VQ-VAE may be over-fitting to Mandarin due to small data size [37].

**Effectiveness of label filtering**

The effectiveness of the proposed label filtering algorithm is evaluated with the HMM(P)-MUBNF representation, which is trained exclusively based on DPGMM-HMM labels, without involving out-of-domain speech data. Algorithm 5.1 requires one tunable parameter $P$, i.e., the percentage of frame labels to be retained. The average ABX error rates attained with different values of $P$ are plotted as in Figure 5.5. $P = 1$ means that all labels are kept, which is the setting used to obtain the results in Tables 5.2 and 5.3.

Under both within-speaker and across-speaker conditions, the optimal values of $P$ are in the range of 0.7 to 0.9. That is, when on average about $10 - 30\%$ of the frame labels are removed, the ABX error rates could be slightly reduced. This indicates that indeed a certain portion of the labels are not reliable. However, if too many labels are removed, e.g., more than 30%, the system performance would degrade significantly, because some good labels are lost.

The proposed label filtering method is very simple in that only the occurrence counts of the labels are considered. Figure 5.5 shows that this criterion is appropriate to a certain extent. However, there may exist infrequent subword units that are meaningful and crucial in conveying linguistic content. In [111,112], it was suggested to reduce the number of DPGMM clusters without ignoring any frame labels. Since these studies were carried out on a different database, direct comparison of system performance can not be made.

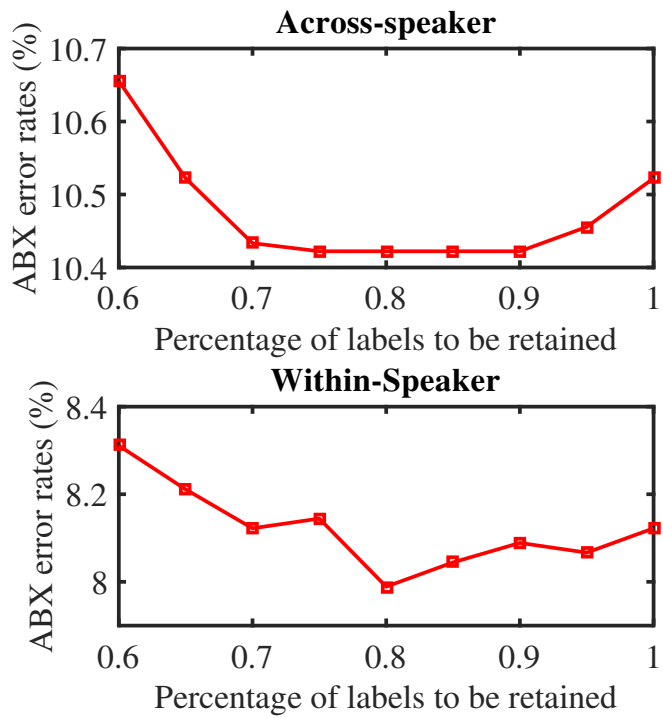

**Figure 5.5:** ABX error rates (%) on HMM(P)-MUBNF representation w.r.t percentage of labels to be retained. The performance is computed by averaging over all languages.

### Comparison of DNN structures

In this part, layer-wise structures of MTL-DNN other than MLP are used. Tables 5.4 and 5.5 compare the system performances obtained by using MLP, LSTM and BLSTM as DNN structures. The feature representations being investigated include MUBNF, HMM(P)-MUBNF and HMM(P)-LI-BNF1, and label filtering is not applied.

It is noted that LSTM and BLSTM do not perform as well as MLP on all three types of BNF representations. Experiments were carried out with different parameter settings on LSTM and BLSTM, and the system performance remained largely unchanged. Figure 5.6 gives the performances of HMM(P)-MUBNF learned by MLP, LSTM and BLSTM for each target language. For English (EN), different DNN structures have similar performance. For French (FR) and Mandarin (MA), the advantage of MLP over (B)LSTM is more prominent. This may be related to that the amount of training data for English is significantly greater than those for French and Mandarin. The advantage of LSTM and BLSTM over MLP in conventional supervised acoustic modeling has been widely recognized and attributed to the capability of capturing temporal characteristics of speech. With limited training data, the benefits of recurrent structures can not be fully exploited. In our systems, contextual information is incorporated via the use of DPGMM-HMM labels and its effectiveness has been demonstrated by the experimental results.

**Table 5.4:** Across-speaker ABX error rate (%) comparison of DNN structures.

| | | English | | | French | | | Mandarin | | | Avg. |
|---|---|---|---|---|---|---|---|---|---|---|---|
| | | 1s | 10s | 120s | 1s | 10s | 120s | 1s | 10s | 120s | |
| MUBNF | MLP | 10.9 | 9.5 | 8.9 | 15.2 | 13.0 | 12.0 | 10.5 | 8.9 | 8.2 | **10.8** |
| | LSTM | 10.4 | 9.6 | 9.0 | 14.6 | 13.3 | 12.3 | 10.9 | 9.3 | 8.6 | 10.9 |
| | BLSTM | 10.4 | 9.6 | 9.0 | 14.7 | 13.3 | 12.1 | 10.7 | 9.3 | 8.6 | 10.9 |
| HMM(P)-MUBNF | MLP | 10.4 | 9.2 | 8.7 | 14.5 | 12.7 | 11.7 | 10.4 | 8.9 | 8.2 | **10.5** |
| | LSTM | 10.0 | 9.3 | 8.6 | 14.3 | 13.1 | 11.8 | 10.7 | 9.3 | 8.6 | 10.6 |
| | BSLTM | 10.1 | 9.4 | 8.9 | 14.2 | 13.0 | 11.9 | 10.8 | 9.4 | 8.7 | 10.7 |
| HMM(P)-LI-BNF1 | MLP | 9.7 | 8.7 | 8.0 | 13.7 | 12.3 | 11.1 | 9.7 | 8.4 | 7.6 | **9.9** |
| | LSTM | 9.6 | 9.1 | 8.1 | 14.1 | 13.3 | 11.6 | 10.2 | 9.1 | 8.0 | 10.3 |
| | BLSTM | 9.5 | 9.0 | 8.2 | 13.7 | 13.0 | 11.6 | 9.7 | 8.7 | 7.8 | 10.1 |

**Table 5.5:** Within-speaker ABX error rate (%) comparison of DNN structures.

| | | English | | | French | | | Mandarin | | | Avg. |
|---|---|---|---|---|---|---|---|---|---|---|---|
| | | 1s | 10s | 120s | 1s | 10s | 120s | 1s | 10s | 120s | |
| MUBNF | MLP | 7.4 | 6.9 | 6.3 | 9.6 | 9.0 | 8.1 | 9.8 | 8.8 | 8.1 | **8.2** |
| | LSTM | 7.4 | 7.1 | 6.8 | 10.0 | 9.5 | 8.7 | 10.4 | 9.5 | 8.7 | 8.7 |
| | BLSTM | 7.4 | 7.1 | 6.7 | 9.9 | 9.5 | 8.9 | 10.4 | 9.4 | 8.7 | 8.7 |
| HMM(P)-MUBNF | MLP | 7.1 | 6.6 | 6.2 | 9.4 | 9.1 | 7.8 | 9.9 | 8.8 | 8.2 | **8.1** |
| | LSTM | 7.2 | 6.8 | 6.4 | 9.9 | 9.4 | 8.7 | 10.4 | 9.5 | 8.8 | 8.6 |
| | BSLTM | 7.3 | 6.9 | 6.5 | 9.6 | 9.5 | 8.4 | 10.5 | 9.4 | 9.0 | 8.6 |
| HMM(P)-LI-BNF1 | MLP | 6.8 | 6.3 | 5.8 | 9.1 | 8.7 | 7.8 | 9.1 | 8.5 | 7.6 | **7.7** |
| | LSTM | 6.7 | 6.6 | 5.9 | 9.5 | 9.4 | 8.2 | 9.6 | 8.9 | 7.9 | 8.1 |
| | BLSTM | 7.0 | 6.6 | 6.1 | 9.3 | 9.2 | 8.2 | 9.4 | 8.7 | 8.0 | 8.1 |

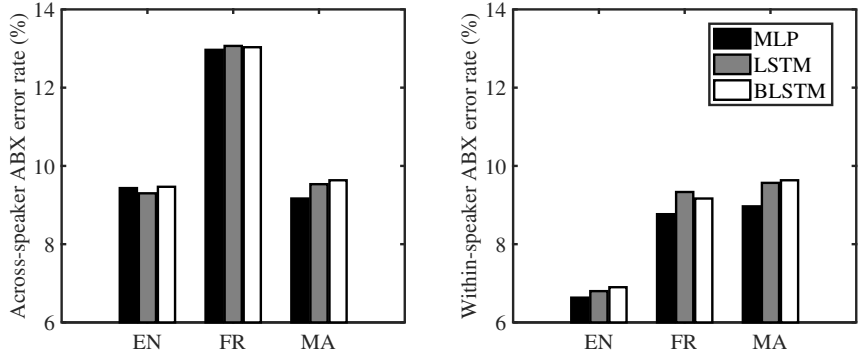

**Figure 5.6:** Average ABX error rates (%) of HMM(P)-MUBNF representation implemented by MLP, LSTM and BLSTM over different utterance lengths.

## 5.4  Summary

This chapter addressed the problem of frame labeling in unsupervised subword modeling. Various approaches are proposed and evaluated on the feature representation learning task of ZeroSpeech 2017. DPGMM-HMM acoustic modeling is applied to capture contextual information in the speech signal, and generate time alignment as the desired frame labels. A label filtering algorithm is proposed to discard unreliable initial labels from DPGMM clustering, so as to benefit DPGMM-HMM frame labeling. Multiple out-of-domain ASRs are utilized to produce language-mismatched phonetically-informed labels as complementary information to the in-domain DPGMM-HMM labels.

The proposed approaches are evaluated by thorough experimental studies. The results demonstrate the advantage of DPGMM-HMM frame labels compared to DPGMM labels. The label filtering algorithm is effective in further improving DPGMM-HMM labels. The system trained with both out-of-domain ASR based labels and DPGMM-HMM labels achieves better performance than that trained with either type of labels only. Combining different types of BNFs by vector concatenation leads to further performance improvement. The best performance achieved by our proposed approaches is 9.7% in terms of across-speaker ABX error rate. It is equal to the performance of the best submitted system in ZeroSpeech 2017 and better than other recently reported systems.

Our proposed approaches are expected to be effective for any combination of languages other than those in ZeroSpeech 2017. Nevertheless, our investigation has suggested that the closeness between target languages and out-of-domain languages and the amount of available training data for individual target languages might have significant impact on the goodness of learned features.

# Chapter 6

# Unsupervised unit discovery

This chapter is focused on unsupervised unit discovery, automatically discovering basic subword units of a language without any transcribed data. In this study, the problem is tackled with the acoustic segment modeling (ASM) approach [53]. As explained in Section 2.2.1, the ASM comprises the sequential steps of initial segmentation, segment labeling and iterative subword modeling. We focus on initial segmentation and segment labeling, and propose to exploit out-of-domain language-mismatched speech resources in these two steps.

In initial segmentation, one or multiple language-mismatched phone recognizers are used to decode target speech into phone sequences with time alignment. The phone boundaries are regarded as segmentation results. In segment labeling, phone posteriorgrams derived from language-mismatched phone recognizers are used as input features to spectral clustering [25] and generate segment-level cluster labels. Language-mismatched phone recognizers provide fine-grained speech representations for out-of-domain languages. They potentially benefits unit discovery for in-domain zero-resource languages.

Linguistic relevance of discovered units is one of the major concerns in unsupervised unit discovery. In the literature, the performance of unsupervised unit discovery is usually evaluated by clustering-based metrics, e.g. purity. These metrics are appropriate for straightforward comparison of the overall efficacy, but not able to provide detailed insights on the fitness of individual clusters and their relation. In our concerned problem, these metrics do not reflect linguistic relevance of discovered acoustic units. To address this problem, a Kullback-Leibler (KL) divergence-based distance metric is defined to analyze linguistic relevance of acoustic units discovered from an unknown language.

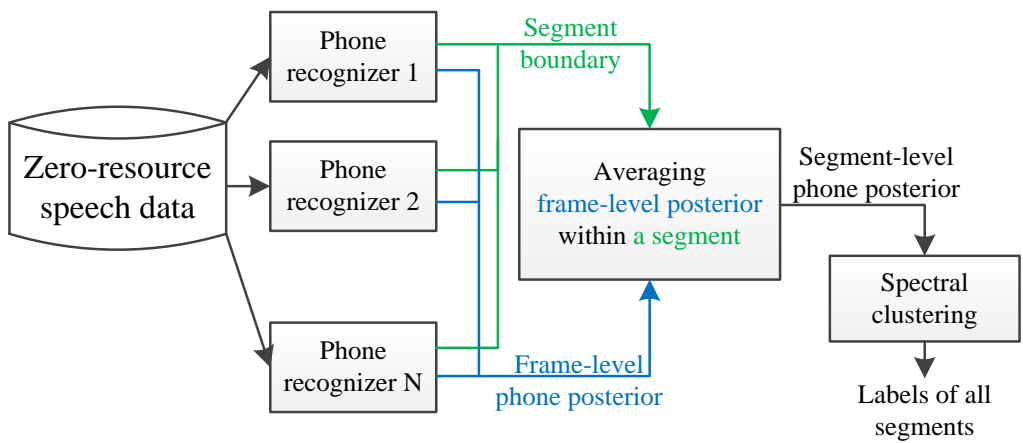

**Figure 6.1:** Exploiting language-mismatched phone recognizers in unsupervised unit discovery.

This chapter is organized as follows. Section 6.1 discusses our approaches to exploiting language-mismatched phone recognizers in unsupervised unit discovery. Section 6.2 introduces the analysis of linguistic relevance of discovered subword units by our proposed KL divergence based method. Experiments are presented in Section 6.1. Section 6.4 draws the summary of this chapter.

# 6.1 Use of language-mismatched phone recognizers in unsupervised unit discovery

## 6.1.1 Initial segmentation

Figure 6.1 illustrates the proposed approach to unsupervised unit discovery. Given untranscribed speech utterances of a target language, one or more phone recognizers pre-trained for other languages are utilized to generate phone boundaries.

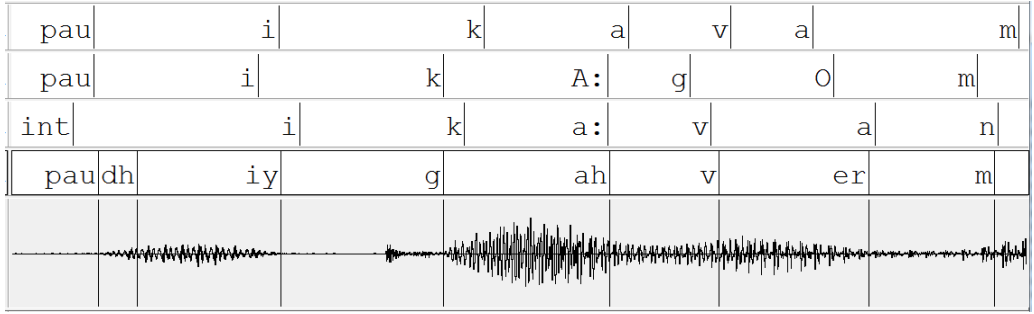

**Figure 6.2:** Example of segmentation of an English utterance using Czech, Hungarian and Russian phone recognizers.

### Single phone recognizer

Phone recognition is the process of decoding a sequence of speech observations. The decoding result comprises a sequence of phone symbols and their time boundaries. With a single phone recognizer, the phone boundaries could be used directly as an initial segmentation of the input utterance. Only the time boundaries are used while the phone identities are ignored.

### Multiple phone recognizers

A single phone recognizer may not be able to provide a good coverage of the phonetic space to support the modeling of a new language, especially when there is a significant mismatch between the two languages concerned. It would be advantageous to make use of multiple phone recognizers from different languages to derive a better informed initial segmentation. Figure 6.2 compares the segmentation results obtained by applying three language-mismatched phone recognizers to an English utterance. The upper three panes show phone-level time alignment given by the Czech, Hungarian and Russian phone recognizers, respectively. The fourth pane shows the manual segmentation of the speech waveform with English phone labels.

The following method is developed to infer a unified speech segmentation from multiple decoding results. Given $N$ phone recognizers, let $\boldsymbol{S^j}$ denote the segment boundaries produced by the $j$-th recognizer,

$$\boldsymbol{S^j} = \{s_1^j, s_2^j, \ldots, s_K^j\}, j = 1, 2, \ldots, N, \tag{6.1}$$

where $s_i^j$ is the location of the $i$-th segment boundary, $i = 1, 2, \ldots, K_j$). The algorithm processes $\{\boldsymbol{S^1}, \ldots, \boldsymbol{S^N}\}$ by merging segment boundaries and eliminating the boundaries of segments shorter than a pre-defined threshold of 30ms. The output of Algorithm 6.1 is denoted as $\boldsymbol{S^*}$.

---

**Algorithm 6.1:** Fusion of segmentation results

---

**Input:** Segment boundaries resulted from $N$ phone recognizers $\boldsymbol{S^1}, \boldsymbol{S^2}, \ldots, \boldsymbol{S^N}$.
**Output:** Fusion of segment boundaries $\boldsymbol{S^*}$.

1: Concatenate all boundaries into

$$\boldsymbol{S_{\text{long}}} = \{s_1^1, s_2^1, \ldots, s_{K_1}^1, s_1^2, s_2^2, \ldots, s_{K_2}^2, \ldots, s_1^N, s_2^N, \ldots, s_{K_N}^N\}. \quad (6.2)$$

2: Sort $\boldsymbol{S_{\text{long}}}$ in ascending order, denoted as

$$\boldsymbol{S_{\text{sort}}} = \{s_1, s_2, \ldots, s_K\}, K = \sum_{j=1}^{N} K_j. \quad (6.3)$$

3: Drop coincided elements in $\boldsymbol{S_{\text{sort}}}$. Denote the output by

$$\boldsymbol{S_{\text{sort\_uni}}} = \{su_1, su_2, \ldots, su_{K\_u}\}. \quad (6.4)$$

4: Eliminate segments represented by $\boldsymbol{S_{\text{sort\_uni}}}$ with duration less than 30ms.
5: **if** $su_m - su_{m-1} = 10\text{ms}$ **then**
6:     drop $su_m$;
7: **else if** $su_m - su_{m-1} = 20\text{ms}$ **then**
8:     $su_{m-1} \leftarrow \frac{su_m + su_{m-1}}{2}$;
9:     drop $su_m$;
10: **else**
11:     Do nothing;
12: **end if**
13: **return** $\boldsymbol{S^*}$;

---

## 6.1.2 Segment labeling and unit discovery

After initial segmentation, each training utterance is divided into a number of variable-length segments. The speech segments from all training utterances are organized into a limited number of clusters, based on their acoustic similarities. Segment labels are generated according to the clustering results.

**Feature representation**

Previous studies showed that posterior features are more robust than conventional spectral features like MFCCs [120]. In this study, segment clustering is performed with phone posteriorgram representation. Let us first consider the case that one phone recognizer is available. Let $C = \{c_1, c_2, \ldots, c_M\}$ denote the $M$ phones covered by the recognizer. The posterior feature vector representing frame $t$ is given as,

$$\boldsymbol{q_t} = \begin{bmatrix} p(c_1|\boldsymbol{o_t}) \\ p(c_2|\boldsymbol{o_t}) \\ \vdots \\ p(c_M|\boldsymbol{o_t}) \end{bmatrix} \tag{6.5}$$

where $p(c_m|\boldsymbol{o_t})$, $m = 1, 2, \ldots, M$, denotes the posterior probability of phone $c_m$ given the observation $\boldsymbol{o_t}$. If $N$ language-mismatched phone recognizers are used for decoding, there would be $N$ phone posterior feature vectors $\boldsymbol{q_t^1}, \boldsymbol{q_t^2}, \ldots, \boldsymbol{q_t^N}$, for each time frame, they are concentrated to form a single feature vector as,

$$\boldsymbol{\hat{q}_t} = \begin{bmatrix} \boldsymbol{q_t^1} \\ \boldsymbol{q_t^2} \\ \vdots \\ \boldsymbol{q_t^N} \end{bmatrix}. \tag{6.6}$$

The dimension of $\boldsymbol{\hat{q}_t}$ is,

$$\hat{M} = \sum_{i=1}^{N} M_i, \tag{6.7}$$

where $M_i$ is the phone number of the $i$-th recognizer. For a given initial segmentation, the segment-level posterior probabilities are obtained as,

$$\boldsymbol{\hat{x}_k} = \frac{1}{e_k - b_k + 1} \sum_{t=b_k}^{e_k} \boldsymbol{\hat{q}_t}, k = 1, 2, \ldots, K, \tag{6.8}$$

where $K$ is the number of segments, $b_k$ and $e_k$ are the starting and ending time frames of the $k$-th segment. The segment-level phone posteriorgram can be expressed in matrix form as,

$$\boldsymbol{X} = \begin{bmatrix} \boldsymbol{\hat{x}_1} & \boldsymbol{\hat{x}_2} & \cdots & \boldsymbol{\hat{x}_K} \end{bmatrix}. \tag{6.9}$$

---

**Algorithm 6.2:** Segment clustering

---

**Input:** $\hat{M} \times K$ phone posteriorgram $\boldsymbol{X}$, cluster number $R$.
**Output:** $R$ clusters and cluster label of each segment.

1: Compute $\boldsymbol{A} = \boldsymbol{X}\boldsymbol{X}^{\mathbf{T}}$
2: Compute $\boldsymbol{L} = \boldsymbol{I} - \boldsymbol{D}^{-\frac{1}{2}}\boldsymbol{A}\boldsymbol{D}^{-\frac{1}{2}}$, where $\boldsymbol{D} = diag\{\boldsymbol{A} \cdot [1, 1, \cdots, 1]\}$
3: Construct matrix $\boldsymbol{Y} = [\boldsymbol{y_1}, \cdots, \boldsymbol{y_R}]$, where $\boldsymbol{y_r}$ is the $r$-th smallest eigenvector of $\boldsymbol{L}$.
4: Normalize each row vector of $\boldsymbol{Y}$ to have unit $l_2$-norm.
5: Apply $k$-means on the $\hat{M}$ rows of $\boldsymbol{Y}$ to find $R$ clusters.
6: Assign the $i$-th phone to cluster $r$ if the $i$-th row vector of $\boldsymbol{Y}$ was assigned to cluster $r$, ($i = 1, 2, \ldots, \hat{M}$, $r = 1, 2, \ldots, R$).
7: Score each segment with the $R$ clusters, label it with the cluster which scores highest.

---

**Clustering algorithm**

The approach of spectral clustering is applied to segment-level phone posteriorgrams. In [25], spectral clustering was applied to Gaussian components in GMM. Here the clustering is performed on the language-mismatched phone classes. In the posteriorgram representation $\boldsymbol{X}$, each row contains the posterior probabilities of a specific phone. Then the problem is to cluster the $\hat{M}$ rows of $\boldsymbol{X}$. Details of the spectral clustering algorithm are provided as in Algorithm 6.2.

After clustering, the speech segments are labeled with respective cluster indices. Each cluster is regarded as a discovered subword unit. The segment labels give a kind of time-aligned pseudo transcriptions that can be used to facilitate supervised acoustic modeling of the target language. In the ideal case, the pseudo transcriptions are in consistency with the ground-truth transcriptions of the target language.

# 6.2 Linguistic relevance of discovered units

## 6.2.1 Shortcomings of existing evaluation metrics

Existing studies usually measure the efficacy of unsupervised unit discovery by clustering-based evaluation metrics, such as purity [25], normalized mutual information (NMI) [25] and average precision (AP) [69]. One drawback is that they do not reflect detailed insights on the fitness of the individual clusters and the relation between the clusters. Let us consider the clustering results produced by two different clustering algorithms (or the same algorithm with different parameter settings). In the first case, the degree of overlap between an automatically learned cluster and its closest ground-truth phone varies greatly from one cluster to another, whereas in the second case, the degrees of overlap are equal across all clusters. Although the two sets of clustering results may give the same purity value, their linguistic implications could be very different.

## 6.2.2   Proposed evaluation metrics

In this study, a new distance metric is proposed to analyze in detail the linguistic relevance of discovered subword units. This metric is based on KL divergence.

### KL divergence

KL divergence, also called information divergence, or relative entropy, is a measure of the difference between two probability distributions [121]. KL divergence between two discrete probability distributions $P$ and $Q$ is defined as,

$$D_{KL}(P||Q) = \sum_i P(i) \log \frac{P(i)}{Q(i)}. \tag{6.10}$$

Equation (6.10) gives a non-symmetric measure. In this study, the symmetric form of KL divergence is adopted as,

$$D_{KL}(P,Q) = D_{KL}(P||Q) + D_{KL}(Q||P) \tag{6.11}$$

$$= \sum_i (P(i) - Q(i)) \cdot \log \frac{P(i)}{Q(i)}. \tag{6.12}$$

KL divergence can be used to model implicit speech variation related to phonetic context, pronunciation variation, speaker characteristics, etc. It has been applied to ASR acoustic modeling [122, 123], data selection [124], cross-lingual TTS and voice conversion [125, 126].

**Distance between discovered unit and ground-truth phone**

The symmetric KL divergence is used to measure the distance between posterior probability distributions between each pair of discovered subword unit and ground-truth phone of the target language. Let $\{g_1, g_2, \ldots, g_K\}$ denote $K$ ground-truth phones, $\{\boldsymbol{v_1}, \boldsymbol{v_2}, \ldots, \boldsymbol{v_{L_k}}\}$ denote posterior probability vectors of $L_k$ frames that are labeled as phone $g_k$ according to ground-truth transcription. $\{\boldsymbol{v_1}, \boldsymbol{v_2}, \ldots, \boldsymbol{v_{L_k}}\}$ is a subset of in total $T$ speech frames $\{\boldsymbol{\hat{q}_1}, \boldsymbol{\hat{q}_2}, \ldots, \boldsymbol{\hat{q}_T}\}$ (refer to Eqt. (6.6)). The centroid of $g_k$ is computed as,

$$\overline{\boldsymbol{v^k}} = \frac{\sum\limits_{i=1}^{L_k} \boldsymbol{v_i}}{L_k}. \tag{6.13}$$

$\overline{\boldsymbol{v^k}}$ is treated as the representative of $g_k$ in the phonetic space. Let $\{u_1, u_2, \ldots, u_R\}$ denote $R$ discovered subword units, $\{\boldsymbol{\mu_1}, \boldsymbol{\mu_2}, \ldots, \boldsymbol{\mu_{N_r}}\} \subset \{\boldsymbol{\hat{q}_1}, \boldsymbol{\hat{q}_2}, \ldots, \boldsymbol{\hat{q}_T}\}$ are the posterior probability vectors of frames assigned to $u_r$ $(r = 1, 2, \ldots, R)$. The distance between $u_r$ and $g_k$ is defined as,

$$D(u_r, g_k) = \frac{\sum\limits_{j=1}^{N_r} D_{KL}(\boldsymbol{\mu_j}, \overline{\boldsymbol{v^k}})}{N_r} \tag{6.14}$$

$$= \frac{\sum\limits_{j=1}^{N_r} \sum\limits_{m=1}^{M} (\mu_j(m) - \overline{v^k}(m)) \cdot \log \frac{\mu_j(m)}{\overline{v^k}(m)}}{N_r}. \tag{6.15}$$

Here, the distortion between two acoustic models $u_r$ and $g_k$ is measured by the KL divergence-based distance between posterior features in the phonetic space. An illustration on the computation process of $D(u_r, g_k)$ is shown in Figure 6.3.

**Closest ground-truth phones**

Let $g_{k^*}(u_r)$ denote the closest ground-truth phone of subword unit $u_r$, where,

$$k^* = \arg\min_k D(u_r, g_k). \tag{6.16}$$

For simplicity, $g_{k^*}(u_r)$ is abbreviated as $g^*(u_r)$. The distance between $u_r$ and $g^*(u_r)$ is denoted as $D^*(u_r)$, and is computed as,

## 6. Unsupervised unit discovery

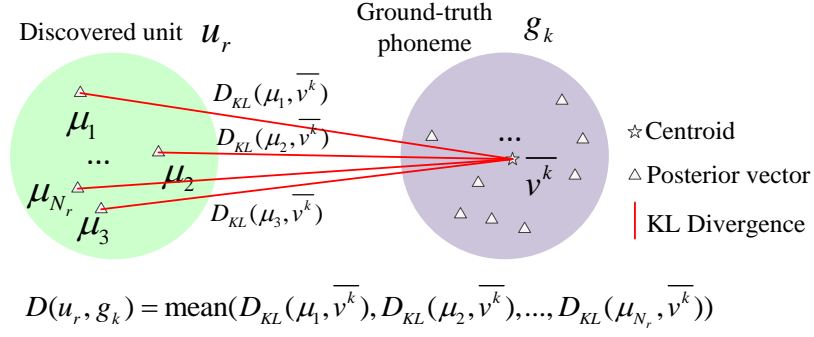

**Figure 6.3:** Illustration on the computation process of $D(u_r, g_k)$.

$$D^*(u_r) = D(u_r, g^*(u_r)). \qquad (6.17)$$

Let $g_{k^{**}}(u_r)$ denote the second closest ground-truth phone of subword unit $u_r$, where,

$$k^{**} = \underset{k \neq k^*}{\arg\min}\, D(u_r, g_k). \qquad (6.18)$$

The distance between $u_r$ and $g^{**}(u_r)$ is denoted as $D^{**}(u_r)$, and is computed as,

$$D^{**}(u_r) = D(u_r, g^{**}(u_r)). \qquad (6.19)$$

The discriminability of the cluster $u_r$ can be measured by

$$\Delta D^*(u_r) = |D^{**}(u_r) - D^*(u_r)|. \qquad (6.20)$$

A small value of $D^*(u_r)$ means that the discovered subword unit matches well with one of the ground-truth phones. Meanwhile, a large value of $\Delta D^*(u_r)$ indicates that $g^*(u_r)$ is discriminatively mapped to $u_r$. The average distance $\overline{D^*(u_r)}$, $\overline{D^{**}(u_r)}$ and $\overline{\Delta D^*(u_r)}$ over all discovered subword units $\{u_r\}$ in a target language are defined as,

$$\overline{D^*(u_r)} = \frac{\sum_{r=1}^{R} D^*(u_r)}{R}, \qquad (6.21)$$

$$\overline{D^{**}(u_r)} = \frac{\sum_{r=1}^{R} D^{**}(u_r)}{R}. \qquad (6.22)$$

$$\overline{\Delta D^*(u_r)} = \left| \overline{D^{**}(u_r)} - \overline{D^*(u_r)} \right|. \qquad (6.23)$$

The distance measure can be further extended to evaluating inherent variability of each ground-truth phones. For the phone $g_k$, the inherent variability $\widetilde{D}(g_k)$ and the average inherent variability over all $\{g_k\}$ are defined as,

$$\widetilde{D}(g_k) = \frac{\sum\limits_{j=1}^{L_k} D_{KL}(\boldsymbol{v_j}, \overline{\boldsymbol{v^k}})}{L_k}, \tag{6.24}$$

$$\overline{\widetilde{D}(g_k)} = \frac{\sum_{k=1}^{K} \widetilde{D}(g_k)}{K}. \tag{6.25}$$

A small value of $\widetilde{D}(g_k)$ indicates that the acoustic-phonetic properties of $g_k$ are highly consistent in the training speech. It must be noted that $D^*(u_r)$ computed for discovered subword units and $\widetilde{D}(g_{k^*})$ for ground-truth phones are comparable, as both of them measure the deviation from a class of posterior feature vectors to the centroid of the same phone class $g_{k^*}$, in the same phonetic space. $\widetilde{D}(g_{k^*})$ is calculated from ground-truth transcription, and independent of the clustering results. Therefore $\widetilde{D}(g_{k^*})$ could be a good reference for $D^*(u_r)$.

The KL divergence metric in this paper is not only applicable to posterior features extracted from phone recognizers, but also to conventional spectral features like MFCCs, or DNN-based representations such as BNFs.

## 6.3 Experiments

### 6.3.1 Database

Experiments on unsupervised unit discovery are carried out with the OGI Multi-language Telephone Speech Corpus (OGI-MTS) [127]. The spontaneous story-telling part of this corpus is used. Five languages are involved: German (GE), Hindi (HI), Japanese (JA), Mandarin (MA) and Spanish (SP). In addition to audio signals, the database provides manual time alignment at phone level for each utterance. Table 6.1 summarizes the amount of audio data and the number of phone units (including a silence unit) in each language.

The OGI-MTS database is chosen for several reasons. The present study is focused on the methodology design for unsupervised unit discovery of an unknown language. The exact identities of the target languages do not matter. The key assumption is that no labeled data is available. On the other hand, ground-truth linguistic knowledge and reliably transcribed and aligned test data are necessary for performance evaluation purpose. Such resources are generally not available for real zero-resource languages. In terms of speaking style, the story-telling speech in OGI-MTS is considered a good match with real-world data that one might be able to collect for a zero-resource language. Furthermore, the use of multiple languages leads to a wider coverage of phonetic variation and makes the experimental study representative and convincing.

**Table 6.1:** Multilingual speech data from the OGI-MTS corpus.

| Language | GE | HI | JA | MA | SP |
|---|---|---|---|---|---|
| Duration (hours) | 1.31 | 0.95 | 0.86 | 0.57 | 1.46 |
| No. phone units | 43 | 46 | 29 | 44 | 38 |

The study in [25] reports experiments on unsupervised unit discovery with the same database. Our results could be compared with [25] so as to better understand the effectiveness of using language-mismatched phone recognizers.

## 6.3.2 Effectiveness of language-mismatched phone recognizers

### Evaluation metric

**Purity** is a commonly used evaluation metric that measures the degree to which the cluster results of a clustering process are in accordance with ground-truth classification labels. Let $G = \{G_1, G_2, \ldots, G_R\}$ denote a set of $R$ clusters, $G' = \{G'_1, G'_2, \ldots, G'_{R'}\}$ a set of $R'$ ground-truth phones. Let $n_{r,r'}$ be the number of frames assigned to the $r$-th cluster and labeled as the $r'$-th phone in the reference transcription. The purity value associated with the $r$-th cluster is defined as,

$$\text{purity(r)} = \frac{\max_{r' \in \{1,2,\ldots,R'\}} n_{r,r'}}{\sum_{r'=1}^{R'} n_{r,r'}}. \tag{6.26}$$

**Table 6.2:** Purity values obtained by exploiting a single recognizer.

|  | Phone recognizer adopted | | | Average |
|  | CZ | HU | RU | |
|---|---|---|---|---|
| GE | 0.409 | 0.378 | 0.408 | 0.398 |
| HI | 0.443 | 0.408 | 0.408 | 0.419 |
| JA | 0.502 | 0.500 | 0.480 | **0.494** |
| MA | 0.381 | 0.412 | 0.345 | 0.379 |
| SP | 0.500 | 0.467 | 0.488 | 0.485 |
| Average | 0.432 | 0.416 | 0.416 | 0.421 |

High purity values are desirable since they indicate a large proportion of the cluster are from the same phone. By averaging purity values of $R$ clusters, the overall purity is computed as,

$$\text{purity} = \frac{\sum_{r=1}^{R} \max_{r' \in \{1,2,...,R'\}} n_{r,r'}}{\sum_{r=1}^{R} \sum_{r'=1}^{R'} n_{r,r'}}. \tag{6.27}$$

In the phone recognizer, silence part of speech is modeled as a phone. As silence segments occur frequently and can be accurately detected, the computed purity values tend to be biased. In this study, the silent labels were removed according to the reference transcriptions at the beginning of the experiments.

## Language-mismatched phone recognizers

Four phone recognizers for Czech (CZ), Hungarian (HU), Russian (RU) and Cantonese (CA) are used as language-mismatched phone recognizers. The CZ, HU and RU recognizers [117] were described as in Section 5.3.1. The numbers of modeled phones are 45, 61 and 52 respectively. The CA recognizer is trained in similar settings as in Section 4.5.1. The total number of modeling units in CA is 73.

## Results with single recognizer

The purity values of clustering results obtained with one of the CZ, HU and RU recognizers are given as in Table 6.2. In this part of experiments, the cluster number $R$ is set to be equal to the number of ground-truth phones in the respective recognizer. The following observations are made:

1. For the same target language, the purity values achieved do not differ much

**Table 6.3:** Purity values obtained by jointly exploiting CZ, HU and RU recognizers w.r.t cluster number $R$. Colored values in the last two columns denote better (red) or worse (blue) performance compared to the proposed approach.

|  | \multicolumn{4}{c|}{$R$} | \multicolumn{2}{c}{Reference} |
|  | 50 | 60 | 70 | 80 | Wang et al. [25] | Single recognizer |
|---|---|---|---|---|---|---|
| GE | 0.418 | 0.419 | **0.428** | 0.427 | 0.403 | 0.398 |
| HI | 0.439 | 0.406 | 0.446 | **0.452** | 0.457 | 0.419 |
| JA | 0.508 | 0.508 | **0.511** | 0.510 | 0.520 | 0.494 |
| MA | 0.384 | 0.386 | 0.376 | **0.388** | 0.367 | 0.379 |
| SP | 0.499 | 0.499 | 0.502 | **0.509** | 0.549 | 0.485 |
| Average | 0.435 | 0.430 | 0.438 | **0.441** | 0.443 | 0.421 |

across different phone recognizers. The purity values depend on the target language. The relative magnitudes of purity of the six target languages are consistent across different recognizers. For examples, the purity values for German are in the range of $0.38 - 0.41$, and those for Japanese are $0.48 - 0.51$;

2. The purity values attained for Japanese were the highest consistently . Theoretically, the number of ground-truth phones in the language affects the purity with a negative correlation. Japanese has only 29 phones, significantly less than the other five languages. Spanish achieves the second highest purity with the second least phone number of 38.

### Results with multiple recognizers

The purity values obtained by the joint use of the CZ, HU and RU recognizers are given as in Table 6.3. In this part, initial segmentation results are generated by applying Algorithm 6.1 to the three recognizers' decoding results. Different clustering numbers $R$ (from 50 to 80) are attempted. Experimental results on the same task in [25] are provided for reference. The last column of the table contains the average purity value achieved by using a single recognizer (i.e., the last column in Table 6.2).

From Table 6.3, the following observations can be made:

1. The best purity value attained with multiple recognizers is higher than that with any single recognizer alone. Among the six target languages, HI shows the most significant improvement and MA benefits the least;

2. The cluster number $R$ has little influence on the result. For best performance,

**Table 6.4:** Purity values obtained by jointly exploiting CZ, HU, RU and CA recognizers w.r.t cluster number $R$. This system is used for analysis of linguistic relevance.

|    | \multicolumn{5}{c}{$R$} | Average |
|----|-------|-------|-------|-------|-------|---------|
|    | 50    | 60    | 70    | 80    | 90    |         |
| GE | 0.428 | 0.433 | 0.439 | 0.437 | 0.437 | 0.435   |
| HI | 0.494 | 0.499 | 0.492 | 0.489 | 0.485 | 0.492   |
| JA | 0.545 | 0.556 | 0.554 | 0.543 | 0.540 | 0.548   |
| MA | 0.414 | 0.425 | 0.434 | 0.426 | 0.426 | 0.425   |
| SP | 0.556 | 0.586 | 0.576 | 0.568 | 0.573 | 0.572   |

$R$ in the range of 70 to 80. In practical applications, the cluster number can either be pre-determined or empirically tuned on development data;

3. With the same database and the same evaluation metric, the proposed use of multiple phone recognizers gives comparable performance to that reported in [25]. For GE and MA, our method achieves better results. A multi-view segment clustering (MSC) algorithm was used in [25], to exploit multiple feature representations simultaneously. Our method is considered to have a simpler implementation.

## 6.3.3 Linguistic relevance of discovered subword units

### Unit discovery system establishment

In this section, the relation between automatically discovered subword units and ground-truth phones is analyzed. A unit discovery system is established beforehand to provide discovered subword units for analysis. This system is developed based mainly on settings described in Section 6.3.2, with only a few exceptions: (1). CA phone recognizer is used in addition to CZ, HU and RU; (2). cluster number $R$ ranges from 50 to 90.

Table 6.4 summarizes purity values of the unit discovery system. Compared with Table 6.3, it is found that the purity values obtained by exploiting four language-mismatched phone recognizers are higher than those by exploiting three recognizers in all of the test languages.

**Table 6.5:** $\overline{D^*(u_r)/D^{**}(u_r)}$ and $\overline{\widetilde{D}(g_k)}$ computed from the discovered subword units.

| | $\overline{D^*(u_r)/D^{**}(u_r)}$ | | | | | $\overline{\widetilde{D}(g_k)}$ |
|---|---|---|---|---|---|---|
| | $R=50$ | $R=60$ | $R=70$ | $R=80$ | $R=90$ | |
| GE | 27.4/30.1 | 27.3/30.0 | 27.3/30.1 | 27.5/30.3 | 27.4/30.3 | 27.8 |
| HI | 27.6/31.5 | 27.6/31.6 | 27.8/31.7 | 28.0/31.7 | 28.1/31.8 | 28.5 |
| JA | 28.5/34.1 | 28.0/33.2 | 28.1/33.2 | 28.3/33.4 | 28.6/34.1 | 27.9 |
| MA | 29.3/31.2 | 29.3/31.5 | 29.0/31.0 | 29.5/31.6 | 29.4/31.3 | 29.1 |
| SP | 28.5/31.9 | 27.8/32.2 | 27.9/32.3 | 28.1/32.5 | 28.0/32.4 | 28.0 |

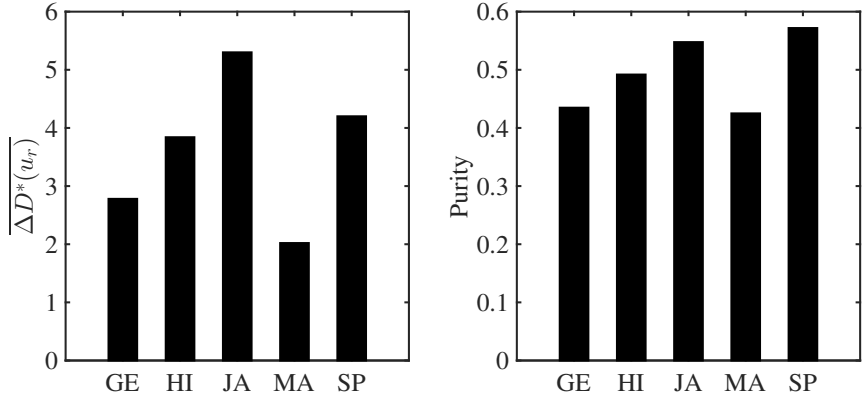

**Figure 6.4:** $\overline{\Delta D^*(u_r)}$ and purity values computed from the discovered subword units.

## Results and analysis

For each discovered subword unit $u_r$, Equations (6.14), (6.16) and (6.17) are used to determine its closest ground-truth phone $g^*(u_r)$ and the distance $D^*(u_r)$ between them. Equation (6.18) is used to determine the second closest ground-truth phone $g^{**}(u_r)$ and the distance $D^{**}(u_r)$. For each ground-truth phone $g_k$, Equation (6.24) is used to calculate the inherent variability $\widetilde{D}(g_k)$. Variables $\overline{D^*(u_r)}, \overline{D^{**}(u_r)}$ and $\overline{\widetilde{D}(g_k)}$, computed by Equations (6.21), (6.22) and (6.25), are summarized in Table 6.5. It is observed that the average KL divergence is not sensitive to the cluster number $R$.

Figure 6.4 compares $\overline{\Delta D^*(u_r)}$ (computed by Equation (6.23)) and average purity values for the five target languages. From Table 6.5 and Figure 6.4, the following observations are made:

1. $\overline{D^*(u_r)}$ is smaller than or approximately equal to $\overline{\widetilde{D}(g_k)}$ for all the target

**Table 6.6:** Uncovered/total number of vowels and consonants ($R = 90$).

|            | GE   | HI   | JA   | MA   | SP   |
|-----------:|------|------|------|------|------|
| Vowels     | 1/18 | 0/13 | 0/7  | 2/17 | 1/11 |
| Consonants | 1/24 | 3/32 | 0/21 | 4/26 | 1/27 |

languages. In other words, the deviation between a discovered subword unit and its closest ground-truth phone is comparable to, if not smaller than, the inherent variability of the phone itself. In fact, the ground-truth phones are labeled based on auditory perception of linguistic experts, whereas the automatically discovered units, as well as the proposed KL divergence metric, is totally data-driven.

2. $\overline{\Delta D^*(u_r)}$ for different languages have the same trend as the purity values. This observation is consistent with our expectation, as a larger $\Delta D^*(u_r)$ implies that the discovered subword unit is mapped to its closest ground-truth phone with a higher confidence, thus naturally leads to a higher purity.

We are interested to understand more about the phonetic coverage of automatically discovered subword units. Each subword unit corresponds to one best-matching ground-truth phone based on Equation (6.16). If a ground-truth phone fails to be selected as the best-matching phone for any of the discovered subword units, it is considered as **not being covered**. On the contrary, if a ground-truth phone is selected as the best-matching phone for at least one discovered unit, it is considered as **being covered**.

Table 6.6 shows the counts of uncovered vowels and consonants for each target language for $R = 90$. It can be seen that most of the linguistically-defined phones could be covered in the process unsupervised unit discovery. Particularly in the case of Japanese, all phones are covered. However, there are quite a few phones of Mandarin that are not covered by the discovered units. The missing vowels and consonants are listed in Table 6.7[1]. It is interesting to see that the majority of missing consonants are unvoiced plosives. These consonants have strong transitory characteristics, i.e., rapidly changing spectral properties. In the segmentation process of unsupervised acoustic modeling, it is assumed that individual frames in the same segment have similar spectral properties. This assumption is not valid for transitory phones. Similarly, the missing vowel /er/, known as Erhuayin [128] in Mandarin, also has transitory properties. This inspires us to investigate alternative features and segment representation which could capture trajectory characteristics of phones.

---

[1] In this table, each phone label is followed by its corresponding IPA transcription. For example, /aa/ is used in OGI-MTS database [127], its IPA transcription is /a/.

**Table 6.7:** Mandarin phones that are not covered by automatically discovered subword units ($R = 90$).

| Vowels | /aa/ (/a/), /er/ (/ɚ/) |
|---|---|
| Consonants | /kh/ (/kʰ/), /ph/ (/pʰ/), /r/ (/ɻ/), /tH/ (/tʰ/) |

**Table 6.8:** Discovered subword units mapped to /uw/ with $R = 50$ and 90.

| # Clusters $R$ | 50 | | 90 | | | |
|---|---|---|---|---|---|---|
| Cluster label $r$ | 33 | 41 | 25 | 49 | 29 | 35 |
| $D^*(u_r)$ /uw/ | **33.9** | **35.9** | 29.1 | 33.1 | 29.7 | 30.4 |
| $D^{**}(u_r)$ /iy/ | **34.1** | — | 34.6 | 35.3 | — | — |
| /ey/ | — | **35.9** | — | — | 35.0 | 31.1 |

It is not expected that the vowel /aa/ is missed. Although /aa/ is not selected as the closest phone to any of the subword units, it is actually identified as the second closest phone to two different discovered subword units. These two units are corresponded to /ae/ (/a/) and /aw/ (/au/) according to the KL divergence. In the transcription of Mandarin speech in OGI-MTS database, /aa/ is used to label the vowel nucleus in the Pinyin Finals /a/ and /ang/, while /ae/ is used to label the vowel nucleus in the Pinyin Final /an/ [129]. The two vowel nuclei are actually very similar in articulation. From this perspective, /aa/ is actually not a missing phone.

It must be noted that the identities of uncovered ground-truth phones depend also on experimental configurations, such as initialization of clustering, cluster number, etc.

For some of the discovered subword units, the value of $D^{**}(u_r)$ is nearly the same as $D^*(u_r)$. In other words, such a discovered subword unit matches equally well with two different phones. This kind of confusion can be alleviated by increasing $R$. Table 6.8 gives an example of confusion between a few Japanese vowels. With $R = 50$, /uw/ (/ɯ/) and /iy/ (/i/) are the closest and second closest phones to cluster 33. Similar observation can be made on /uw/ and /ey/ (/e/) to cluster 41. When $R$ is increased to 90, the confusion is significantly alleviated. A larger $R$ leads to smaller-size as well as finer clusters, therefore the learned clusters containing segments of multiple ground-truth phones tend to split and form linguistically more explicit subword units.

## 6.4   Summary

This chapter presents our efforts to the task of unsupervised unit discovery. A new approach is proposed to assist unit discovery by exploiting out-of-domain language-mismatched phone recognizers in initial segmentation and segment labeling. While existing segmentation approaches rely on spectral discontinuities of the acoustic signal, our approach uses multiple phone recognizers to decode and segment speech. The recognizers provide posteriorgram representation for segment clustering and labeling.

Investigation on linguistic relevance of automatically discovered subword units is another research focus of this chapter. A symmetric KL divergence metric is defined and used to measure the distance between each pair of subword unit and ground-truth phone.

Experiments are carried out with the OGI Multi-language Telephone Speech (OGI-MTS) Corpus. Experimental results demonstrate that out-of-domain phone recognizers are effective in segmentation and segment labeling. The results are insensitive to the cluster number, i.e. the number of discovered subword units. Increasing the number of phone recognizers is beneficial to unit discovery performance. Our best performance in terms of purity is comparable to those reported in [25], while our approach is relatively simple in implementation.

Experimental results also show that our KL divergence-based evaluation metric is consistent with purity. The deviation between a discovered unit and its closest ground-truth phone is comparable to the inherent variability of the phone. While in general the unit discovery results have a good coverage of linguistically-defined phones, there are a few exceptions, e.g. /er/ in Mandarin, probably due to limited feature representation capability of the adopted unit discovery system. The confusion of a discovered unit between ground-truth phones can be alleviated with a large cluster number. Further investigation is needed to apply alternative features and segment representations to better capture trajectory characteristics of phones.

# Chapter 7

# Conclusion and Future Work

## 7.1 Conclusion and contribution

This research investigates unsupervised acoustic modeling for zero-resource languages. It is assumed that that only untranscribed speech data is available, and linguistic knowledge about the target language is absent. This problem is essential in spoken language technology applications, particularly for many languages that have very limited or no linguistic resources.

There are two research problems being tackled in this study. The first problem is automatic discovery of fundamental speech units (e.g. subword units) of a language. The second problem concerns the learning of frame-level feature representations that are robust to linguistically-irrelevant variations. Various approaches are proposed to exploit out-of-domain language resources to improve the modeling of in-domain zero-resource languages.

Towards the goal of unsupervised subword modeling, this thesis has made contributions in the following aspects.

- Speaker adaptation approaches are applied and evaluated extensively on learning speaker-invariant features in the unsupervised scenario. The proposed methods include fMLLR estimation with out-of-domain ASR, disentangled speech representation learning, and speaker adversarial training. Combinations of these approaches are also investigated.

- New approaches to frame labeling to improve the efficacy of supervised model

training. The approaches include DPGMM-HMM frame labeling and out-of-domain ASR decoding. A label filtering algorithm is proposed to improve DPGMM-HMM frame labeling by removing possibly erroneous labels. Multiple types of frame labels are jointly applied under a multi-task learning (MTL) framework to achieve informative learning of feature representation.

Towards unsupervised unit discovery, this thesis has made contributions in the following aspects.

- Language-mismatched phone recognizers are exploited in the acoustic segment modeling (ASM) framework for unsupervised unit discovery. The recognizers are utilized to generate initial segmentation and perform segment labeling.

- A symmetric KL divergence metric is proposed for the analysis of linguistic relevance of discovered subword units.

In Chapter 4, speaker adaptation approaches to learning speaker-invariant features are presented. The fMLLR approach exploits an out-of-domain ASR to estimate speaker-specific fMLLR transforms. Disentangled speech representation learning trains an FHVAE to disentangle phonetic information and speaker variation. Speaker adversarial training adds an adversarial task into the MTL-DNN model, forcing the hidden representation to carry less speaker information. Experimental results on ZeroSpeech 2017 demonstrate the effectiveness of all the approaches. The fMLLR approach achieves the most significant performance improvement compared to the baseline, demonstrating the efficacy of out-of-domain ASR systems in speaker adapted feature learning. Reconstructed MFCC features by disentangled representation learning are shown to capture less speaker-dependent information than original ones, and beneficial to improving DNN-BNF based subword modeling. The hidden representation of speaker adversarial training contains less speaker-dependent information compare to that without adversarial training. Combining out-of-domain ASR based adaptation and adversarial training contributes to further improvement, in which our best performance (10.3%/7.8%) is achieved. Adversarial training is not complementary to disentangled representation learning.

In Chapter 5, frame labeling approaches to improving unsupervised subword modeling are studied. The advantage of DPGMM-HMM frame labeling over conventional DPGMM is that DPGMM-HMM could model contextual information in speech frames. The proposed label filtering algorithm discards unreliable labels from DPGMM clustering. Frame labeling is also achieved by out-of-domain ASR decoding. Experimental results on ZeroSpeech 2017 demonstrate the advancement of DPGMM-HMM labels over DPGMM labels, and that label filtering could further improve the DPGMM-HMM approach. The system trained with both out-of-domain ASR based labels and DPGMM-HMM labels achieves better performance than that trained with either type of labels only. Combining different types of BNFs by vector concatenation leads to further performance improvement. The best performance achieved by our proposed approaches is 9.7% in terms of across-speaker ABX error rate. It is equal to the performance of the best submitted system in ZeroSpeech 2017 and better than other recently reported systems.

In Chapter 6, the use of out-of-domain language-mismatched phone recognizers in unsupervised unit discovery is presented. Multiple language-mismatched phone recognizers are used to decode and segment target speech utterances, and generate phone posteriorgram for segment clustering and labeling. Our proposed KL divergence based metric for analyzing linguistic relevance of discovered subword units is studied. Experimental results on OGI-MTS corpus demonstrate the efficacy of out-of-domain language-mismatched phone recognizers in unit discovery. Increasing the number of recognizers is beneficial to performance improvement. Our best system in terms of purity is comparable to the study in [25], while our approach is relatively simpler in implementation. Experiments also show that KL divergence-based evaluation metric is consistent with purity. The deviation between a discovered unit and its closest ground-truth phone is comparable to the inherent variability of the phone. While in general the unit discovery results have a good coverage of linguistically-defined phones, there are a few exceptions, e.g. /er/ in Mandarin, probably due to limited feature representation capacity in the current discovery system. The confusion between ground-truth phones can be alleviated with a large cluster number.

## 7.2   Future work

We suggest the following directions for future work:

**Pair-wise learning in unsupervised subword modeling**

It has been shown by previous studies that pair-wise information is beneficial to learning speaker-invariant and subword-discriminative feature representation. There has been no study on investigating and comparing pair-wise learning approach with other approaches in speaker-invariant feature learning. It is believed that our best system could be further improved by incorporating pair-wise learning, by using a cascade system, or adding a triplet loss to the cross-entropy loss function in our MTL-DNN framework.

**Unsupervised lexical modeling and ASR**

The present study is focused on discovering and modeling basic subword units of zero-resource languages. Future work may move a step further to lexical modeling and language modeling, and to develop an integrated ASR system for zero-resource languages. Recently, there are studies on unsupervised lexical modeling [22], unsupervised ASR [12, 13, 22], and unsupervised large-vocabulary ASR [130]. The Bayesian model [130] and the generative adversarial network (GAN) [12] are two possible frameworks. It is believed that our findings on unsupervised subword modeling and unit discovery serve as a basis to improving unsupervised lexical modeling and ASR.

# Appendix A

# Published work

**Journal article**

1. **Siyuan Feng** and Tan Lee. Exploiting Cross-Lingual Speaker and Phonetic Diversity for Unsupervised Subword Modeling. In *IEEE/ACM Trans. Audio, Speech, Lang. Process.*, vol 27, no. 12, pp. 2000-2011, 2019.

**Peer-reviewed conference proceedings**

1. **Siyuan Feng** and Tan Lee. 2019. Improving Unsupervised Subword Modeling via Disentangled Speech Representation Learning and Transformation. In *Proc. INTERSPEECH*, 2019, pp. 1093-1097. [Oral]

2. **Siyuan Feng**, Tan Lee and Zhiyuan Peng. 2019. Combining Adversarial Training and Disentangled Speech Representation for Robust Zero-Resource Subword Modeling. In *Proc. INTERSPEECH*, 2019, pp. 281–285. [Oral]

3. **Siyuan Feng** and Tan Lee. 2018. Exploiting speaker and phonetic diversity of mismatched language resources for unsupervised subword modeling. In *Proc. INTERSPEECH*, 2018, pp. 2673–2677. [Oral]

4. **Siyuan Feng** and Tan Lee. 2018. Improving cross-lingual knowledge transferability using multilingual TDNN-BLSTM with language-dependent pre-final layer. In *Proc. INTERSPEECH*, 2018, pp. 2439–2443. [Poster]

5. **Siyuan Feng** and Tan Lee. 2017. On the linguistic relevance of speech units learned by unsupervised acoustic modeling. In *Proc. INTERSPEECH*, 2017, pp. 2068–2072. [Oral]

A. Published work

6. **Siyuan Feng**, Tan Lee and Haipeng Wang. 2016. Exploiting language-mismatched phoneme recognizers for unsupervised acoustic modeling. In *Proc. ISCSLP*, 2016, pp. 1–5. [Oral]

7. Zhiyuan Peng*, **Siyuan Feng*** and Tan Lee. 2018. Adversarial Multi-Task Deep Features and Unsupervised Back-End Adaptation for Language Recognition. In *Proc. ICASSP*, 2019, pp. 5961–5965. [Poster]

8. Ying Qin, Tan Lee, **Siyuan Feng** and Anthony Pak Hin Kong. 2018. Automatic speech assessment for people with aphasia using TDNN-BLSTM with multi-task learning. In *Proc. INTERSPEECH*, 2018, pp. 3418–3422. [Poster]

9. Man-Ling Sung, **Siyuan Feng** and Tan Lee. 2018. Unsupervised pattern discovery from thematic speech archives based on multilingual bottleneck features. In *Proc. APSIPA ASC*, 2018, pp. 1448–1455. [Oral]

10. Yuanyuan Liu, Ying Qin, **Siyuan Feng**, Tan Lee and P.C. Ching. 2018. Disordered speech assessment using Kullback-Leibler divergence features with multi-task acoustic modeling. In *Proc. ISCSLP*, 2018, pp. 61–65. [Oral]